\pdfoutput=1
\documentclass[english,10pt]{article}

\usepackage[left=.86in,right=.86in,bottom=.86in,top=.86in]{geometry}

\usepackage[utf8]{inputenc}
\usepackage{babel}
\usepackage{slantsc}
\usepackage{array}
\usepackage{amsmath}
\usepackage{amsthm}
\usepackage{amssymb}
\usepackage{graphicx}
\usepackage{enumerate}
\usepackage{mathtools}
\usepackage{natbib}
\usepackage{bbm}
\usepackage{tikz}
\usepackage{multirow}
\usetikzlibrary{tikzmark}
\usepackage{subfigure}
\usepackage{xspace}
\usepackage{bigints}
\usepackage{booktabs}
\usepackage[flushleft]{threeparttable}
\usepackage{pdflscape}

\usepackage{pgf}
\usepackage{pgfplots}
\pgfplotsset{compat=newest} 
\pgfplotsset{plot coordinates/math parser=false} 
\newlength\figureheight 
\newlength\figurewidth 

\usepackage{heuristica}
\usepackage[heuristica,vvarbb,bigdelims]{newtxmath}
\usepackage[T1]{fontenc}
\usepackage{textcomp} 
\usepackage{fix-cm}

\newtheorem{theorem}{Theorem}
\newtheorem{proposition}{Proposition}

\newtheorem{corollary}{Corollary}

\newtheorem{lemmaapp}{Lemma}[section]

\newtheorem{propapp}{Proposition}[section]

\newtheorem{claimapp}{Claim}[section]
\newtheorem*{theorem*}{Theorem}
\theoremstyle{definition}
\newtheorem{definition}{Definition}
\newtheorem{example}{Example}
\newtheorem{assumption}{Assumption}

  \newtheorem*{rep@theorem}{\rep@title}
\newcommand{\newreptheorem}[2]{%
\newenvironment{rep#1}[1]{%
 \def\rep@title{#2 \ref{##1}}%
 \begin{rep@theorem}}%
 {\end{rep@theorem}}}
\newreptheorem{lemma}{Lemma}

\usepackage[labelsep=period]{caption}

\DeclareMathOperator*{\diag}{diag}

\DeclareMathOperator*{\argmin}{arg\,min}

\DeclareMathOperator*{\tr}{tr}

\makeatletter
\newcommand{\subalign}[1]{%
  \vcenter{%
    \Let@ \restore@math@cr \default@tag
    \baselineskip\fontdimen10 \scriptfont\tw@
    \advance\baselineskip\fontdimen12 \scriptfont\tw@
    \lineskip\thr@@\fontdimen8 \scriptfont\thr@@
    \lineskiplimit\lineskip
    \ialign{\hfil$\m@th\scriptstyle##$&$\m@th\scriptstyle{}##$\crcr
      #1\crcr
    }%
  }
}
\makeatother

\definecolor{PennBlue}{RGB}{001,031,091}
\definecolor{PennRed}{HTML}{6e2f35}
\usepackage{hyperref}
\hypersetup{
	pdfborder = {0 0 0},
    colorlinks,
    citecolor=PennRed,
    filecolor=PennRed,
    linkcolor=PennRed,
    urlcolor=PennRed,
    breaklinks=true
}

\AtBeginDocument{}

\usepackage[blocks]{authblk}
\title{\fontsize{18}{18} 
\selectfont{Simple Models and Biased Forecasts}\thanks{This paper subsumes and substantially extends the second half of my job-market paper, titled ``Macroeconomics with Learning and Misspecification: A General Theory and Applications.'' I am grateful to Daron Acemoglu, Marios Angeletos, Larry Christiano, Ian Dew-Becker, Th\'{e}o Durandard, Marty Eichenbaum, Drew Fudenberg, Laura G\'{a}ti, Lars Hansen, Ben H\'{e}bert, Ali Jadbabaie, Diego K\"{a}nzig, Chen Lian, Alessandro Pavan, Giorgio Primiceri, Bruno Strulovici, Alireza Tahbaz-Salehi, Harald Uhlig, Iv\'{a}n Werning, and numerous seminar and conference participants for their helpful comments.
}
}
\author{\vspace{0.1in}\hspace{0.0in}\fontsize{13}{13}\selectfont  {Pooya Molavi}\thanks{ Northwestern University, \href{mailto:pmolavi@kellogg.northwestern.edu}{pmolavi@kellogg.northwestern.edu}.}
}
\date{\today
}

\AtBeginDocument{%
  \mathchardef\mathcomma\mathcode`\,
  \mathcode`\,="8000 
}
{\catcode`,=\active
  \gdef,{\mathcomma\discretionary{}{}{}}
}

\vfuzz \hfuzz
\allowdisplaybreaks

\usepackage[protrusion=true,expansion=true,final,babel]{microtype}
\begin{document}

\maketitle

\begin{abstract}
\fontsize{10}{10}\selectfont \baselineskip0.5cm
\noindent This paper proposes a framework in which agents are constrained to use simple models to forecast economic variables and characterizes the resulting biases. It considers agents who can only entertain state-space models with no more than $d$ states, where $d$ measures the intertemporal complexity of a model. Agents are boundedly rational in that they can only consider models that are too simple to capture the true process, yet they use the best model among those considered. Using simple models adds persistence to forward-looking decisions and increases the comovement among them. This mechanism narrows the gap between business-cycle theory and data. In a new neoclassical synthesis model, the assumption that agents use simple models fits the data much better than the rational-expectations hypothesis. Moreover, simple models simultaneously resolve the Barro-King and forward guidance puzzles while improving the propagation of TFP shocks.
\end{abstract}

\thispagestyle{empty}
\newpage 
\setcounter{page}{1}\fontsize{11.5}{11.5}\selectfont\baselineskip0.64cm

\section{Introduction}
When faced with the difficult task of forecasting in a complex world, people tend to rely on simple models and past experiences. Yet the rational-expectations hypothesis assumes that agents can forecast the future as if they knew the true model of the economy. The unrealistic nature of the rational-expectations assumption would not be of great concern if the predictions of the standard macro models were robust to alternative specifications of expectations. However, the answers to many important questions in macroeconomics, ranging from the power of forward guidance to the response of the economy to aggregate supply and demand shocks, are sensitive to how agents form their expectations.

This paper studies the forecasting biases arising from individuals' reliance on simple models and the macroeconomic implications of those biases. I introduce a framework in which the true data-generating process features complex intertemporal relationships among variables, but agents can only entertain stochastic models with a bound on their intertemporal complexity. Specifically, they only consider stochastic processes that can be represented using a $d$-dimensional state variable, where $d$ is a parameter that measures the complexity of an agent's model. Agents are boundedly rational in that they can only entertain models that are too simple to capture the true process, but they find the best $d$-dimensional approximation to it.

The framework has sharp predictions for agents' forecasts and forward-looking actions. I show that agents misperceive \emph{intertemporal} statistical relationships among time-series variables. In particular, while agents can accurately forecast the most persistent components of those observables, they miss the dynamics of the other components. This bias increases the \emph{persistence} and \emph{comovement} of agents' forward-looking choices. Agents’ forecasts and forward-looking actions are anchored to the most persistent state variables in the economy. This increases the persistence of those forecasts and actions. Furthermore, different agents---with different payoffs, facing different decisions, and using models with different dimensions---all agree on the most persistent components. Since forward-looking decisions of such agents are influenced by a limited set of common components, those decisions comove more than they would under rational expectations.

This paper focuses on bounded rationality manifested as a reduction in intertemporal complexity. I make two main assumptions, which allow me to abstract from errors that agents might make when confronting other forms of complexity. First, I assume that agents are capable of entertaining any stochastic model that has a $d$-dimensional linear-Gaussian state-space representation. Second, the best model is defined as the stochastic process that minimizes the Kullback--Leibler divergence from the true process. While this approach deviates from the utility-based notion of model optimality prevalent in the rational-inattention literature, it aligns with the emerging literature on model misspecification in game theory (e.g., \cite{Esponda2014}). These simplifying assumptions result in a useful linear-invariance property: Expectations formed using simple models respect linear \emph{intratemporal} relationships among observables. Moreover, these assumptions enhance the framework's tractability, rendering it applicable even in large-scale macro models without adding to the computational burden or introducing additional degrees of freedom (beyond $d$).

I apply the framework to study the propagation of aggregate shocks in a general-equilibrium economy. Many business-cycle models have difficulty generating empirically plausible degrees of persistence and comovement in endogenous variables in response to shocks. A common solution pursued in the literature is to introduce auxiliary frictions such as habit formation and adjustment costs (e.g., \citet*{Christiano2005}). However, the resulting DSGE models often require implausibly large frictions to generate realistic business cycles. This paper proposes a novel and parsimonious alternative that relies on agents' bounded rationality. By replacing rational expectations with simple models, this approach increases persistence and comovement in a way that is universally applicable across applications. 

The paper's main applied contribution is to quantify the extent to which the additional persistence and comovement delivered by simple models can improve the empirical fit of business-cycle models. It does so in the context of a standard model economy, which combines elements of the New Keynesian and real business cycle models. The model economy features price and wage rigidities, endogenous capital formation subject to neoclassical adjustment costs, and realistic monetary and fiscal policies. However, it does not contain any of the add-ons often introduced in DSGE models to increase persistence and comovement: external habit formation in consumption, investment-adjustment costs, price and wage indexation, endogenous capital utilization, or a monetary policy that responds to the level and growth rate of the output gap. 

Without these add-ons, and under rational expectations, the model economy does not generate realistic impulse-response functions (IRFs). The IRFs to productivity shocks are essentially monotonic, not hump-shaped. Small monetary expansions lead to unrealistically large increases in output, consumption, investment, and inflation, a manifestation of the forward-guidance puzzle. Investment-demand shocks lead to a negative comovement between consumption and investment, the \cite{barro1984time} puzzle. These observations suggest that standard rational-expectations New Keynesian models cannot provide a satisfactory account of business cycles. 

Replacing rational expectations with simple models addresses these shortcomings. The responses to productivity shocks become hump-shaped. Expansionary monetary-policy shocks lead to much smaller increases in output, consumption, investment, and inflation---well within the range of estimated responses to identified monetary-policy shocks. Investment shocks generate a positive comovement between investment and consumption. Overall, the framework of simple models emerges as a plausible and parsimonious substitute for DSGE add-ons. Nonetheless, the extent to which simple models can narrow the gap between the business-cycle theory and data is ultimately an empirical question.

To answer this question, I use Bayesian estimation techniques. The parameters of the model economy are estimated separately under rational expectations and under the assumption that agents use one-dimensional simple models. A comparison of marginal likelihoods finds overwhelming evidence (by more than 150 log points) in favor of simple models. Perhaps more strikingly, no single standard DSGE add-on can increase the marginal likelihood by as much as simple models. The key to the empirical success of the boundedly rational model is its ability to reduce the impact of aggregate demand shocks on inflation and to generate comovement in quantities in response to those shocks. Posterior variance decomposition confirms this explanation. At business-cycle frequencies, demand shocks account for the vast majority of the variance in output, consumption, and investment, and a negligible portion of the variance in inflation or interest rates. 

\paragraph*{Related Literature.} This paper belongs to the  literature in macroeconomics on deviations from full-information rational expectations (FIRE)---see \citet*{Woodford2013} for a survey. The literatures on dispersed information, e.g., \cite{lucas1972expectations}, noisy information, e.g., \citet*{orphanides2003monetary} and \citet*{angeletos2009incomplete}, sticky information, e.g., \cite{Mankiw2002}, or costly attention, e.g., \cite{Sims2003}, \cite{woodford2003imperfect}, \cite{Mackowiak2009}, and \citet*{Gabaix2014}, deviate from the FIRE benchmark by imposing imperfect knowledge of the payoff-relevant variables.\footnote{See also \citet*{nimark2008dynamic}, \citet{lorenzoni2009theory},  \citet*{alvarez2015monetary},  \cite{Angeletos-Lian}, and \citet*{Angeletos-Huo-2021}.} This paper abstracts from the difficulty of observing a large cross-section of variables and instead focuses on the difficulty of comprehending complex time-series (or intertemporal) relationships. The predictions of this framework also distinguish it from the literature mentioned above: In my model, agents fully uncover cross-sectional relationships among variables, but their expectations could deviate from rational expectations even if the economy has a single exogenous shock. 

The paper also contributes to the literature that quantifies the consequences of bounded rationality using DSGE models. \cite{slobodyan2012learning} evaluate the empirical performance of a medium-scale DSGE model with agents who do adaptive learning and show that the adaptive learning model fits the data better than the rational-expectations model. \cite{Mackowiak2015} consider a model with rationally inattentive agents, whereas \citet*{Confidence} consider an island economy where agents experience shocks to their higher-order expectations. \citet*{chahrour2021sectoral} consider a multisector economy in which the deviation from FIRE arises from selective reporting by public media. And \citet*{bianchi2021diagnostic} estimate a New Keynesian model in which agents have diagnostic expectations. This paper differs from these earlier contributions in its focus on intertemporal complexity and dimension reduction as bounded rationality. Furthermore, the quantitative macro model considered here is a standard New Keynesian model with a single belief parameter: 
the dimension $d$ of agents' models. The remaining parameters are standard preference, technology, and policy parameters, which are estimated using standard Bayesian techniques.

A large literature studies the question of whether households, firms, and professional forecasters under- or over-extrapolate from new information. \cite{coibion2015information} provide evidence of under-extrapolation in consensus forecasts for professional forecasters. \citet*{bordalo2020overreaction} show that individual forecasts of professional forecasters over-extrapolate from recent news. \citet*{angeletos2021imperfect} find evidence of under-extrapolation at short horizons and over-extrapolation at longer horizons, whereas \citet*{afrouzi2020overreaction} find evidence of over-extrapolation in a lab setting. More recently, \cite{broer2020forecaster} find evidence for both under- and over-extrapolation depending on the aggregate variable being studied. In parallel with this empirical literature, many papers have proposed theoretical models of under- and over-extrapolation. Natural expectations, e.g.,  \citet*{Fuster2010} and \citet*{Fuster2012}, and diagnostic expectations, e.g., \citet*{Bordalo2018}, are examples of models where agents over-extrapolate from the recent past. Cognitive discounting of \cite{Gabaix2020} and level-$k$ thinking, e.g., \citet*{garcia2019low} and \cite{farhi2019monetary}, are examples of models that feature under-extrapolation. The framework proposed in this paper is neither a model of under-extrapolation nor of over-extrapolation; agents who use simple models always under-extrapolate some observables and over-extrapolate others.

This paper also contributes to the literature that studies the properties of pseudo-true models. The term pseudo-true model originates in the pioneering work of \cite{Sawa}, who proposes using the Kullback--Leibler divergence as a model-selection criterion when models are misspecified. Agents in the restricted-perceptions equilibrium of \citet*{Bray1982} and \citet*{Bray1986}, \cite{Rabin2010}'s model of the gambler's fallacy, the natural-expectations framework of \citet*{Fuster2010} and \citet*{Fuster2012}, the Berk--Nash equilibrium of \citet*{Esponda2014,Esponda2016}, and the constrained rational-expectations equilibrium of \citet*{CREE} all use pseudo-true models to forecast payoff-relevant variables. Agents in \cite{Smith1998} also have a misspecified model of the economy since they believe that current and future prices do not depend on anything but the first few moments of the wealth distribution. However, despite this long history, surprisingly few general results on the properties of pseudo-true models have appeared in the literature. Such results are almost exclusively derived (with the notable exception of \cite{Rabin2010}) in settings where the set of models is sufficiently restricted that the pseudo-true model can be estimated using OLS regression and the bias in agents' forecasts reduces to the omitted-variable bias. I contribute to this literature by characterizing the set of pseudo-true state-space models of a given dimension.

The state-space models used in this paper are relatives of dynamic-factor models, e.g.,  \cite{stock2011dynamic,Stock2016}. However, the two offer two distinct ways of decomposing time-series data. Dynamic factor models decompose data into common factors and idiosyncratic disturbances, whereas state-space models decompose it into persistent and transitory components. The two approaches thus suggest two different simplifications of large time-series data: using a small number of common factors in the former case and a small number of persistent states in the latter.\footnote{The sets of time series that can be represented by dynamic-factor and state-space models are not nested. Instead, any finite dynamic-factor model has a state-space representation, and any finite state-space model has a dynamic-factor representation. See \cite{forni2001generalized} for a representation result for the (generalized) dynamic factor models.}

Finally, in a follow-up paper, \citet*{molavi2021model} use a closely related framework to study the implications of model misspecification for asset prices and returns. They show that constraining the complexity of investors' models leads to return and forecast-error predictability and provides a parsimonious account of several puzzles in the asset-pricing literature. 

\paragraph*{Outline.} The rest of the paper is organized as follows: Section \ref{sec:framework} presents the framework of simple models and formally defines and discusses the notion of fit used in the paper. Section \ref{sec:characterization} contains the paper's characterization results for simple models. Section \ref{sec:predictions} discusses the implications of using simple models for agents' forecasts and choices. Section \ref{sec:DSGE} presents a business-cycle application. Section \ref{sec:conclusion} concludes. Omitted details and additional results for the business-cycle application are relegated to three appendices. Some additional theoretical results as well as the proofs can be found in the online appendices.

\section{Framework}\label{sec:framework}

In this section, I present the general framework and the main behavioral assumption of the paper.

\subsection{Environment}
Time is discrete and is indexed by $t\in\mathbb{Z}$. An agent observes a sequence of variables over time and uses her past observations to forecast their future values. I let $y_t\in\mathbb{R}^n$ denote the time-$t$ value of the vector of observables, or simply the \emph{observable}. Vector $y_t$ follows a mean-zero stochastic process $\mathbb{P}$ with the corresponding expectation operator $\mathbb{E}[\cdot]$. I start by taking $\mathbb{P}$ as a primitive, but the process will be an endogenous outcome of agents' actions in the business-cycle application studied in Section \ref{sec:DSGE}.

I make several technical assumptions on the true process. First, $\mathbb{P}$ is purely non-deterministic, stationary, and ergodic, and has a finite second moment. Second, there exists a subspace $\mathcal{W}$ of $\mathbb{R}^n$ (possibly equal to $\mathbb{R}^n$ itself) such that $y_t$ is supported on $\mathcal{W}$ with density $\mathbbm{f}$.\footnote{This assumption is weaker than the assumption that $\mathbb{P}$ has full support over $\mathbb{R}^{n}$ because it allows for the possibility that the true process is degenerate. This additional level of generality will be useful in applications where the elements of $y_t$ may be linearly dependent.} Finally, the true process has finite entropy rate, i.e., $\lim_{t\to\infty}\frac{1}{t}\mathbb{E}\left[-\log \mathbbm{f}(y_1,\dots,y_t)\right]<\infty$. These assumptions are all quite weak. For instance, they are satisfied if $y_t$ follows a stationary vector ARMA process with Gaussian innovations.

The agent has perfect information about the past realizations of the observable; her time-$t$ information set is given by $\{y_{t},y_{t-1},\dots\}$. However,  she may use a misspecified model to map her information to her forecasts. This model misspecification leads to deviations in the agent's forecasts from those that arise in the rational-expectations benchmark. 

\subsection{Simple Models}
As the paper's main behavioral assumption, I assume that the agent is constrained to use state-space models with a small number of state variables to forecast the vector of observables. She can only entertain models of the form
\begin{equation}\label{eq:simple_model}
\begin{aligned}
    & z_t = A z_{t-1} + w_t,\\
    & y_{t} = B'z_t + v_t,
\end{aligned}
\end{equation}
where $z_t$ is the $d$-dimensional vector of \emph{subjective latent states}, $A\in\mathbb{R}^{d\times d}$, $w_t\in\mathbb{R}^d$ is i.i.d. $\mathcal{N}(0,Q)$, $B\in\mathbb{R}^{d\times n}$, $v_t\in\mathbb{R}^{n}$ is i.i.d. $\mathcal{N}(0,R)$, and $w_t$ and $v_t$ are independent. While the integer $d$ is a primitive of the model that parameterizes the dimension of the agent's models, matrices $A$, $B$, $Q$, and $R$ are parameters that are determined endogenously by maximizing the fit to the true process. Formally, I define a \emph{$d$-state model} as a stationary stochastic process over $\{y_t\}_{t=-\infty}^\infty$ that has a representation of the form \eqref{eq:simple_model} such that (i) the dimension of vector $z_t$ is $d$, (ii) $A$ is a convergent matrix, (iii) $Q$ is positive definite, and (iv) $R$ is positive semidefinite.\footnote{A matrix is \emph{convergent} if all of its eigenvalues are smaller than one in magnitude. $A$ being convergent and $Q$ being positive definite are sufficient for a model $(A,B,Q,R)$ to define a stationary ergodic process.} I let $P^\theta$ denote the $d$-state model parameterized by the collection of matrices $\theta\equiv(A,B,Q,R)$, let $E^\theta[\cdot]$ denote the corresponding expectation operator, and let $\Theta_d$ denote the set of all $d$-state models.\footnote{One can define the set of $d$-state models without any reference to the latent state $z_t$. Stochastic process $P$ for $\{y_t\}_{t=-\infty}^\infty$ with expectation operator $E$ is a $d$-state model if $E[y_ty_{t-l}']=CA^{l-1}\overline{C}'$ for all $l=1,2,\dots$, some convergent $d\times d$ matrix $A$, and some $C,\overline{C}\in\mathbb{R}^{n\times d}$. See, for instance, \cite{faurre1976stochastic} or \citet[Chapter~7]{katayama2005subspace}. I opt for the definition that uses the subjective latent state since $z_t$ will have an intuitive interpretation as agents' view of the state of the economy in the macro application I consider in this paper.} Whenever there is no risk of confusion, I use the term $d$-state model to refer both to the stochastic process $P^\theta$ for $y_t$ and the parameters $\theta=(A,B,Q,R)$ of its state-space representation.

The integer $d$ captures the agent's sophistication in modeling the stochastic process for the vector of observables, with larger values of $d$ indicating agents who can entertain more complex models. When $d$ is sufficiently large, the agent can approximate the unconditional and conditional second moments of any purely non-deterministic covariance-stationary process arbitrarily well using a model in her set of models. On the other hand, when $d$ is small relative to the number of states required to model the true process, no model in the agent's set of models will provide a good approximation to $\mathbb{P}$. The agent then necessarily ends up with a misspecified model of the true process and biased forecasts---regardless of which model in the set $\Theta_d$ she uses to make her forecasts. Characterizing this bias is the focus of the next section of the paper. 

My preferred rationale for the constraint on the number of states is to capture the agent's bounded rationality, but the constraint can also arise from the agent's rational fear of overfitting. Models with a large number of parameters and many degrees of freedom are prone to overfitting. Such concerns may lead rational agents to prefer more parsimonious statistical models, especially if they only have a short time series to draw upon when estimating the parameters of their model. In the remainder of the paper, I abstract away from any issues arising from small samples and instead consider the long-run limit where the sampling error vanishes.

\subsection{The Notion of Fit}
I assume that the agent forecasts using a model in the family of $d$-state models that provides the best fit to the true process. I use the Kullback--Leibler divergence rate of process $P^\theta$ from the true process $\mathbb{P}$ as the measure of the fit of model $\theta$.\footnote{The mean-squared forecast error is another commonly used notion of fit. In Online Appendix \ref{app:MSE}, I define the weighted mean-squared forecast error and show that it is equivalent to the Kullback--Leibler divergence rate under an appropriate choice of the weighting matrix.} The \emph{Kullback--Leibler divergence rate} (KLDR) of $P^\theta$ from $\mathbb{P}$ is denoted by $\text{KLDR}(\theta)$ and defined as follows. Recall that the true process is supported on a subspace $\mathcal{W}$ of $\mathbb{R}^n$. If $P^\theta$ is also supported on $\mathcal{W}$, then
\[
\text{KLDR}(\theta)\equiv \lim_{t\to\infty}\frac{1}{t}\mathbb{E}\left[\log\left(\frac{ \mathbbm{f}(y_1,\dots,y_t)}{f^\theta(y_1,\dots,y_t)}\right)\right],
\]
where $f^\theta$ denotes the density of $P^\theta$; if $P^\theta$ is not supported on $\mathcal{W}$, then $\text{KLDR}(\theta)\equiv +\infty$.

The Kullback--Leibler divergence rate is the natural generalization of Kullback--Leibler (KL) divergence to stationary stochastic processes. In the i.i.d. case, the KL divergence of a candidate model from the true model captures the difficulty of rejecting the candidate model in favor of the true model using a likelihood-ratio test. That is why the KL divergence is commonly used as a measure of a model's fit.\footnote{See, for instance, \cite{hansen2008robustness}.}  Similarly, $\text{KLDR}(\theta)$ captures the rate at which the power of a test for separating a stochastic process $P^\theta$ from the true process $\mathbb{P}$ approaches one as $t\to\infty$.\footnote{See, for instance, \cite{Shalizi2009}.} The KLDR is also tightly linked to asymptotics of Bayesian learning, as discussed in the following subsection.

Model $\theta\in\Theta_d$ is a \emph{pseudo-true} $d$-state model if $\text{KLDR}(\theta)\leq \text{KLDR}(\tilde{\theta})$ for all $\tilde{\theta}\in\Theta_d$. If the agent's set of models contains a model $\theta$ such that $f^\theta(y_1,\dots,y_t)=\mathbbm{f}(y_1,\dots,y_t)$ almost everywhere and for all $t$, then any pseudo-true $d$-state model is observationally equivalent to the true process.\footnote{Processes $P$ and $\tilde{P}$ are \emph{observationally equivalent} if all their finite-dimensional marginal distributions are identical.} The set of models $\Theta_d$ is then correctly specified. When no such $d$-state model exists, $\text{KLDR}(\theta)>0$ for any model $\theta\in\Theta_d$, and the set of models is misspecified. The following proposition states that the pseudo-true models are observationally equivalent to the true process when the set of models is correctly specified:
\begin{proposition}\label{prop:true}
Suppose the set $\Theta_d$ of $d$-state models is correctly specified. Then any pseudo-true $d$-state model $P^\theta$ is observationally equivalent to the true process $\mathbb{P}$.
\end{proposition}

The paper's focus is the misspecified case, where $d$ is small relative to the number of states required to capture the true process. This statement is about $d$ being smaller than the ``true $d$,'' not necessarily smaller than $n$, the dimension of $y_t$. However, it is often natural to also think of $d$ as much smaller than $n$. Approximating the true process by a pseudo-true $d$-state model then corresponds to using a parsimonious time-series model to capture the essential features of a large data set. Unless otherwise specified, I assume throughout the paper that $d\leq n$. However, the paper's characterization results easily generalize to the $d> n$ case.

\subsection{Learning Foundation}\label{app:learning}
Pseudo-true models arise naturally as the long-run outcome of learning by Bayesian agents with misspecified priors. Consider an agent who starts with prior $\mu_0$ with full support over the points in the set $\mathbb{R}^d\times \Theta_d$, each corresponding to an initial value of the subjective states $z_0$ and a $d$-state model $\theta$, which describes how states and the observable evolve over time. Suppose the agent observes $y_t$ over time and updates her belief using Bayes' rule. Let $\mu_t$ denote the agent's time-$t$ Bayesian posterior over $\mathbb{R}^d\times \Theta_d$. \cite{Berk1966}'s theorem establishes that, in the limit $t\to\infty$, the agent's posterior will assign a probability of one to the set of pseudo-true models.\footnote{While \cite{Berk1966} only covers the case of i.i.d. observations and parametric models, the result has been extended much more generally. \cite{Milhaud1998} and \cite{Kleijn2006} substantially extend \cite{Berk1966} by providing conditions for the weak convergence of posterior distributions and considering infinite-dimensional models. \cite{Shalizi2009}'s extension of Berk's theorem covers the case of non-i.i.d. observations and hidden Markov models.} 

This result offers an ``as if'' interpretation of the pseudo-true $d$-state models. One can assume that the agent has a subjective prior---which may be different from the true distribution---and updates her belief in light of new information  using Bayes' law. By Berk's theorem, as long as the agent's prior is supported on the set of $d$-state models, she will forecast the observable in the long run \emph{as if} she were using a pseudo-true $d$-state model. Focusing on pseudo-true models allows me to abstract away from learning dynamics and focus on the asymptotic bias caused by misspecification.\footnote{One can alternatively consider agents who estimate the parameters of their $d$-state models using a quasi-maximum-likelihood estimator. Such agents also will asymptotically forecast \emph{as if} they relied on the pseudo-true $d$-state models. See, for instance, Theorem 2 of \cite{douc2012}.}

The set of pseudo-true $d$-state models is independent of the agent's preferences. Instead, it only depends on the number of states the agent can entertain and the true stochastic process. The independence of the agent's pseudo-true models from her preferences is evident given the ``as if'' interpretation discussed above. Two agents who start with identical priors, observe the same sequence of observations, and update their beliefs using Bayes' rule will end up with identical posteriors at any point in time---irrespective of their preferences. Berk's theorem goes a step further by establishing that, in the long run, the posterior only depends on the support of the prior (not its fine details) and the distribution of observations (not their realizations).

The independence of the agent's pseudo-true models from her preferences has a significant consequence: The set of pseudo-true $d$-state models is generically disjoint from the set of $d$-state models that maximize the agent's payoff. However, this disparity is a feature, not a bug, of a positive theory of bounded rationality. While finding the payoff-maximizing model requires knowledge of the true process, one arrives at the set of pseudo-true models simply by following Bayes' rule---no knowledge of the true process is necessary. Following Bayes' rule would have led the agent to the truth had her model been correctly specified, but it can lead her astray in the presence of model misspecification. 

Agents' use of pseudo-true models should therefore be viewed as a positive statement---not a normative one. A pseudo-true $d$-state model is not what an agent should use for forecasting in order to maximize her payoff. It is what she will use to forecast in the long run if she starts with a prior over the set of $d$-state models and updates her belief using Bayes' rule.

\section{Characterization of Pseudo-True Models}\label{sec:characterization}
In this section, I characterize the set of pseudo-true $d$-state models, beginning with the case where $d=1$. As a preliminary step, I discuss a useful property of the pseudo-true models, which is of independent interest.

\subsection{The Invariance Property}
I begin with a result that shows the invariance of the pseudo-true $d$-state models to linear transformations of the observable. Consider an agent who, instead of observing vector $y_t\in\mathbb{R}^n$, observes vector $\tilde{y}_t=Ty_{t}\in\mathbb{R}^m$, where $T$ denotes an $m \times n$ matrix. As long as $T$ is a rank-$n$ matrix, $y_t$ and $\tilde{y}_t$ convey the exact same information. Thus, one might expect that the agent's beliefs when she observes $y_t$ are consistent with her beliefs when she instead observes $\tilde{y}_t$. 

The following definition formalizes the notion that two probability distributions are consistent with each other given a linear transformation of the observable. Let $T\in\mathbb{R}^{m\times n}$ be a matrix and $P$ be a probability distribution over infinite sequences in $\mathbb{R}^n$. The probability distribution over infinite sequences in $\mathbb{R}^m$ induced by $T$ and $P$ is denoted by $T(P)$ and defined as $T(P)(\mathcal{Y})\equiv P\left(\{y_t\}_{t=-\infty}^\infty: \{Ty_t\}_{t=-\infty}^\infty\in\mathcal{Y}\right)$ for any measurable set $\mathcal{Y}\subseteq {\mathbb{R}^m}^{\mathbb{Z}}$.\footnote{The probability distribution induced by a mapping is formally known as the \emph{pushforward measure}.} If the observable $y_t$ follows the stochastic process $\mathbb{P}$, then its linear transformation $\tilde{y}_t=Ty_t$ follows the transformed process $T(\mathbb{P})$. The following result establishes that transforming the observable by a rank-$n$ matrix leads the set of pseudo-true models to be transformed accordingly:

\begin{theorem}[linear invariance]\label{thm:linear_invariance}
Suppose $T\in\mathbb{R}^{m\times n}$ is a rank-$n$ matrix. Then $P^\theta$ is a pseudo-true $d$-state model given true model $\mathbb{P}$ if and only if $T\left(P^\theta\right)$ is a pseudo-true $d$-state model given true model $T\left(\mathbb{P}\right)$.
\end{theorem}

The result shows that an agent using simple models can discern all linear intratemporal relationships among the observables while facing significant constraints in understanding complex intertemporal relationships. While arguably stark, this dichotomy highlights the paper's premise that forecasting is challenging because it requires forecasters to recognize stochastic patterns that unfold over time. The result makes it possible to abstract from the cognitive costs of acquiring information about a large cross-section of variables and the mistakes individuals make when dealing with cross-sectional complexity, allowing me to instead concentrate on time-series complexity.

The linear invariance property makes the predictions of the framework invariant to the exact specification of the variables included in the vector of observables. The agent's pseudo-true models and forecasts only depend on the observables' information content, not on how that information is presented. For instance, whether the agent observes the nominal interest rate and the inflation rate or the real interest rate and the inflation rate is immaterial to how she forms her expectations. Likewise, the agent's expectations remain unchanged if the vector of observables is augmented with linear combinations of variables already in her information set.

The theorem thus suggests that it is without loss to assume that the vector of observables is free of redundant variables. Define the lag-$l$ autocovariance matrix of the true process as follows:
\begin{equation}
    \Gamma_{l}\equiv \mathbb{E}[y_t y'_{t-l}].
\end{equation}
When $y_t$ includes redundant variables, the variance-covariance matrix $\Gamma_0$ is singular, and the true process is degenerate.\footnote{A probability distribution on a space is said to be \emph{degenerate} if it is supported on a manifold of lower dimension.} In such cases, a lower-dimensional vector $\tilde{y}_t$ and a full-rank matrix $T$ exist such that $\mathbb{E}[\tilde{y}_t\tilde{y}_t']$ is non-singular and  $y_t=T\tilde{y}_t$. Therefore, by Theorem~\ref{thm:linear_invariance}, the pseudo-true models given $y_t$ can be found by first finding the pseudo-true models given $\tilde{y}_t$ and then applying transformation $T$. This observation implies that there is no loss of generality in assuming that the variance-covariance matrix $\Gamma_0$ is non-singular and that the agent only considers subjective models with non-singular variance-covariance matrices.\footnote{Whenever the true variance-covariance matrix $\Gamma_0$ is non-singular, any subjective model with a singular variance-covariance matrix is dominated in terms of the fit to the true process by every subjective model with a non-singular variance-covariance matrix. Therefore, no subjective model with a singular variance-covariance matrix can be a pseudo-true model.} I maintain these assumptions throughout the rest of the paper. 

\subsection{Pseudo-True One-State Models}
I start the analysis of pseudo-true models by considering the case where the agent can only entertain one-state models. In this case, a complete characterization of the agent's pseudo-true models is possible. The insights from the single-state case generalize to the $d$-state case, as discussed later in this section.

The agent's pseudo-true one-state forecasts turn out to depend on the true process only through the unconditional variance and the autocorrelation structure of the vector of observables. The autocorrelations are measured by a novel set of  objects, which I refer to as autocorrelation matrices. I define the lag-$l$ \emph{autocorrelation matrix} of the observable under the true process as follows:\footnote{Here and throughout the paper, I follow the usual convention that, for a symmetric positive definite matrix $X$, the square-root matrix $X^{\frac{1}{2}}$ is the unique symmetric positive definite matrix that satisfies $X^{\frac{1}{2}}X^{\frac{1}{2}}=X$.}
\begin{equation}
    C_l \equiv \frac{1}{2}\Gamma_0^{\frac{-1}{2}}\left(\Gamma_l + \Gamma_l'\right)\Gamma_0^{\frac{-1}{2}}.
\end{equation}

The concept of autocorrelation matrices naturally extends the idea of autocorrelation functions. If the observable $y_t$ is a scalar, $C_l$ simplifies to the standard autocorrelation function at lag~$l$. However, when the observable is an $n$-dimensional vector, $C_l$ is an $n\times n$ real symmetric matrix with eigenvalues inside the unit circle.\footnote{See Lemma \ref{lem:spectral_radius_of_autocorrelation} in the Online Appendix for a proof.} Autocorrelation matrices capture the extent of serial correlation in the vector of observables. When the spectral radius of $C_l$ is close to zero for all $l$, the process is close to being i.i.d., whereas when the spectral radius of $C_l$ is close to one, then the process is close to being unit root.\footnote{The spectral radius $\rho(X)$ of matrix $X$ denotes the maximum among the magnitudes of eigenvalues of $X$.} 

With the definition of autocorrelation matrices at hand, I can state the general characterization result for the $d=1$ case:
\begin{theorem}\label{thm:1_state_general}
Under any pseudo-true one-state model $\theta$, the agent's $s$-period-ahead forecast is given by 
\begin{equation}
E^\theta_t[y_{t+s}]= {a}^{s}(1-\eta)q{p}'\sum_{\tau=0}^\infty {a}^\tau {\eta}^\tau y_{t-\tau},\label{eq:1-state-s_ahead_forecast_body}
\end{equation}
where $a$ and $\eta$ are scalars in the $[-1,1]$ and $[0,1]$ intervals, respectively, that maximize $\lambda_{\text{max}}(\Omega(\tilde{a},\tilde{\eta}))$, the largest eigenvalue of the $n\times n$ real symmetric matrix
\[
\Omega(\tilde{a},\tilde{\eta}) \equiv -\frac{\tilde{a}^2(1-\tilde{\eta})^2}{1-\tilde{a}^2\tilde{\eta}^2}I + \frac{2(1-\tilde{\eta})(1-\tilde{a}^2\tilde{\eta})}{1-\tilde{a}^2\tilde{\eta}^2}\sum_{\tau=1}^\infty \tilde{a}^\tau\tilde{\eta}^{\tau-1} C_\tau,
\]
and $p=\Gamma_0^{\frac{-1}{2}}u$ and $q=\Gamma_0^{\frac{1}{2}}u$, where $u$ is an eigenvector of $\Omega(a,\eta)$ with eigenvalue $\lambda_{\text{max}}(\Omega(a,\eta))$, normalized so that $u'u=1$.
\end{theorem}

The endogenous variables $a$, $\eta$, $p$, and $q$ have intuitive meanings. The scalar $a$ represents the persistence of the subjective latent state. If $a=0$, the subjective state is i.i.d., whereas if $a=1$, it follows a unit-root process.\footnote{The theorem does not rule out the possibility that $|a|=1$, in which case the corresponding state-space model might not be stationary ergodic. However, Lemma \ref{lemapp:1-state-stationary-ergodic} in the Online Appendix establishes that any pseudo-true one-state model inherits the stationarity and ergodicity of the true process.} The scalar $\eta$ captures the perceived noise in the agent's observations of the subjective state. When $\eta$ is small, the agent believes recent observations to be highly informative of the value of the subjective state. As a result, her expectations respond more to recent observations and discount old observations more. The vector $p$ determines the agent's relative attention to different components of the vector of observables. When $p_i$ is larger than $p_j$, the agent puts more weight on $y_{i,t-\tau}$ relative to $y_{j,t-\tau}$ for all $\tau$ when forming her estimate of the subjective state. Finally, the vector $q$ captures the relative sensitivity of the agent's forecasts of different observables to changes in her estimate of the subjective state. When $q_i$ is larger than $q_j$, then a change in the estimated value of the state at time $t$ leads the agent to change her forecast of $y_{i,t+s}$ by more than her forecast of $y_{j,t+s}$ for all $s$.

It follows standard Kalman filter results that the agent's forecasts take the form of equation~\eqref{eq:1-state-s_ahead_forecast_body} for \emph{some} $a$, $\eta$, $p$, and $q$. The substance of the result is rather characterizing the $(a,\eta,p,q)$ tuple that lead to a model with minimal KLDR from the true process. The theorem suggests a tractable way of computing the pseudo-true one-state forecasts in any stationary and ergodic environment given only the knowledge of the true autocorrelation matrices.

The theorem significantly reduces the computational complexity of finding the set of pseudo-true models. It concentrates out all parameters in the agent's models except for two scalars. As a result, the optimization problem simplifies from a problem over a $2n$-dimensional non-compact manifold to a much simpler problem over a two-dimensional compact rectangle.\footnote{The set of all $d$-state models is a non-compact manifold of dimension $2nd$ \citep*{Gevers1984}. Additionally, the KLDR is a non-convex function of $\theta=(A,B,Q,R)$.} Furthermore, since the size of the problem is independent of $n$, it can be solved efficiently in any application, regardless of the dimension of the vector of observables. 

The next result characterizes the perceived variance-covariance matrix of the observable under the pseudo-true one-state models:
\begin{theorem}\label{thm:1-state-covariance}
Given any pseudo-true one-state model $\theta$, the subjective variance-covariance of the vector of observables, $E^\theta[y_ty_t']$, coincides with the true variance-covariance matrix, $\Gamma_0\equiv\mathbb{E}[y_ty_t']$.
\end{theorem}

The theorem hinges on two main assumptions: First, there are no constraints on the agent's set of models other than the bound on the number of subjective state variables; put differently, matrices $A$, $B$, $Q$, and $R$ of representation \eqref{eq:simple_model} are unrestricted other than the constraint on their dimension. This flexibility allows the agent to represent any cross-sectional correlation pattern by an appropriate selection of matrices $A$, $B$, $Q$, and $R$. Second, the agent uses a model that minimizes the KLDR from the true process. This leads her to a set of such matrices that perfectly capture the true cross-sectional correlations.

Theorems \ref{thm:1_state_general} and \ref{thm:1-state-covariance} fully characterize the pseudo-true one-state models in terms of the true variance-covariance matrix $\Gamma_0$ and the tuple $(a,\eta,p,q)$, which in turn only depends on the true autocorrelation matrices $\{C_l\}_{l=1}^\infty$. Any unconditional or conditional moment of the pseudo-true one-state model can, in turn, be found in terms of $\Gamma_0$ and $(a,\eta,p,q)$.

\subsection{Pseudo-True One-State Models Under Exponential Ergodicity}
The pseudo-true one-state models can be found in closed form given a class of true stochastic processes that naturally arise in applications. The appropriate class turns out to be the following:
\begin{definition}\label{def:exp_erg}
A stationary ergodic process $\mathbb{P}$ is \emph{exponentially ergodic} if $\rho(C_l)\leq \rho(C_1)^{l}$ for all $l\geq 1$, where $\rho(C_l)$ denotes the spectral radius of $C_l$.
\end{definition}
Exponential ergodicity is stronger than ergodicity. Ergodicity requires that the serial correlation at lag $l$ decays to zero as $l\to\infty$. Exponential ergodicity requires the rate of decay to be faster than $\rho(C_1)$. Although exponentially-ergodic processes only constitute a subset of the class of stationary ergodic processes, many standard processes are exponentially ergodic. For instance, the vector of observables follows an exponentially-ergodic process if it is a spanning linear combination of $n$ independent AR(1) shocks. 

The following result characterizes the agent's pseudo-true one-state forecasts when the true process is exponentially ergodic. It links the agent's forecasts to the eigenvalues and eigenvectors of the true autocorrelation matrix at lag one:
\begin{theorem}\label{thm:1-state-exp-erg-closed-form}
Suppose the true process is exponentially ergodic. Under any pseudo-true one-state model $\theta$, the agent’s $s$-period-ahead forecast is given by
\begin{equation}
E^\theta_t[y_{t+s}] = {a}^{s}q{p}'y_t,
\end{equation}
where $a$ is an eigenvalue of $C_1$ largest in magnitude, $u$ denotes the corresponding eigenvector normalized so that $u'u=1$, and $p=\Gamma_0^{\frac{-1}{2}}u$ and $q=\Gamma_0^{\frac{1}{2}}u$.
\end{theorem}

A remarkable feature of the characterization in Theorem \ref{thm:1-state-exp-erg-closed-form} is that the agent's forecasts only depend on the last realization of the observable (and not its lags). In other words, the pseudo-true one-state model is \emph{Markovian} if the true process is exponentially ergodic. This property might come as a surprise in light of the fact that in the correctly-specified case forecasts obtained using the stationary Kalman filter generically use the entire history of the observable. The seeming discrepancy between the two results is due to misspecification of the agent's set of models in Theorem \ref{thm:1-state-exp-erg-closed-form}, as illustrated by the following example:

\begin{example}
Suppose the observable is scalar and follows an $\text{AR}(\infty)$ process: $y_{t+1} = \sum_{\tau=1}^\infty\phi_\tau y_{t+1-\tau}$.\footnote{Such a representation exists for generic processes in the class of mean-zero, purely non-deterministic, and stationary processes.} It is then immediate that the one-step-ahead forecast of the observable under the true, correctly-specified model is given by
\[
\mathbb{E}_t[y_{t+1}]=\sum_{\tau=1}^\infty\phi_\tau y_{t+1-\tau}.
\]
Contrast this with what an agent can do when she is constrained to use (misspecified) one-state models. Under any such model $\theta$, the agent's one-step-ahead forecast takes a similar form:
\[
E_t^\theta[y_{t+1}]=\sum_{\tau=1}^\infty\alpha_\tau y_{t+1-\tau}.
\]
However, the restriction to one-state models constrains coefficients $\{\alpha_\tau\}_{\tau=1}^\infty$  to be given by $\alpha_\tau=(1-\eta)a^\tau\eta^{\tau-1}$ for some $a\in[-1,1]$, some $\eta\in[0,1]$, and all $\tau$. Therefore, the pseudo-true one-state model is the model that picks $\{\alpha_\tau\}_{\tau=1}^\infty$ to minimize the KLDR subject to the constraint that $\alpha_\tau=(1-\eta)a^\tau\eta^{\tau-1}$ for all $\tau$. The agent wants to set $\alpha_\tau$ to a value that is related to the correlation of $y_{t+1}$ and $y_{t-\tau}$, but the constraint prevents her from fine-tuning the $\{\alpha_\tau\}_{\tau=1}^\infty$ coefficients. When the true process is exponentially ergodic, $y_{t+1}$ is much more correlated with $y_t$ than it is with lags of $y_t$. Then, the best such a constrained agent can do is to fine-tune the coefficient of $y_t$ and entirely disregard its lags. In other words, the constrained minimizer of the KLDR is Markovian even though the unconstrained minimizer is not.
\end{example}

The next example illustrates the use of Theorem \ref{thm:1-state-exp-erg-closed-form} in the context of a commonly-used process:
\begin{example}\label{eg:diagonal}
Suppose the true process $\mathbb{P}$ has the following representation:
\begin{equation}\label{eq:true_model_VAR_body}
\begin{aligned}
    & f_{t} = Ff_{t-1} + \epsilon_t,\\
    & y_t = H'f_t,
\end{aligned}
\end{equation}
where $\epsilon_t\sim \mathcal{N}(0,\Sigma)$, $F=\diag(\alpha_1,\alpha_2,\dots,\alpha_n)$, $\Sigma=\diag\left(\sigma_1^2,\sigma_2^2,\dots,\sigma_n^2\right)$, $H\in\mathbb{R}^{n\times n}$ is an invertible square matrix, and $1>|\alpha_1|>|\alpha_2|>\dots>|\alpha_n|>0$. It is easy to verify that $\rho(C_l)=|\alpha_1|^l=\rho(C_1)^l$; that is, the true process is exponentially ergodic. Therefore, Theorem \ref{thm:1-state-exp-erg-closed-form} can be used to characterize the pseudo-true one-state forecasts. The persistence, noise, relative attention, and relative sensitivity are, respectively, given by $a=\alpha_1$, $\eta=0$, $p=(H'VH)^{\frac{1}{2}}H^{-1}V^{-1}e_1$, and $q=(H'VH)^{\frac{-1}{2}}H'V e_1$, where $V \equiv (I-F^2)^{-1}\Sigma$ is the variance-covariance matrix of $f_t$ and $e_1$ denotes the first coordinate vector.\footnote{See the proof of Lemma \ref{lem:Akaike} for a derivation.}

The agent's forecasts take a particularly simple form when $H$ is the identity matrix, i.e., $y_{it}=f_{it}$ for $i=1,\dots,n$. Then, $p$ and $q$ are both multiples of the first coordinate vector $e_1$, and the agent's forecasts simplify to 
\begin{align*}
    & E_t^\theta[y_{1,t+s}] = \alpha_1^s y_{1t} = \mathbb{E}_t[y_{1,t+s}],\\
    & E_t^\theta[y_{i,t+s}] = 0, \qquad \forall i\neq 1.
\end{align*}
The agent's forecast of the most persistent element of the vector of observables coincides with its rational-expectations counterpart, but she forecasts every other element of the observable as if it were i.i.d.
\end{example}

A noteworthy feature of the pseudo-true model in Example \ref{eg:diagonal} is that the persistence parameter $a$ does not depend on the volatilities of the underlying AR(1) processes. The agent uses the subjective latent state to track the most persistent component of $y_t$, even if the most persistent component has a small variance. However, this result should not come as a surprise given the linear-invariance result: One can always equalize the volatilities of different components of $y_t$ by an appropriate linear transformation of the observable without altering the persistence of the subjective latent state in the agent's pseudo-true model. Therefore, the persistence parameter cannot depend on the volatilities.

The example also illustrates that the agent exhibits a form of \emph{persistence bias}. She forecasts the most persistent component of the vector of observables as accurately as under rational expectations but misses the dynamics of the other components. The intuition for the result is easiest to see when the most persistent component is close to being unit root. In that case, poorly tracking the most persistent component would lead to persistent mistakes in the agent's forecasts. The persistence of those mistakes would make them costly from the point of view of KLDR minimization. Therefore, any pseudo-true model tracks the component close to unit root as best possible, even if doing so results in errors in forecasting the other components. In Section \ref{subsec:persistence}, I generalize the insight of this example by formally establishing persistence bias in a more general context.\footnote{\citet*{Bidder-Dew-Becker2016} and \citet*{DEWBECKER2019461} propose an alternative reason agents might focus on tracking the most persistent components of a payoff-relevant variable. \citet*{Bidder-Dew-Becker2016} show that long-run risk is the worst case scenario for ambiguity-averse agents. \citet*{DEWBECKER2019461} show that, as a result, ambiguity-averse agents will learn most about dynamics at the lowest frequencies.}

Example \ref{eg:diagonal} can be generalized by relaxing the assumption that matrices $F$ and $\Sigma$ are diagonal and allowing for non-Gaussian innovations. The key requirement for the process to be exponentially ergodic is that matrix $H$ in representation \eqref{eq:true_model_VAR_body} is
full rank. This assumption can be seen as a full-information (or spanning) assumption. If the agent observes an observable of the form \eqref{eq:true_model_VAR_body} with a full-rank matrix $H$, then she has enough information to forecast the observable as well as in the full-information rational-expectations benchmark—even if she fails to do so due to the constraint on her set of models. See Online Appendix \ref{app:exp_erg} for a detailed discussion of this generalization.

\subsection{Pseudo-True $d$-State Models}\label{sec:d-state}
I end this section by discussing how the insights from the $d=1$ case generalize when $d>1$. To characterize the pseudo-true $d$-state models, one needs to find models $\theta=(A,B,Q,R)$ that minimize the KLDR from the true process. Doing so requires minimizing a non-convex function over a non-compact set, consisting of all the matrices $A$, $B$, $Q$, and $R$ of appropriate dimensions. This problem does not lend itself to an analytical solution without further restrictions.

To address this issue, I restrict the models the agent considers to be Markovian. A $d$-state model~$\theta$ is \emph{Markovian} if $P^\theta$ satisfies the Markov property, i.e., $P^\theta(y_{t+1}|y_t,y_{t-1},\dots)=P^\theta(y_{t+1}|y_t)$. An agent who believes the observable follows a Markovian $d$-state model believes that (a) the current realization of the observable contains all the information required for forecasting, and (b) all the relevant information contained in $y_t$ can be summarized by a $d$-dimensional state variable. The following proposition provides a necessary and sufficient condition for a model to be Markovian:
\begin{proposition}\label{prop:Markovian}
Let $\text{Var}^\theta_t(y_{t+1})$ denote the variance-covariance matrix of $y_{t+1}$ given model $\theta$ and conditional on the history $\{y_{\tau}\}_{\tau\leq t}$ of the observable, and let $\text{Var}^\theta(y_{t+1}|z_t)$ denote the corresponding variance-covariance matrix conditional on the time-$t$ realization of the subjective latent state. $ \text{Var}^\theta_t(y_{t+1})\succeq\text{Var}^\theta(y_{t+1}|z_t)=B'QB+R$ for any $d$-state model $\theta$, with equality if and only if $\theta$ is Markovian.\footnote{I use the usual convention that $X\succeq Y$ for symmetric positive semidefinite matrices $X$ and $Y$ if $X-Y$ is positive semidefinite.}
\end{proposition}

The proposition highlights an intuitive property of Markovian models. Note that the agent can observe the history $\{y_{\tau}\}_{\tau\leq t}$ of the observable but not the subjective latent state $z_t$. The first part of the proposition shows that the agent cannot forecast any better than if she knew the realization of the latent subjective state. In other words, the forecast error given $z_t$ provides a lower bound on the forecast error given $\{y_{\tau}\}_{\tau\leq t}$. The second part of the proposition shows that the agent can achieve this lower bound when her model of the world is Markovian. She can then forecast as well as an agent who knows the latent subjective state, because all the relevant information in the latent state can be extracted from the realized history of the observable. Markovian models can thus be seen as models that feature \emph{full information}.

Markovian models constitute only a subset of the class of all state-space models of a given dimension. However, the pseudo-true one-state models happen to be Markovian when the true process is exponentially ergodic, as shown by the following corollary of Theorem \ref{thm:1-state-exp-erg-closed-form}:
\begin{corollary}
If the true process is exponentially ergodic, then any pseudo-true one-state model is Markovian.
\end{corollary}
The result shows that constraining the agent to Markovian models is without loss when $d=1$ and the true process is exponentially ergodic. Even with the flexibility to choose non-Markovian models, an agent who is attempting to minimize the KLDR from an exponentially ergodic process settles on a Markovian model. Whether this result continues to hold for $d$-state models with $d>1$ remains an open question. However, I can still make progress by taking the restriction to Markovian models as an assumption and characterizing the resulting pseudo-true models. A Markovian $d$-state model $\theta$ is a \emph{pseudo-true Markovian $d$-state model} if $\text{KLDR}(\theta)\leq \text{KLDR}(\tilde{\theta})$ for any Markovian $d$-state model~$\tilde{\theta}$. 

The pseudo-true Markovian models have a number of appealing properties. 
They satisfy a version of the linear-invariance result of Theorem \ref{thm:linear_invariance}. They share Bayesian and quasi-maximum-likelihood learning foundations with other pseudo-true models. Perhaps most importantly, they can be fully characterized in closed-form in some useful cases:
\begin{theorem}\label{thm:d-state-m-i-o}
Suppose either $d=1$ or the lag-one autocovariance matrix is symmetric. Then the following statements hold:
\begin{itemize}
\item[(a)] Under any pseudo-true Markovian $d$-state model $\theta$, the agent’s $s$-period-ahead forecast is given by
\begin{equation}
E^{\theta}_t[y_{t+s}] = \sum_{i=1}^d{a_i}^{s}q_i{p'_i}y_t,
\end{equation}
where $a_1,\dots,a_d$ are $d$ eigenvalues of $C_1$ largest in magnitude (with the possibility that some of the $a_i$ are equal), $u_i$ denotes an eigenvector corresponding to $a_i$ normalized such that $u_i'u_k=\mathbbm{1}_{\{i=k\}}$ for all $i$ and $k$, $p_i\equiv\Gamma_0^{\frac{-1}{2}}u_i$, and $q_i\equiv\Gamma_0^{\frac{1}{2}}u_i$. 
\item[(b)] Under any pseudo-true Markovian $d$-state model $\theta$, the subjective variance-covariance of the vector of observables, $E^\theta[y_ty_t']$, coincides with the true variance-covariance matrix, $\Gamma_0\equiv\mathbb{E}[y_ty_t']$.
\end{itemize}
\end{theorem}
The result shows that the insights from the analysis of one-state simple models broadly carry over to $d$-state ones. In particular, agents who are restricted to Markovian $d$-state models exhibit a form of persistence bias. They focus on perfectly forecasting the $d$ most persistent components of the vector of observables at the expense of the other components. Moreover, agents who are constrained to use Markovian $d$-state models uncover the true variance-covariance matrix of the observable.

The theorem also suggests that state-space models can be estimated consistently by principal component analysis (PCA). This conclusion is reminiscent of a central result in the theory of dynamic factor models on the consistency of the principal components estimator for the common components.\footnote{See, for instance, \citet{stock2002forecasting}.} However, Theorem \ref{thm:d-state-m-i-o} is different along several dimensions. First, it concerns state-space models, not dynamic factor models. Second, the estimator suggested by the theorem uses the principal components of the lag-one autocorrelation matrix, while the PCA estimator of dynamic factor models is constructed from the principal components of the variance-covariance matrix. Lastly, Theorem \ref{thm:d-state-m-i-o} suggests that the PCA estimator is consistent (at least under the theorem's assumptions) even if the number of states is misspecified.\footnote{An estimator for a misspecified model is consistent if the estimate converges to a pseudo-true model as the sample size goes to infinity.} I am aware of no similar result on the consistency of the PCA estimator for dynamic factor models when the number of common factors is misspecified.

\section{Behavioral Implications}\label{sec:predictions}
In this section, I apply the characterization results from the previous section to develop the behavioral implications of the simple models framework. Throughout the section, I maintain the assumption that at least one of the following is satisfied for every agent: (a) the agent is constrained to use one-state models and the true process is exponentially ergodic; (b) the agent is constrained to use Markovian one-state models; or (c) the agent is constrained to use Markovian $d$-state models and the lag-one autocovariance matrix, $\Gamma_1$, is symmetric.

To elaborate on the behavioral implications of the framework, I embed it in a reduced-form economy. Consider a finite set of agents, indexed by $j=1,\dots,J$. In every period $t$, each agent $j$ takes a purely forward-looking decision $x_{jt}$, which depends on her forecasts via the best-response function
\begin{equation}
    	x_{jt} = E_{jt}\left[\sum_{s=1}^\infty c_{js}'y_{t+s}\right],\label{eq:reduced_form}
\end{equation}
where $y_t\in\mathbb{R}^n$ is as before the vector of observables, $E_{jt}[\cdot]$ denotes agent $j$'s subjective forecasts, and $c_{js}\in\mathbb{R}^{n}$ are preference parameters satisfying $\sum_{s=1}^\infty \Vert c_{js}\Vert_2<\infty$ for all $j$.\footnote{The assumption that each agent takes a single action is without loss of generality. The analysis would be identical if one instead assumed that agent $j$ makes multiple choices in each period, with the $k$th action of agent $j$ given by $x_{jkt} = E_{jt}\left[\sum_{s=1}^\infty c_{jks}'y_{t+s}\right]$.} I continue to take the true process $\mathbb{P}$ as a primitive of the economy and assume that agent $j$ can entertain state-space models with no more than $d_j$ states. In Online Appendix \ref{app:PE_GE}, I provide an analysis suggesting that the partial equilibrium insights would generalize to a general equilibrium economy, in which $\mathbb{P}$ itself is an endogenous outcome of agents' choices.

The reduced-form specification in \eqref{eq:reduced_form} allows the derivation of sharp theoretical results, which highlight the role of simple models and biased forecasts and are independent of the specifics of agents' decision problems. These results are valid up to first order for purely forward-looking decisions that depend non-linearly on the forecasts of the observable. They also hold arbitrarily well when decisions are sufficiently forward-looking (e.g., when the discount factor is close to one). In the next section, I further develop the implications of the general framework in the context of a microfounded general equilibrium macro model. 

\subsection{Persistence Bias}\label{subsec:persistence}
Decomposing the observable into its more and less persistent components will be useful for the subsequent discussions:
\begin{proposition}\label{prop:decomposition}
Let $a_i$ denote the $i$th largest eigenvalue of the first autocorrelation matrix, $C_1$, in magnitude, and let $u_i$ denote the corresponding eigenvector, normalized such that $u_i'u_k=\mathbbm{1}_{\{i=k\}}$ for all $i$ and $k$. The observable can be decomposed as follows:
\begin{equation}\label{eq:decomposition}
y_t = \sum_{i=1}^n y_{t}^{(i)}q_i,
\end{equation}
where $y_{t}^{(i)}\equiv {p_i}'y_t$, $p_i\equiv\Gamma_0^{\frac{-1}{2}}u_i$, $q_i\equiv\Gamma_0^{\frac{1}{2}}u_i$, $u_i$ is as in Theorem \ref{thm:d-state-m-i-o}, and scalars $y_{t}^{(i)}$ all have unit variance. If $\rho_i$ denotes the lag-one autocorrelation of $y_t^{(i)}$, then $|\rho_1|\geq |\rho_2|\geq\dots\geq |\rho_n|$.
\end{proposition}

This proposition represents the observable in terms of the basis vectors $\{q_i\}_{i=1}^n$, with $y_{t}^{(i)}$ denoting the components (or coordinates) of $y_t$ with respect to this basis. The components of $y_t$ are sorted by their persistence, with $y_{t}^{(1)}$ representing the \emph{most persistent component} and $y_{t}^{(n)}$ the \emph{least persistent component} of the observable. This decomposition is valid for arbitrary stationary stochastic processes and is independent of agents' forecasting and decision problems. However, the way agents' choices respond to changes in the observable neatly aligns with the decomposition in \eqref{eq:decomposition}. This is shown in the following corollary of Theorems \ref{thm:1-state-exp-erg-closed-form} and \ref{thm:d-state-m-i-o}:
\begin{corollary}[persistence bias]\label{cor:persistence-bias}
Agent $j$'s time-$t$ forecasts and forward-looking actions only respond to changes in the $d_j$ most persistent components of $y_t$.
\end{corollary}

Agents who use pseudo-true $d$-state models treat the most persistent and least persistent components of $y_t$ in qualitatively different ways. A change in the current value of the observable can be decomposed into changes in the components $y_{t}^{(i)}$ of $y_t$. Agents do not change their forecasts in response to changes in the least persistent components of $y_t$. Consequently, their forward-looking actions also remain unresponsive to current changes in these less persistent components. 

It is worth noting that agents' forecasts and actions are  unresponsive to changes in the less persistent components of the observable only on impact. In general, different components of $y_t$ do not evolve independently. Therefore, a change in the current value of $y_t^{(i)}$ could lead to changes in the values of $y_{t+s}^{(j)}$ for some $j\neq i$ and $s>0$. This can result in a delayed response of agents' forecasts and actions to changes in the observable's less persistent components. 

\subsection{Increased Comovement}
Constraining agents to use simple models increases the comovement between their forward-looking choices. The argument for this prediction is best seen by considering agents $j$ and $k$, both of whom are constrained to use one-state models. Because of persistence bias, the agents' time-$t$ actions can be written as time-invariant linear functions of the observable's most persistent component. More specifically, $x_{jt}=g_{j}^{(1)}y_t^{(1)}$ and $x_{kt}=g_{k}^{(1)}y_t^{(1)}$, where $g_{j}^{(1)}$ and $g_{k}^{(1)}$ are constants that depend on the true process and the agents' preferences. Thus, one agent's actions can be expressed as a constant multiple of the other agent's actions. In other words, the agents' actions comove perfectly. The following proposition formalizes and extends this conclusion:
\begin{proposition}\label{prop:d-state-comovement}
Let $D\equiv \max_{j}d_j$ denote the largest value of $d_j$ among agents and $x_t\equiv(x_{jt})_{j}\in\mathbb{R}^{J}$ denote the vector containing agents' time-$t$ actions. Given generic true processes, $x_t$ has the factor structure
\[
x_t = Gy_t^{(1:D)},
\]
where $G$ is a $J\times D$ matrix of loadings and $y_t^{(1:D)}$ is the $D$-dimensional vector consisting of the $D$ most persistent components of the observable.\footnote{The result requires all pseudo-true models of a given dimension to be observationally equivalent, a condition that holds for generic true processes.}
\end{proposition}

The proposition establishes that the $J$-dimensional vector of all the forward-looking actions of all the agents in the economy moves with the $D$ factors collected in $y_t^{(1:D)}$. The number of factors depends solely on the complexity of agents' models, while the composition of these factors depends on the properties of the true process. The loadings of actions on different factors depend on the preference parameters $c_{js}$. If $D$ is much smaller than $J$ (which is often a reasonable assumption), then a large number of actions comove with movements in a small number of factors.

Agents' actions exhibit comovement not only among those using models with the same value of $d$ but also among those using models of different dimensions. To see the intuition for this result, consider agents $j$ and $k$ who use models of dimensions $d_j$ and $d_k>d_j$, respectively. While the agents disagree on the number of state variables needed to forecast the observable, they agree on what $d_j$ of those state variables ought to be. The $d_j$ states used by agent $j$ are a subset of the $d_k$ states used by agent $k$ (up to linear transformations). This strong form of comovement is a unique prediction of the framework of simple models. This result relies on the fact that pseudo-true $d$-state models rank the components of $y_t$ consistently across $d$: As $d$ increases, pseudo-true forecasts condition on additional components of $y_t$ but without altering the components already being used to forecast. 

A low-dimensional factor structure is one natural expression of comovement. Another commonly used comovement measure is the Pearson correlation coefficient between two variables. The following corollary of Proposition \ref{prop:d-state-comovement} shows that constraining any two agents to one-state models increases the correlation between their actions:
\footnote{Note that this corollary does not generalize beyond the one-state case; constraining agents to models with $d_j,d_k>1$ states might actually \emph{decrease} the correlation between their actions compared to the rational-expectations benchmark.}
\begin{corollary}
Consider actions $j$ and $k$, both of the form \eqref{eq:reduced_form}, taken by agents $j$ and $k$ with $d_j=d_k=1$. Generically,
\[
1=\left|\text{Corr}\left(x_{jt}^{1d},x_{kt}^{1d}\right)\right| > \left|\text{Corr}\left(x_{jt}^{\text{RE}},x_{kt}^{\text{RE}}\right)\right|,
\]
where $x_{jt}^{1d}$ and $x_{jt}^{\text{RE}}$ denote agent $j$'s time-$t$ action when using a pseudo-true one-state model and the true model, respectively.
\end{corollary}

The time-$t$ actions of agents using a pseudo-true one-state model depend solely on the current realization of $y_t^{(1)}$, the most persistent component of $y_t$. Consequently, an econometrician who analyzes those actions will conclude that the actions are driven by a single shock to the economy. This conclusion holds regardless of the specifics of preferences, technology, or market structure. It holds both in partial equilibrium and in general equilibrium, as suggested by the analysis in Online Appendix \ref{app:PE_GE}. However, the single shock recovered by the econometrician is not a true shock. It is an endogenous index whose statistical properties depend on the primitives of the economy, the stochastic properties of the shocks that hit it, and the parameters of policy rules.

\subsection{Under- and Over-Extrapolation}
The framework proposed in this paper is neither a model of under-extrapolation nor of over-extrapolation. Instead, agents who use simple models forecast using a parsimonious model that provides an approximation to the true process and balances forecast errors across different horizons and variables. Consequently, simple models do not lead to mistakes that invariably go in the same direction. In fact, agents who use a pseudo-true $d$-state model under-extrapolate some variables and over-extrapolate others.
\begin{proposition}\label{prop:d-state-over-under-reaction}
Let $y_t^{(1)}$ denote the most persistent component of $y_t$ and $y_t^{(n)}$ denote its least persistent component. If the true process is exponentially ergodic and $d_j<n$, then:
\begin{itemize}
    \item[(a)] Agent $j$ overestimates the magnitude of $y_t^{(1)}$'s autocorrelation at all lags.     \item[(b)] Agent $j$ underestimates the magnitude of $y_t^{(n)}$'s autocorrelation at all lags.  
\end{itemize}
\end{proposition}
The following example illustrates the proposition:
\begin{example}
Suppose the vector of observables is given by $y_t=(y_{1t},y_{2t})'\in \mathbb{R}^2$, and each element of $y_t$ follows an independent $\text{ARMA}(1,1)$ process
\begin{align*}
    & y_{1t} = \phi_1 y_{1,t-1}+\epsilon_{1t} + \vartheta_1\epsilon_{1,t-1},\\
    & y_{2t} = \phi_2 y_{2,t-1}+\epsilon_{2t} + \vartheta_2\epsilon_{2,t-1},
\end{align*}
where $\phi_1,\phi_2,\vartheta_1,\vartheta_2\in(0,1)$ are constants, and $\epsilon_{1t}$ and $\epsilon_{2t}$ are i.i.d. mean-zero random variables with finite variances. Additionally, assume that $\phi_1>\phi_2$ and $\frac{(\phi_1+\vartheta_1)(1+\phi_1\vartheta_1)}{1+2\phi_1\vartheta_1+\vartheta_1^2} >  \frac{(\phi_2+\vartheta_2)(1+\phi_2\vartheta_2)}{1+2\phi_2\vartheta_2+\vartheta_2^2}$. This assumption ensures that $y_{1t}$ has a higher autocorrelation than $y_{2t}$ at all lags.

The lag-$l$ autocorrelation matrix is given by
\[
C_l = \begin{pmatrix} \displaystyle\frac{(\phi_1+\vartheta_1)(1+\phi_1\vartheta_1)}{1+2\phi_1\vartheta_1+\vartheta_1^2}\phi_1^{l-1} & 0 \\ 0 & \displaystyle\frac{(\phi_2+\vartheta_2)(1+\phi_2\vartheta_2)}{1+2\phi_2\vartheta_2+\vartheta_2^2}\phi_2^{l-1}
\end{pmatrix}.
\]
The $i$th largest eigenvalue of $C_1$ in magnitude is $\frac{(\phi_i+\vartheta_i)(1+\phi_i\vartheta_i)}{1+2\phi_i\vartheta_i+\vartheta_i^2}$, and the corresponding eigenvector is $u_i=e_i$, where $e_i$ denotes the $i$th standard coordinate vector. Since the two elements of $y_t$ are independent, the variance-covariance matrix, $\Gamma_0$, is diagonal. Therefore, $y_t^{(1)}q_1=y_{1t}e_1$ and $y_t^{(2)}q_2=y_{2t}e_2$, i.e., the most persistent component of $y_t$ is its first component in the standard coordinates and its least persistent component is its second component in the standard coordinates. The spectral radius of the lag-$l$ autocorrelation matrix satisfies
\[
\rho(C_l) = \frac{(\phi_1+\vartheta_1)(1+\phi_1\vartheta_1)}{1+2\phi_1\vartheta_1+\vartheta_1^2}\phi_1^{l-1}\geq \left(\frac{(\phi_1+\vartheta_1)(1+\phi_1\vartheta_1)}{1+2\phi_1\vartheta_1+\vartheta_1^2}\right)^l = \rho(C_1)^l,
\]
with the inequality strict for $l>1$. That is, the true process is exponentially ergodic.

The pseudo-true one-state model is described by Theorem \ref{thm:1-state-exp-erg-closed-form}. Under any such model, $y_{1t}$ follows an AR(1) process with persistence parameter $a=\frac{(\phi_1+\vartheta_1)(1+\phi_1\vartheta_1)}{1+2\phi_1\vartheta_1+\vartheta_1^2}$ and $y_{2t}$ is i.i.d. over time. The pseudo-true lag-$l$ autocorrelation of $y_{1t}$ is equal to $a^l$, while the pseudo-true lag-$l$ autocorrelation of $y_{2t}$ is zero for any $l\geq 1$. On the other hand, the true lag-$l$ autocorrelation of $y_{it}$ is given by $\frac{(\phi_i+\vartheta_i)(1+\phi_i\vartheta_i)}{1+2\phi_i\vartheta_i+\vartheta_i^2}\phi_i^{l-1}$ for $i=1,2$. Therefore, an agent who uses a pseudo-true one-state model overestimates the autocorrelation of $y_{1t}$ at all lags (strictly so for lags $l>1$) while strictly underestimating the autocorrelation of $y_{2t}$ at all lags.
\end{example}

Agents who use simple models over-extrapolate the most persistent components of the observable and under-extrapolate the least persistent ones. For observables with intermediate persistence, the pattern could be under- or over-extrapolation depending on the variable and the horizon being considered. These predictions set this paper's framework apart from models that hardwire under- or over-extrapolation.

\section{A Business-Cycle Application}\label{sec:DSGE}

In this section, I use the framework of simple models to study how bounded rationality changes the response of an economy to supply, demand, and policy shocks and the propagation of those shocks to endogenous variables. I do so in the context of a standard business-cycle model economy, which combines elements of the New Keynesian and real business cycle models. This exercise demonstrates that the macroeconomic model's empirical fit is improved when bounded rationality is introduced in the form of dimensionality reduction. Moreover, simple models can serve as a parsimonious substitute for add-ons, such as external habit formation, investment-adjustment costs, and endogenous capital utilization, which are used to increase the persistence and comovement of endogenous variables.

\subsection{The Model Economy}
The model economy is a New Keynesian economy with price and wage rigidities and endogenous capital formation. Alternatively, the model can be viewed as a DSGE model \`{a} la \citet*{Christiano2005}, \citet*{Smets2007}, and \citet*{justiniano2010investment} but without the following add-ons: (i) external habit formation in consumption, (ii) investment-adjustment costs, (iii) price and wage indexation, (iv) endogenous capital utilization, (v) a monetary policy that responds to the level and growth rate of the output gap. The first three add-ons increase the persistence of consumption, investment, inflation, and wages. The last two enhance the comovement properties of the economy. 

To focus on the core mechanisms, I initially consider a version of the model with only three shocks: a total-factor productivity (TFP) shock, an investment shock, and a monetary-policy shock. The TFP shock is the main supply shock in DSGE models, while the investment shock is the demand shock that explains the largest variance shares of real variables at business-cycle frequencies \citep{justiniano2010investment}. Monetary-policy shocks contribute little to the variance of nominal and real variables at business-cycle frequencies. However, they have clear empirical counterparts that can be identified using vector autoregression (VAR), narrative, and high-frequency approaches. Comparing model-implied impulse-response functions (IRFs) to monetary-policy shocks with their empirical counterparts provides a powerful test of the model economy's internal propagation mechanism \citep{Christiano2005}. I focus on these three shocks to clearly illustrate how bounded rationality alters the behavior of the model economy. Later, I enrich the economy with a full suite of standard DSGE shocks, estimate it using Bayesian techniques---both given simple models and under rational expectations---and perform Bayesian model selection.

The economy is populated by seven groups of agents: final-good producers, intermediate-goods producers, investment firms, employment agencies, households, labor unions, and the government. In what follows, I describe each group's problem in detail.

\subsubsection*{Final-good producers}
The final good $Y_t$ is produced by competitive firms by combining a continuum of intermediate goods, indexed by $i$, according to the CES production function
\begin{equation*}
Y_t  = \left[\int_0^1 {Y_t(i)}^{\frac{1}{1+\lambda_{p}}}di\right]^{1+\lambda_{p}},
\end{equation*}
where $\lambda_{p}$ denotes the elasticity of substitution. Profit maximization and the zero-profit condition imply that the price of the final good is given by the price index
\begin{equation*}
P_t = \left[\int_0^1 P_t(i)^{\frac{1}{\lambda_{p}}} di\right]^{\lambda_{p}},
\end{equation*}
where $P_t(i)$ denotes the price of intermediate good $i$. The demand for good $i$ is given by the isoelastic demand schedule
\begin{equation*}
Y_t(i) = \left(\frac{P_t(i)}{P_t}\right)^{-\frac{1+\lambda_{p}}{\lambda_{p}}}Y_t.
\end{equation*}

\subsubsection*{Intermediate-goods producers}
A monopolist produces intermediate good $i$ according to the production function
\begin{equation*}
Y_t(i) = \max\left\{a_t K_t(i)^\alpha \left(\gamma^t L_t(i)\right)^{1-\alpha}-\gamma^t F,0\right\},
\end{equation*}
where $K_t(i)$ and $L_t(i)$ denote the capital and labor inputs of the monopolist, respectively, $F$ is a fixed cost of production, chosen so that profits are zero along the balanced-growth path, $\gamma$ denotes the exogenous rate of labor-augmenting technological progress, and $a_t$ is a stationary TFP shock, which follows the AR(1) process $\log a_t = \rho_a \log a_{t-1} + \varepsilon_{at}$ with $\varepsilon_{at}$ i.i.d. $\mathcal{N}(0,\sigma_a^2)$. 

Intermediate-goods producers are subject to nominal frictions \`{a} la Calvo. Each period the price of a randomly-selected fraction $\xi_p$ of intermediate goods grows at rate $\pi$,  where $\pi$ denotes the value of gross inflation rate along the balanced-growth path. The remaining intermediate-goods producers choose their prices $P_t(i)$ optimally by maximizing the present-discounted value of future profits,
\begin{equation*}
E_{pt}\left[\sum_{s=0}^\infty \xi_p^s\beta^s\Lambda_{t+s}\Big(\pi^sP_t(i)Y_{t+s}(i)-W_{t+s}L_{t+s}(i)-r_{t+s}K_{t+s}(i)\Big)\right],
\end{equation*}
subject to the demand curve 
\[
Y_{t+s}(i)=\left(\frac{\pi^sP_t(i)}{P_{t+s}}\right)^{-\frac{1+\lambda_{p}}{\lambda_{p}}}Y_{t+s},
\]
where $\Lambda_t$ is the marginal utility of nominal income, $W_t$ is the nominal wage, $r_t$ is the rental rate of capital, and $E_{pt}$ denotes the time-$t$ forecasts of intermediate-goods producers about the path $\{\Lambda_{t+s},W_{t+s},r_{t+s},P_{t+s},Y_{t+s},a_{t+s}\}_{s\geq 1}$ of variables they take as given.

\subsubsection*{Investment firms}
The capital stock of the economy is owned by competitive investment firms. They take the rental rate of capital and the price of the final good as given and maximize the present-discounted value of profits
\begin{equation*}
E_{it}\left[\sum_{s=0}^\infty \beta^s\Lambda_{t+s}\left(r_{t+s}K_{t+s} - P_{t+s} I_{t+s}\right)\right],
\end{equation*}
subject to the capital accumulation equation
\begin{equation*}
K_{t+1} = (1-\delta)K_{t} + \mu_t\left(I_{t}-S_k\left(\frac{I_{t}}{K_{t}}\right)K_{t}\right),
\end{equation*}
where $I_t$ is investment, $K_t$ denotes the physical capital, $E_{it}$ denotes the time-$t$ forecasts of investment firms, $S_k(\cdot)$ represents the adjustment cost function, and $\mu_t$ is the investment shock, which follows the AR(1) process $\log \mu_t = \rho_\mu \log \mu_{t-1} + \varepsilon_{\mu t}$ with $\varepsilon_{\mu t}$ is i.i.d. $\mathcal{N}(0,\sigma_\mu^2)$. I assume that the adjustment cost satisfies $S_k=S_k'=0$ and $S_k''=\varsigma_k>0$ along the balanced-growth path.\footnote{Note that the adjustment cost is a neoclassical cost \`{a} la \cite{hayashi1982tobin}, not the investment-adjustment cost common in the DSGE literature. The investment-adjustment cost specification leads to an investment Euler equation with a backward-looking term, whereas investment will have no backward-looking term in the current specification.} I also assume there is no spot market for installed capital.\footnote{This assumption is immaterial under rational expectations. However, this may no longer be the case away from rational expectations: When there is no spot market for capital, investment depends on agents' expectations about the infinite future path of returns to capital; when a spot market exists, investment only depends on agents' expectations of the rental rate of capital and its price in the next period.}

\subsubsection*{Employment agencies}
There is a continuum of households, indexed by $j$, each of which is a monopolistic supplier of a specialized type of labor. A competitive employment agency combines specialized labor into a homogeneous labor input using the CES function
\begin{equation*}
L_t =\left[\int_0^1 {L_t(j)}^{\frac{1}{1+\lambda_{w}}}dj\right]^{1+\lambda_{w}},
\end{equation*} 
where $\lambda_{w}$ denotes the elasticity of substitution among differentiated types of labor. Profit maximization by employment agencies and the zero-profit condition imply that the price of the homogeneous labor input is given by the wage index
\begin{equation*}
W_t = \left[\int_0^1 {W_t(j)}^{\frac{1}{\lambda_{w}}}\right]^{\lambda_{w}},
\end{equation*}
and the demand for the labor of type $j$ is given by the isoelastic labor-demand curve
\begin{equation*}
L_t(j) = \left(\frac{W_t(j)}{W_t}\right)^{-\frac{1+\lambda_{w}}{\lambda_{w}}}L_t.
\end{equation*}

\subsubsection*{Households}
Households supply labor, consume the final good, and save in  a short-term nominal government bond. Their wages are subjective to nominal rigidities \`{a} la Calvo. However, as is common in the literature, I assume that a competitive insurance agency fully insures households against fluctuations in their labor income resulting from nominal frictions. Consequently, the equilibrium labor income of each household is equal to $W_t L_t$, the average labor income in the economy.

Each household takes the labor income and the stream of profits from the ownership of firms as given and chooses consumption and saving in government bonds to maximize the utility function
\begin{equation*}
E_{ct}\left[\sum_{s=0}^\infty \beta^s \left(\log(C_{t+s}) - \varphi \frac{{L_{t+s}(j)}^{1+\nu}}{1+\nu}\right)\right],
\end{equation*}
subject to a no-Ponzi condition and the nominal budget constraint
\begin{equation*}
P_t C_t + T_t + B_t \leq R_{t-1} B_{t-1} + W_t L_t+\Pi_t,
\end{equation*}
where $C_t$ is consumption, $T_t$ denotes lump-sum taxes, $B_t$ is the holding of one-period government bonds, $R_t$ is the gross nominal interest rate, $\Pi_t$ denotes profits from the ownership of firms, $\nu$ is the inverse Frisch elasticity of labor supply, and $\varphi$ is a constant that determines the steady-state working hours. The operator $E_{ct}$ denotes the time-$t$ forecasts of households about the path $\{L_{t+s},W_{t+s},P_{t+s},T_{t+s},R_{t+s},\Pi_{t+s}\}_{s\geq 1}$ of aggregate and idiosyncratic observables that enter their decision problem.

\subsubsection*{Labor unions}
Wages are set by a continuum of labor unions, also indexed by $j$, each representing a household. Each period, a randomly-selected fraction $\xi_w$ of unions cannot freely set the wage of the household they represent. The nominal wages of those households grow at the rate $\gamma \pi$.\footnote{Since there is technological progress, absent this assumption, there would be no balanced-growth path without wage dispersion. Note that this is different than the assumption of wage indexation common in the DSGE literature: Wages are not indexed to the current inflation rate but to its steady-state value.} The remaining fraction of labor unions sets the optimal wage $W_t(j)$ by maximizing
\begin{equation*}
E_{wt}\left[\sum_{s=0}^\infty \xi_w^s\beta^s\left(-\varphi\frac{L_{t+s}(j)^{1+\nu}}{1+\nu}+\Lambda_{t+s}(\gamma\pi)^sW_t(j)L_{t+s}(j)\right)\right],
\end{equation*}
subject to the labor demand curve
\[
L_{t+s}(j) = \left(\frac{(\gamma\pi)^s W_t(j)}{W_{t+s}}\right)^{-\frac{1+\lambda_{w}}{\lambda_{w}}}L_{t+s},
\] 
where $E_{wt}$ denotes the time-$t$ forecasts of labor unions about the variables they take as given.

\subsubsection*{The government}
The monetary policy sets the nominal interest rate following a Taylor rule
\begin{equation*}
\frac{R_t}{R} = \left(\frac{R_{t-1}}{R}\right)^{\rho_R}\left(\frac{\pi_t}{\pi}\right)^{(1-{\rho_R})\phi_\pi}m_t,
\end{equation*}
where $\pi_t \equiv P_t/P_{t-1}$, and $R$ and $\pi$ are the steady-state gross nominal interest rate and inflation rate, respectively.\footnote{In the New Keynesian literature, it is often assumed that the monetary authority responds both to changes in the inflation rate and to changes in the output gap. However, the right notion of the output gap is not clear here: It can be defined relative to the flexible price allocation in which agents re-estimate their models, the one in which agents' models are unchanged, or the rational-expectations flexible-price allocation. I bypass the question of how the output gap ought to be defined by instead assuming that the monetary authority only responds to deviations in the inflation rate.} \footnote{The steady-state gross nominal interest rate $\pi$ can also be seen as the central bank's inflation target.}
$m_t$ is a monetary policy shock that follows the AR(1) process $\log m_t = \rho_m \log m_{t-1} + \varepsilon_{mt}$ with $\varepsilon_{mt}$ is i.i.d. $\mathcal{N}(0,\sigma^2_m)$.

Government spending $G_t$ is exogenous. In the baseline specification, I assume that government spending grows at the same rate as GDP, that is, $G_t = g\gamma^t$ for some $g$. The government finances spending by issuing short-term nominal bonds and levying lump-sum taxes on households. The nominal government budget constraint is given by 
\begin{equation*}
R_{t-1}B_{t-1} + P_t G_t - T_t = B_t,
\end{equation*}
where $T_t$ denotes nominal taxes. Taxes follow a tax rule that ensures that the real value of public debt $B_t/P_t$ grows at rate $\gamma$, the deterministic growth rate of the economy.\footnote{Ricardian equivalence does not necessarily hold when agents use simple models. Therefore, both the timing of taxes and the value of the outstanding public debt might affect the response of the economy to shocks. See also \cite{eusepi2018fiscal}, where the authors use an adaptive learning framework to study the effects of the level of public debt on the transmission of monetary policy.}

\subsection{Equilibrium}
The analysis proceeds in two steps. The first is to characterize the \emph{temporary equilibrium}, which imposes individual optimality and market clearing but not rational expectations.\footnote{The notion of temporary equilibrium goes back to \cite{Grandmont1977}. See \cite{Woodford2013} for a discussion of temporary equilibria in the context of modern monetary models and \cite{farhi2019monetary} for an application to heterogeneous-agent New Keynesian economies.} The second step is to supplement the temporary equilibrium with the model of expectation formation and characterize the resulting (full) equilibrium.

The first step of the analysis is relatively straightforward. I start by deriving agents' first-order optimality conditions. I then characterize the balanced-growth path along which inflation is constant and equal to the monetary authority's inflation target, and output, consumption, investment, government spending, capital stock, real wages, and the public debt all grow at rate $\gamma$, the deterministic growth rate of labor productivity. Finally, I log-linearize the optimality conditions around the balanced-growth path. 

Extra care is necessary when working with the optimality conditions without imposing rational expectations. Away from rational expectations, one agent's optimality conditions cannot be simplified using equilibrium conditions that are not necessarily respected by the agent's expectation operator---conditions such as other agents' optimality conditions.\footnote{\cite{preston2005learning} is the first to make this point in the context of adaptive-learning models. See, also, \citet*[p.~272]{woodford2003interest} for a discussion.} For instance, the optimality conditions of firms resetting their prices cannot be combined with the law of motion of the price index to obtain the usual recursive version of the Phillips curve. Likewise, households' optimality conditions cannot be combined with the government budget constraint to obtain the usual recursive consumption-Euler equation. 

I instead simplify each agent's first-order optimality conditions using only equations that are fully understood by the agent. Combining households' first-order optimality conditions with their budget constraints and no-Ponzi conditions yields the following version of the permanent-income hypothesis:
\begin{align}
\hat{c}_t = -\hat{R}_t + E_{ct} \left[\sum_{s=1}^\infty \beta^{s}\left(\frac{1-\beta}{\beta}\frac{x}{c}\hat{x}_{t+s}-\frac{1-\beta}{\beta}\frac{\tau}{c}\hat{\tau}_{t+s}-\frac{x-\tau}{c}\hat{R}_{t+s}+\frac{1}{\beta}\frac{x-\tau}{c}\hat{\pi}_{t+s}\right)\right],\label{eq:TE_c_hat_body}
\end{align}
where lowercase letters with hats denote log-deviations from the balanced-growth path and lowercase letters without hats denote steady-state values, $\pi$ refers to inflation rate, $x$ refers to households' total pre-tax real income, and $\tau$ is their real tax burden. Likewise, the solution to investment firms' optimality conditions yields the following investment equation:
\begin{align}
\hat{i}_t =  &\,\hat{k}_t + \frac{1}{\varsigma_k}\left(\hat{\mu}_t-\hat{\lambda}_t\right) + E_{it}\left[\sum_{s=1}^\infty \beta^{s}\left(\frac{1-\beta}{\varsigma_k\beta}\hat{\lambda}_{t+s}+\frac{1}{\varsigma_k}\left(\frac{1}{\beta}-\frac{1-\delta}{\gamma}\right)\hat{\rho}_{t+s}+\frac{1}{\varsigma_k}\left(1-\frac{1-\delta}{\gamma}\right)\hat{\mu}_{t+s}\right)\right],\label{eq:TE_i_hat_body}
\end{align}
where $\hat{\rho}$ and $\hat{\lambda}$ denote log-deviations of the rental rate of capital and the stochastic discount factor, respectively, from their values along the balanced-growth path. Meanwhile, intermediate-goods firms' optimality conditions yield the following price Phillips curve:
\begin{align}
\hat{\pi}_t = &\,\kappa\left(\alpha \hat{\rho}_t + (1-\alpha) \hat{w}_t -\hat{a}_t\right) + E_{pt}\left[\sum_{s=1}^\infty\xi_p^s\beta^s\left(\frac{1-\xi_p}{\xi_p}\hat{\pi}_{t+s}+\kappa\left(\alpha \hat{\rho}_{t+s} + (1-\alpha) \hat{w}_{t+s} -\hat{a}_{t+s}\right)\right)\right],\label{eq:TE_pi_hat_body}
\end{align}
where $\kappa$ is a constant. Finally, the solution to labor unions' wage-setting problem yields the following wage Phillips curve:
\begin{equation}
\hat{w}_t = \frac{(1+\beta)\kappa_w}{1+\beta-\xi_w\beta}\hat{\ell}_{t}+\frac{\hat{w}_{t-1}-\hat{\pi}_t}{1+\beta-\xi_w\beta}+\frac{1+\beta}{1+\beta-\xi_w\beta}E_{wt}\left[\sum_{s=1}^\infty \xi_w^s\beta^s\Big(\frac{\nu_w\kappa_w}{1-\xi_w\beta}\hat{\pi}_{t+s} +\kappa_w\hat{\ell}_{t+s} + \nu_w\kappa_w\hat{w}_{t+s}\Big)\right],\label{eq:TE_w_hat_body}
\end{equation}
where $\nu_w$ and $\kappa_w$ are constants and $\hat{\ell}_{t} \equiv \nu\hat{L}_{t} - \hat{\lambda}_{t} - \hat{w}_t$. 
Equations \eqref{eq:TE_c_hat_body}--\eqref{eq:TE_w_hat_body} are the model economy's only forward-looking temporary-equilibrium conditions. The remaining conditions are static and do not feature agents' subjective expectation operator. The full list of temporary-equilibrium conditions can be found in Appendix \ref{appendix:DSGE_TE}.

Equations \eqref{eq:TE_c_hat_body}--\eqref{eq:TE_w_hat_body} determine consumption, investment, inflation, and the real wage as functions of current and past observables as well as agents' expectations of future values of observables. They are valid under arbitrary expectations---as long as agents all understand their own problems and their expectations satisfy the law of iterated expectations. Together with the remaining temporary-equilibrium conditions and the specification of agents' expectations, they fully determine the equilibrium of the economy.

I next describe agents' subjective expectations. For simplicity, I assume that households, investment firms, intermediate-goods producers, and labor unions all face identical constraints on the models they can entertain, thus ending up with identical subjective expectations. Every agent has perfect foresight about the balanced-growth path of the economy. Agents also all have full information about the log-deviations of all endogenous and exogenous variables from the balanced-growth path. In particular, agents' time-$t$ information set is given by the history $\{\omega_{s}\}_{s\leq t}$ of vector 
\[
\omega_s \equiv \left(\hat{a}_s, \hat{m}_s,\hat{\mu}_s,\hat{k}_s,\hat{y}_s,\hat{x}_s,\hat{c}_s, \hat{i}_s, \hat{L}_s, \hat{\rho}_s, \hat{\pi}_s, \hat{R}_s, \hat{w}_s, \hat{\tau}_s \right)',
\]
consisting of the time-$s$ realizations of TFP, monetary-policy, and investment shocks as well as log-deviations of capital stock, GDP, pre-tax income, consumption, investment, hours, rental rate of capital, inflation, nominal interest rate,  wages, and taxes.\footnote{In equilibrium, some elements of $\omega_t$ are linear combinations of other elements of $\omega_t$. By the linear invariance result, dropping the redundant variables from $\omega_t$ does not change any of the equilibrium outcomes. Similarly, I can add additional variables to $\omega_t$ that, in equilibrium, are linearly dependent on variables already included in $\omega_t$ without changing anything.} Instead of imposing rational expectations, I assume that agents are constrained to use one-dimensional state-space models of the form \eqref{eq:simple_model} to forecast future values of $\omega$.

The equilibrium definition is straightforward. An equilibrium consists of a stochastic process $\mathbb{P}^*$ for $\{\omega_t\}_t$ and a model $\theta^*$ for agents such that (i) $\mathbb{P}^*$ is derived from market-clearing conditions and optimal behavior by all agents in the economy given subjective model $\theta^*$, and (ii) $\theta^*$ is a pseudo-true one-state model given the stochastic process $\mathbb{P}^*$.\footnote{In earlier work \citep{CREE}, I referred to this equilibrium notion as the constrained rational-expectations equilibrium.} 

\subsection{Impulse-Response Functions}
I begin my analysis of the business-cycle economy by studying its impulse-response functions (IRFs) to TFP, investment, and monetary-policy shocks---both under rational expectations and under the assumption that agents use pseudo-true one-state models. I use the same calibration of primitive parameters and shock processes in both variants. This makes it possible to transparently see how bounded rationality changes the internal propagation mechanism of the economy.\footnote{Figure \ref{fig:IRFs} plots the posterior on IRFs from a Bayesian estimation of the fully flexible version of the model (enriched with a full suite of DSGE shocks). The posterior concentrates its mass on parameter configurations with implied IRFs that are qualitatively similar to those obtained in this subsection.}

The model parameters are calibrated as follows: A period represents a quarter. I set $\beta=0.99$, $\gamma=1.005$, $\delta=0.025$, $g/y=b/y=0$, $\nu=1$, $\alpha=1/3$, $\lambda_p=0.5$, $\rho_R=0.8$, and $\phi_\pi=1.5$. The persistence parameters of the shocks are set to $\rho_a=0.95$ for TFP, $\rho_m=0.4$ for monetary policy, and $\rho_\mu=0.7$ for investment shocks. The standard deviations of shocks are set to $\sigma_a=1$ for TFP, $\sigma_m=0.5$ for monetary policy, and $\sigma_\mu=2$ for investment shocks. These values are in the ballpark of both the values chosen in the literature and those obtained from Bayesian estimation of the model. The values of three parameters---price rigidity $\xi_p$, wage rigidity $\xi_w$, and capital-adjustment cost $\varsigma_k$---have a great influence on the persistence of endogenous variables and the propagation of shocks. 
I assume flexible wages and set $\xi_p=0.6$ and $\varsigma_k=0.5$. These conservative values are picked to highlight the fact that the model economy can generate realistic IRFs without relying on counterfactually large degrees of nominal and real frictions often assumed in business-cycle models.

There are no free parameters for agents' expectations (other than $d$, which I have set equal to one). Agents' models, beliefs, and forecasts are all pinned down by structural parameters of the economy and the stochastic processes of the shocks. In equilibrium, agents' forecasts of elements of vector $\omega_s$ are given by
\[
E_t[\omega_{t+s}] = {a^*}^sq^*{p^*}'\omega_t,
\]
where the perceived persistence is given by
\[
a^* = 0.997,
\]
and the relative-attention vector $p^*$ and the relative-sensitivity vector $q^*$ are given by\footnote{Vector $p^*$ is identified only up to a set of linear transformations. Since $\omega_s$ contains redundant variables, $\tilde{p}'\omega_s={p^*}'\omega_s$ for all $s$ and a set of vectors $\tilde{p}$ belonging to a subspace. By the linear invariance result, all such vectors lead to identical forecasts and actions for agents at all times.}
\renewcommand{\arraystretch}{1.7}
\[
\begin{tabular}{c@{\hskip 5pt}c@{\hskip 5pt}c@{\hskip 10pt}c@{\hskip 10pt}c@{\hskip 10pt}c@{\hskip 10pt}c@{\hskip 10pt}c@{\hskip 10pt}c@{\hskip 10pt}c@{\hskip 10pt}c@{\hskip 10pt}c@{\hskip 10pt}c@{\hskip 10pt}c@{\hskip 10pt}c}
 & & \( \hat{a} \) & \( \hat{m} \) & \( \hat{\mu} \) & \( \hat{k} \) & \( \hat{y} \) & \( \hat{x} \) & \( \hat{c} \) & \( \hat{i} \) & \( \hat{L} \) & \( \hat{\rho} \) & \( \hat{\pi} \) & \( \hat{R} \) & \( \hat{w} \)\\
\( {p^*}' \) & = & \big(0.01 & 0.02 & 0.01 & 0.92 & 0.07\big), & & & & & & & &  \\
\( {q^*}' \) & = & \big(0.47 & $-0.00$ & 0.19 & 1.00 & 1.02 & 1.08 & 1.08 & 0.85 & $-0.19$ & $-0.31$ & 0.14 & 0.13 & 0.88\big). \\
\end{tabular}
\]

Agents forecast following a three-step procedure. First, they project the vector of observables on the relative-attention vector $p^*$ to form their estimate $\hat{z}_t\equiv {p^*}'\omega_t$ of the current value of the latent subjective state---I refer to $\hat{z}_t$ as agents' ``nowcast.'' Then, they form their forecasts of future values of the subjective state given its perceived persistence: $E_t[z_{t+s}]={a^*}^s\hat{z}_t$. Finally, they multiply their forecasts of the subjective state by the relative-sensitivity vector $q^*$ to form their forecasts of observables: $E_t[\omega_{t+s}]=q^*E_t[z_{t+s}]$.

Agents perceive the subjective state as highly persistent but not unit root ($a^*=0.997$). The nowcast is much more sensitive to changes in the capital stock ($p^*_k=0.92)$ than to changes in GDP ($p^*_y=0.07)$, and it barely responds to innovations in the three exogenous shocks ($|p^*|\leq 0.02)$. This is a manifestation of persistence bias: In equilibrium, the capital stock is more persistent than output and TFP, monetary-policy, and investment shocks. Agents' forecasts of capital stock, output, gross income, consumption, investment, and the real wage move almost one-for-one with changes in agents' nowcast ($|q^*|\geq 0.88)$, whereas their forecasts of TFP, investment shock, hours, and the rental rate of capital exhibit much less sensitivity. The observables whose forecasts are least sensitive to new information ($|q^*|\leq 0.14)$ are the nominal variables: agents' expectations of monetary-policy shock, inflation, and nominal interest rate are somewhat ``anchored'' to their steady-state values.

\providecommand{\mathdefault}[1]{#1}

\begin{figure}[htbp!]
    \centering
    \begin{minipage}{\linewidth}
        \centering
        \resizebox{\linewidth}{!}{\input{simple_IRF_a_white.pgf}}
        \captionof{figure}{Impulse-response functions to a positive TFP shock.}\label{fig:IRF_TFP}
        \vspace{0.5em}
        \footnotesize
        \raggedright
        \textit{Notes:} Baseline calibration. One-dimensional simple models in solid blue. Rational expectations in dashed red. Responses to a one-percent increase in TFP. Output, consumption, and investment in percents; inflation in percentage points.
    \end{minipage}
\end{figure}

Figure \ref{fig:IRF_TFP} plots the IRFs to a positive TFP shock. With one-dimensional simple models, the IRFs of real variables mimic the hump-shaped responses found in models with features that serve to increase the sluggishness of aggregate variables, such as consumption-habit formation and investment-adjustment costs. The response of output on impact is 77\% smaller with simple models than under rational expectations (RE). The corresponding figures for consumption and investment are 79\% and 73\%, respectively. 
The responses of real variables peak after one quarter under RE. With simple models, the peak ranges from six quarters after impact for consumption to eight quarters after impact for investment.  Simple models thus provide a novel account of the hump-shaped responses of aggregate variables to TFP shocks in empirical studies, which does not rely on auxiliary frictions.\footnote{For a meta-analysis of the responses of aggregate variables to technology shocks, see  \citet[pp.~135--151]{ramey2016macroeconomic}.} \footnote{With simple models, the IRFs to TFP shocks are hump-shaped even when there are no nominal rigidities, no adjustment costs, and TFP is i.i.d. over time, suggesting a resolution to \citet*{cogley1993impulse}'s observation that the RBC model has a weak propagation mechanism. See the earlier version of this paper \citep{molavi2023simple} for the IRFs in the standard RBC model, which has no nominal or real frictions.} The fact that quantities respond less with simple models requires more of the increase in TFP to be absorbed by a fall in prices. The result is a larger decrease in inflation in response to the increase in TFP. Nevertheless, the response of inflation is more muted and more transitory than those of real variables (both under RE and with simple models).

\begin{figure}[htbp!]
    \centering
    \begin{minipage}{\linewidth}
        \centering
        \resizebox{\linewidth}{!}{\input{simple_IRF_m_white.pgf}}
        \captionof{figure}{Impulse-response functions to an expansionary monetary-policy shock.}\label{fig:IRF_mp}
        \vspace{0.5em}
        \footnotesize
        \raggedright
        \textit{Notes:} Baseline calibration. One-dimensional simple models in solid blue. Rational expectations in dashed red. The shock is normalized to reduce the nominal rate by 25 basis points on impact in each variant. Output, consumption, and investment in percents; inflation in percentage points.        
    \end{minipage}
\end{figure}

Figure \ref{fig:IRF_mp} plots the IRFs to an expansionary  monetary-policy shock. Agents' use of simple models dampens the responses of real variables by about 98\% on impact. However, here, simple models also dampen the response of inflation  on impact by 98.6\%. Bounded rationality also increases the persistence of responses to monetary-policy shocks. The response of all aggregate variables decrease monotonically and rapidly under rational expectations. In contrast, the responses are hump-shaped and significantly more persistent when agents use simple models. 

The responses of aggregate variables to monetary-policy shocks are counterfactually strong under rational expectations. This is because changes in future real interest rates have the same effect on current output as changes in current real interest rates. With nominal rigidities and rational expectations, monetary-policy shocks change agents' forecasts of future real interest rates. These expected changes pass through almost one-for-one to aggregate output and consumption.\footnote{This is the forward guidance puzzle of \citet*{del2023forward}. An earlier version of this paper \citep{molavi2023simple} showed that replacing rational expectations with simple models greatly reduces the power of forward guidance  in an  estimated three-equation New Keynesian model and offers a quantitatively plausible resolution to the forward-guidance puzzle. See \cite{Angeletos-Lian}, \cite{garcia2019low},  and \cite{farhi2019monetary} for other resolutions based on relaxations of the rational-expectations assumption.} The additional persistence resulting from agents' use of simple models allows the model to generate realistic IRFs to monetary-policy shocks---despite the fact that the economy has no wage rigidity, wage or price indexation, habit formation, or investment-adjustment costs, and it only has moderate degrees of price rigidity and capital-adjustment costs.

\begin{figure}[htbp!]
    \centering
    \begin{minipage}{\linewidth}
        \centering
        \resizebox{\linewidth}{!}{\input{simple_IRF_mu_white.pgf}}
        \captionof{figure}{Impulse-response functions to a positive investment shock.}\label{fig:IRF_investment}
        \vspace{0.5em}
        \footnotesize
        \raggedright
        \textit{Notes:} Baseline calibration. One-dimensional simple models in solid blue. Rational expectations in dashed red. Responses to a one-percent increase in the demand for investment. Output, consumption, and investment in percents; inflation in percentage points.
    \end{minipage}
\end{figure}

Figure \ref{fig:IRF_investment} plots the IRFs to an investment shock. Under rational expectations, the economy produces a negative comovement between consumption and investment in response to the shock. This is due to \cite{barro1984time}'s observation that investment shocks increase the marginal productivity of investment and the rate of return, incentivizing households to save more and postpone consumption. DSGE models overturn this prediction through frictions such as nominal-wage rigidity, non-time-separable preferences, endogenous capital utilization, and investment-adjustment costs. The framework of simple models offers an alternative solution that relies on the anchoring of expectations. Since $p^*_\mu$ is close to zero in equilibrium, agents' expectations do not move much in response to the positive investment shocks. This dampens the initial response of forward-looking variables such as consumption to the shock. As time passes, the investment boom increases the capital stock and aggregate output, leading to an increase in the value of agents' nowcast. This improvement in agents' nowcast makes them optimistic about their future income, thus leading to an increase in consumption through the permanent-income equation \eqref{eq:TE_c_hat_body}.

These impulse-response functions offer suggestive evidence that simple models could improve the empirical fit of the business-cycle model economy. However, to establish this conclusively requires the use of a statistical model-selection criterion. I do so by estimating the parameters of the economic model separately under rational expectations and under simple models and performing Bayesian model selection.

\subsection{Bayesian Inference}\label{subsec:Bayesian_estimation}

The model is estimated with Bayesian estimation techniques using seven key macroeconomic quarterly US time series as observable variables: the log differences of real GDP, real consumption, real investment, and the real wage; log hours worked; the log difference of the GDP deflator; and the federal funds rate. The construction of the time series closely follows \cite{justiniano2010investment}. In particular, consumption includes services and non-durables but excludes consumer durables, whereas investment includes consumer durables. The sample period is 1954:III--2007:IV. Online Appendix \ref{app:data} includes more details on the data used to construct the likelihood function as well as the prior densities and posterior estimates of model parameters.

Since the likelihood uses seven macroeconomic series, the theoretical economic model needs seven exogenous shocks to avoid issues with stochastic singularity. I enrich the model with four additional shocks: an intertemporal preference shock, price- and wage-markup shocks, and a government spending shock, ending up with the seven shocks commonly assumed in the DSGE literature. The intertemporal preference and government spending shocks follow AR(1) processes with Gaussian innovations. Following \cite{Smets2007} and \cite{justiniano2010investment}, I assume that markup shocks follow ARMA(1,1) processes with Gaussian innovations. The moving average component of these shocks help capture high-frequency fluctuations in price and wage inflation.

I partition the model parameters into two groups. The first group consists of $\beta$, $\gamma$, $\delta$, $\varphi$, $F$, $g/y$, and $b/y$. These parameters are set using level information not used in the Bayesian estimation step. I set $\beta=0.99$, which implies a steady-state annualized real interest rate of about four percent. I set $\gamma=1.005$, which implies an annual real GDP growth rate of two percent. I set $\delta=0.025$, implying an annual rate of depreciation on capital equal to 10 percent. No value is picked for $\varphi$ because the value of $\varphi$ does not affect anything other than the steady-state working hours. The fixed cost of production $F$ is set to guarantee that profits are zero along the balanced-growth path. I set the steady-state ratio of government spending to GDP $g/y$ to $0.21$ and the steady-state ratio of public debt to GDP $b/y$ to $0.39$. These values correspond to the average ratios of government spending to GDP and public debt to GDP, respectively, in the sample used in Bayesian estimation.

The priors on the remaining parameters are fairly diffuse and in line with those adopted in \cite{Smets2007} and \cite{justiniano2010investment}. Following those papers, the intertemporal preference, the price-markup, and the wage-markup shocks are normalized to enter with a unit coefficient in the consumption, inflation, and wage equations, respectively. The prior distributions of all the persistence parameters are beta, with mean 0.6 and standard deviation 0.15.  The priors on the standard deviations of innovations are disperse and chosen to generate volatilities for the endogenous variables broadly in line with the data. Specifically, the priors for the standard deviations of innovations are inverse gamma, with mean 0.5 and standard deviation 1.0.  

\begin{table}[htbp!]
    \centering
    \caption{Posterior variance decomposition at business-cycle frequencies.\label{tab:var_decompose}}
    \begin{threeparttable}
        \resizebox{\linewidth}{!}{\renewcommand{\arraystretch}{0.8}
\begin{tabular}{lccccccc}
\toprule
Series\textbackslash shock & Government & Preference & Investment & Technology & Price markup & Wage markup & Monetary \\
\midrule
GDP & \parbox[t]{20mm}{\centering 14.2 \vspace{-1mm} \\ {\small [12.0, 15.7]}\vspace{2mm}} & \parbox[t]{20mm}{\centering 14.5 \vspace{-1mm} \\ {\small [12.3, 15.9]}\vspace{2mm}} & \parbox[t]{20mm}{\centering 65.5 \vspace{-1mm} \\ {\small [62.5, 69.6]}\vspace{2mm}} & \parbox[t]{20mm}{\centering 0.0 \vspace{-1mm} \\ {\small [0.0, 0.0]}\vspace{2mm}} & \parbox[t]{20mm}{\centering 0.2 \vspace{-1mm} \\ {\small [0.1, 0.3]}\vspace{2mm}} & \parbox[t]{20mm}{\centering 0.1 \vspace{-1mm} \\ {\small [0.0, 0.1]}\vspace{2mm}} & \parbox[t]{20mm}{\centering 5.5 \vspace{-1mm} \\ {\small [4.2, 6.3]}\vspace{2mm}} \\
Consumption & \parbox[t]{20mm}{\centering 0.0 \vspace{-1mm} \\ {\small [0.0, 0.0]}\vspace{2mm}} & \parbox[t]{20mm}{\centering 70.2 \vspace{-1mm} \\ {\small [61.9, 73.6]}\vspace{2mm}} & \parbox[t]{20mm}{\centering 12.3 \vspace{-1mm} \\ {\small [8.0, 22.1]}\vspace{2mm}} & \parbox[t]{20mm}{\centering 0.1 \vspace{-1mm} \\ {\small [0.0, 0.1]}\vspace{2mm}} & \parbox[t]{20mm}{\centering 1.0 \vspace{-1mm} \\ {\small [0.5, 1.3]}\vspace{2mm}} & \parbox[t]{20mm}{\centering 0.1 \vspace{-1mm} \\ {\small [0.1, 0.2]}\vspace{2mm}} & \parbox[t]{20mm}{\centering 16.2 \vspace{-1mm} \\ {\small [11.6, 19.1]}\vspace{2mm}} \\
Investment & \parbox[t]{20mm}{\centering 0.0 \vspace{-1mm} \\ {\small [0.0, 0.0]}\vspace{2mm}} & \parbox[t]{20mm}{\centering 0.0 \vspace{-1mm} \\ {\small [0.0, 0.0]}\vspace{2mm}} & \parbox[t]{20mm}{\centering 99.5 \vspace{-1mm} \\ {\small [99.3, 99.7]}\vspace{2mm}} & \parbox[t]{20mm}{\centering 0.0 \vspace{-1mm} \\ {\small [0.0, 0.0]}\vspace{2mm}} & \parbox[t]{20mm}{\centering 0.0 \vspace{-1mm} \\ {\small [0.0, 0.0]}\vspace{2mm}} & \parbox[t]{20mm}{\centering 0.1 \vspace{-1mm} \\ {\small [0.0, 0.1]}\vspace{2mm}} & \parbox[t]{20mm}{\centering 0.4 \vspace{-1mm} \\ {\small [0.2, 0.6]}\vspace{2mm}} \\
Hours & \parbox[t]{20mm}{\centering 7.4 \vspace{-1mm} \\ {\small [6.5, 8.2]}\vspace{2mm}} & \parbox[t]{20mm}{\centering 7.5 \vspace{-1mm} \\ {\small [6.6, 8.4]}\vspace{2mm}} & \parbox[t]{20mm}{\centering 36.9 \vspace{-1mm} \\ {\small [33.2, 39.9]}\vspace{2mm}} & \parbox[t]{20mm}{\centering 45.2 \vspace{-1mm} \\ {\small [41.8, 48.2]}\vspace{2mm}} & \parbox[t]{20mm}{\centering 0.1 \vspace{-1mm} \\ {\small [0.1, 0.2]}\vspace{2mm}} & \parbox[t]{20mm}{\centering 0.0 \vspace{-1mm} \\ {\small [0.0, 0.0]}\vspace{2mm}} & \parbox[t]{20mm}{\centering 2.9 \vspace{-1mm} \\ {\small [2.2, 3.4]}\vspace{2mm}} \\
Inflation & \parbox[t]{20mm}{\centering 0.1 \vspace{-1mm} \\ {\small [0.1, 0.1]}\vspace{2mm}} & \parbox[t]{20mm}{\centering 0.1 \vspace{-1mm} \\ {\small [0.1, 0.1]}\vspace{2mm}} & \parbox[t]{20mm}{\centering 3.6 \vspace{-1mm} \\ {\small [0.1, 6.6]}\vspace{2mm}} & \parbox[t]{20mm}{\centering 7.6 \vspace{-1mm} \\ {\small [4.0, 9.8]}\vspace{2mm}} & \parbox[t]{20mm}{\centering 76.6 \vspace{-1mm} \\ {\small [66.3, 85.1]}\vspace{2mm}} & \parbox[t]{20mm}{\centering 12.0 \vspace{-1mm} \\ {\small [6.4, 14.8]}\vspace{2mm}} & \parbox[t]{20mm}{\centering 0.0 \vspace{-1mm} \\ {\small [0.0, 0.0]}\vspace{2mm}} \\
Real wage & \parbox[t]{20mm}{\centering 0.0 \vspace{-1mm} \\ {\small [0.0, 0.0]}\vspace{2mm}} & \parbox[t]{20mm}{\centering 0.0 \vspace{-1mm} \\ {\small [0.0, 0.0]}\vspace{2mm}} & \parbox[t]{20mm}{\centering 4.8 \vspace{-1mm} \\ {\small [3.1, 5.7]}\vspace{2mm}} & \parbox[t]{20mm}{\centering 1.2 \vspace{-1mm} \\ {\small [0.4, 1.8]}\vspace{2mm}} & \parbox[t]{20mm}{\centering 13.8 \vspace{-1mm} \\ {\small [9.2, 16.9]}\vspace{2mm}} & \parbox[t]{20mm}{\centering 80.2 \vspace{-1mm} \\ {\small [76.1, 85.5]}\vspace{2mm}} & \parbox[t]{20mm}{\centering 0.0 \vspace{-1mm} \\ {\small [0.0, 0.0]}\vspace{2mm}} \\
Interest rate & \parbox[t]{20mm}{\centering 0.0 \vspace{-1mm} \\ {\small [0.0, 0.0]}\vspace{2mm}} & \parbox[t]{20mm}{\centering 0.0 \vspace{-1mm} \\ {\small [0.0, 0.0]}\vspace{2mm}} & \parbox[t]{20mm}{\centering 0.3 \vspace{-1mm} \\ {\small [0.0, 0.5]}\vspace{2mm}} & \parbox[t]{20mm}{\centering 0.5 \vspace{-1mm} \\ {\small [0.1, 0.6]}\vspace{2mm}} & \parbox[t]{20mm}{\centering 4.1 \vspace{-1mm} \\ {\small [2.1, 5.2]}\vspace{2mm}} & \parbox[t]{20mm}{\centering 0.9 \vspace{-1mm} \\ {\small [0.3, 1.2]}\vspace{2mm}} & \parbox[t]{20mm}{\centering 94.3 \vspace{-1mm} \\ {\small [92.3, 96.8]}\vspace{2mm}} \\
\bottomrule
\end{tabular}} 
        \begin{tablenotes}
        \footnotesize
        \item \parbox[t]{\textwidth}{\textit{Notes:} One-dimensional simple models. Business-cycle frequencies correspond to periodic components with cycles between 6 and 32 quarters, as in \cite{stock1999business}. Variance decomposition is performed at the posterior mode. 68~percent HPDIs computed using Laplace's approximation in brackets. HPDI bounds need not add up to one.}
        \end{tablenotes}
    \end{threeparttable} 
\end{table}

Table \ref{tab:var_decompose} reports the contribution of each shock to the variance of each macroeconomic variable at business cycle-frequencies. The first three columns make clear that the three aggregate demand shocks account for the largest share of the fluctuations in aggregate quantities: 94\% for output, 82\% for consumption, more than 99\% for investment, and 52\% for hours. The only other shock with significant contribution to fluctuations in these variables is the technology shock, which explains 45\% of fluctuations in hours but almost no part of the fluctuations in the other three. The aggregate demand shocks are non-inflationary: together they explain less than 4\% of fluctuations in inflation. This does not rely on the monetary policy aggressively leaning against them: the three aggregate demand shocks explain a negligible part of the fluctuations in the nominal interest rate.\footnote{These findings also hold in a (time-domain) forecast-error variance decomposition, as seen in Figure \ref{fig:var_decompose}.}

These findings are consistent with \citet*{angeletos2020business}'s anatomy of the US business cycles in the post-war period. They identify non-inflationary aggregate demand shocks as the main drivers of business cycles. These are shocks that lead to increases in output, consumption, investment, and hours with essentially no effect on inflation and no movement in TFP. Standard DSGE shocks cannot simultaneously meet all these requirements under rational expectations: Technology shocks move TFP while expansionary aggregate demand shocks generate inflation through the New Keynesian Phillips curve. Simple models increase the persistence of subjective expectations, thus significantly weakening the impact of the feedback embedded in the New Keynesian Phillips curve from expectations to current inflation. This allows demand-driven fluctuations in aggregate quantities with essentially no movement in productivity, inflation, or the nominal interest rate.

The observations made in this section give some credence to the idea that models featuring boundedly rational agents can better account for various aspects of business-cycle fluctuations. I end this section by comparing the fit of the estimated model to the data under rational expectations and simple models.  I do so by comparing the marginal likelihoods of the Bayesian posteriors, reported in columns two and three of Table \ref{tab:marg_likelihood}. The marginal likelihood is more than 150 log points higher with simple models than rational expectations, implying overwhelming posterior odds in favor of the former. Columns four to seven report the marginal likelihood given alternative specifications of the economic model, which feature additional add-ons but maintain rational expectations. While many of these add-ons can improve the fit, none of them do so as much as bounded rationality---although consumption-habit formation comes close. This finding suggests that the framework introduced in this paper is not only  favored by data over rational expectations but also can act as a parsimonious substitute for the ad hoc frictions commonly assumed in DSGE models.

\newcolumntype{C}[1]{>{\centering\arraybackslash}m{#1}}

\begin{table}[htbp!]
    \centering
    \caption{Marginal likelihoods.\label{tab:marg_likelihood}}
    \begin{threeparttable}
    \resizebox{\linewidth}{!}{
    \renewcommand{\arraystretch}{0.95}
    \begin{tabular}{C{2.5cm}C{2.5cm}C{2.0cm}C{2.0cm}C{2.0cm}C{2.0cm}C{2.0cm}C{2.0cm}}
    \toprule
     & Simple models & \multicolumn{6}{c}{Rational expectations} \\
      \cmidrule(lr){2-2} \cmidrule(lr){3-8}
     Add-on & --- & --- & Indexation & Utilization & Investment adjustment & Taylor & Habit\\
     \midrule
    Log marginal likelihood  & $-$1319.5 & $-$1470.4  & $-$1470.8 & $-$1468.5 & $-$1457.1 & $-$1400.0 & $-$1321.8\\
    \bottomrule
    \end{tabular}}
    \begin{tablenotes}
    \footnotesize
    \item \parbox[t]{\textwidth}{\textit{Notes:} Log marginal likelihoods are computed using Laplace's approximation. The specifications in columns two and three only feature price and wage rigidities. The specifications in columns four to eight each feature a single addition: price and wage indexation, endogenous capital utilization, investment-adjustment costs (instead of neoclassical capital-adjustment costs), a Taylor rule that responds to the level and growth rate of the output gap, and external habit formation in consumption, respectively.}
    \end{tablenotes}
    \end{threeparttable} 
\end{table}

\section{Conclusion}\label{sec:conclusion}
This paper suggests a novel approach to modeling bounded rationality. This approach is portable across different applications. I illustrated the use of the framework in a medium-scale new neoclassical synthesis economy. The additional persistence and comovement arising from agents' use of simple models simultaneously resolves several puzzles (the weak propagation of TFP shocks, the inability of demand shocks to generate the right comovement between investment and consumption, and the forward guidance puzzle) while improving the fit of the model to data.

An earlier version of this paper \citep{molavi2023simple} integrated simple models into three other workhorse models in macroeconomics, including the Diamond--Mortensen--Pissarides (DMP) model. Replacing rational expectations with simple models in the DMP model adds persistence to the unemployment rate, number of vacancies, and job-finding rate in response to both productivity and separation shocks, thus improving the internal propagation mechanism of the DMP model. Moreover, it enables the DMP model to generate negative comovement between the unemployment rate and vacancies in response to separation shocks---reversing a counterfactual prediction of the model under rational expectations.

This paper focuses on representative-agent macro models to allow for a transparent discussion of how simple models work. However, one can easily incorporate simple models into modern heterogeneous-agent macro models. This allows one to examine how bounded rationality in the face of intertemporal complexity affects predictions of such models. This can be done because neither the additional degrees of freedom nor the computational burden of finding the equilibrium with simple models scale with the size of the macro model. This paper's approach can also be extended to allow for heterogeneity in $d$ without having to contend with the complications associated with heterogeneous-belief macro models (such as the ``infinite regress'' problem).

Throughout the paper, I took dimension $d$ of agents' models as a primitive parameter. This parameter can be identified using expectations data.\footnote{See \citet*{molavi2021model} for a discussion of how this can be done in a closely related framework.} I leave the problem of estimating $d$ to future research.

\newpage
\appendix
\numberwithin{equation}{section}
\numberwithin{figure}{section}
\numberwithin{table}{section}

\section*{Omitted Details and Additional Results for the Business-Cycle Application}

\section{Temporary Equilibrium}\label{appendix:DSGE_TE}
In this appendix, I list the equations that characterize the log-linearized temporary equilibrium of the business-cycle economy studied in Section \ref{sec:DSGE}. The derivations are straightforward, and so, are omitted for brevity. These temporary-equilibrium conditions impose individual optimality and market clearing conditions but not rational expectations. They are valid under arbitrary specifications of expectations---as long as agents all understand their own problems, and their expectations satisfy linear-invariance and the law of iterated expectations.
I provide the conditions that characterize the fully flexible model used in Bayesian estimation. The baseline specification can be obtained by setting the values of parameters $\xi_w$, $g/y$, $b/y$, $\sigma_\psi$, $\sigma_g$, $\sigma_p$, and $\sigma_w$ equal to zero.

The steady-state values are given by
\begin{equation*}
\begin{aligned}
& \rho = \frac{\gamma}{\beta} - (1-\delta),\\
& w = \left[\frac{1}{1+\lambda_p}\alpha^\alpha(1-\alpha)^{1-\alpha}\frac{1}{\rho^\alpha}\right]^{\frac{1}{1-\alpha}},\\
& \frac{k}{L} = \frac{w}{\rho}\frac{\alpha}{1-\alpha},\\
& \frac{F}{L}=\left(\frac{k}{L}\right)^\alpha - \rho\frac{k}{L}-w,\\
& \frac{y}{L} = \left(\frac{k}{L}\right)^\alpha-\frac{F}{L},\\
& \frac{i}{L} = \left(\gamma-(1-\delta)\right)\frac{k}{L},\\
& \frac{c}{L} = \frac{y}{L} - \frac{i}{L}-\frac{g}{y}\frac{y}{L},\\
& \frac{x}{L} = \frac{y}{L} - \frac{i}{L},\\
& \frac{\tau}{L} = \left(\frac{g}{y} + \frac{1-\beta}{\beta}\frac{b}{y}\right)\frac{y}{L}.
\end{aligned}
\end{equation*}

The log-linear permanent-income equation is given by
\begin{align}
\hat{c}_t = \hat{\psi}_t -\hat{R}_t + E_{ct} \left[\sum_{s=1}^\infty \beta^{s}\left(\frac{1-\beta}{\beta}\frac{x}{c}\hat{x}_{t+s}-\frac{1-\beta}{\beta}\frac{\tau}{c}\hat{\tau}_{t+s}-\frac{1-\beta}{\beta}\hat{\psi}_{t+s}-\frac{x-\tau}{c}\hat{R}_{t+s}+\frac{1}{\beta}\frac{x-\tau}{c}\hat{\pi}_{t+s}\right)\right].\label{eq:TE_c_hat}
\end{align}
Households' pre-tax income is given by
\begin{equation}
\hat{x}_t = \frac{y}{x}\hat{y}_t - \frac{i}{x}\hat{i}_t.\label{eq:TE_x_hat}
\end{equation}
The intertemporal preference shock follows the exogenous process
\begin{equation}
\hat{\psi}_t = \rho_\psi \hat{\psi}_{t-1} + \varepsilon_{\psi t},\qquad \varepsilon_{\psi t}\sim\mathcal{N}(0,\sigma^2_\psi).\label{eq:TE_psi_hat}
\end{equation}
Investment is given by
\begin{align}
\hat{i}_t =  \hat{k}_t + \frac{1}{\varsigma_k}\left(\hat{\mu}_t-\hat{\psi}_t+\hat{c}_t\right) + E_{it}\left[\sum_{s=1}^\infty \beta^{s}\left(\frac{1-\beta}{\varsigma_k\beta}\hat{\psi}_{t+s}-\frac{1-\beta}{\varsigma_k\beta}\hat{c}_{t+s}+\frac{1}{\varsigma_k}\left(\frac{1}{\beta}-\frac{1-\delta}{\gamma}\right)\hat{\rho}_{t+s}+\frac{1}{\varsigma_k}\left(1-\frac{1-\delta}{\gamma}\right)\hat{\mu}_{t+s}\right)\right],\label{eq:TE_i_hat}
\end{align}
where the investment shock follows the AR(1) process
\begin{equation}
\hat{\mu}_t = \rho_\mu \hat{\mu}_{t-1} + \varepsilon_{\mu t},\qquad \varepsilon_{\mu t}\sim\mathcal{N}(0,\sigma^2_\mu).\label{eq:TE_mu_hat}
\end{equation}
Capital stock evolves according to 
\begin{equation}
\hat{k}_{t} = \frac{1-\delta}{\gamma}\hat{k}_{t-1} + \left(1-\frac{1-\delta}{\gamma}\right)\left(\hat{i}_{t-1}+\hat{\mu}_{t-1}\right).\label{eq:TE_k_hat}
\end{equation}
Government spending follows the exogenous process
\begin{equation}
\hat{g}_t = \rho_g \hat{g}_{t-1} + \varepsilon_{gt},\qquad \varepsilon_{g t}\sim\mathcal{N}(0,\sigma^2_g),\label{eq:TE_g_hat}
\end{equation}
and GDP is given by 
\begin{equation}
\hat{y}_t = \frac{c}{y}\hat{c}_t + \frac{i}{y}\hat{i}_t + \frac{g}{y}\hat{g}_t.\label{eq:TE_y_hat}
\end{equation}
Inflation is given by
\begin{align}
\hat{\pi}_t = \hat{\lambda}_{pt} + \kappa\left(\alpha \hat{\rho}_t + (1-\alpha) \hat{w}_t -\hat{a}_t\right) & + E_{pt}\left[\sum_{s=1}^\infty\xi_p^s\beta^s\left(\frac{1-\xi_p}{\xi_p}\hat{\pi}_{t+s}+\hat{\lambda}_{p,t+s}+\kappa\left(\alpha \hat{\rho}_{t+s} + (1-\alpha) \hat{w}_{t+s} -\hat{a}_{t+s}\right)\right)\right],\label{eq:TE_pi_hat}
\end{align}
where $\kappa \equiv \frac{(1-\xi_p)(1-\xi_p\beta)}{\xi_p}$ is a constant, TFP follows the exogenous process
\begin{equation}
\hat{a}_t = \rho_a \hat{a}_{t-1} + \varepsilon_{at},\qquad \varepsilon_{a t}\sim\mathcal{N}(0,\sigma^2_a),\label{eq:TE_a_hat}
\end{equation}
and the price markup shock follows the exogenous process
\begin{equation}
\hat{\lambda}_{pt} = \rho_p \hat{\lambda}_{p,t-1} + \varepsilon_{pt},\qquad \varepsilon_{pt}\sim\mathcal{N}(0,\sigma^2_p).\label{eq:TE_lambda_p_hat}
\end{equation}
The real wage is given by
\begin{align}
\hat{w}_t = &\,\frac{1+\beta}{1+\beta-\xi_w\beta}\left(\hat{\lambda}_{wt}+\kappa_w\hat{\ell}_{t}\right)+\frac{1}{1+\beta-\xi_w\beta}\left(\hat{w}_{t-1}-\hat{\pi}_t\right)\nonumber\\ & +\frac{1+\beta}{1+\beta-\xi_w\beta}E_{wt}\left[\sum_{s=1}^\infty \xi_w^s\beta^s\Big(\frac{\nu_w\kappa_w}{1-\xi_w\beta}\hat{\pi}_{t+s} +\hat{\lambda}_{w,t+s} +\kappa_w\hat{\ell}_{t+s} + \nu_w\kappa_w\hat{w}_{t+s}\Big)\right],\label{eq:TE_w_hat}
\end{align}
where $\kappa_w \equiv \frac{(1-\xi_w)(1-\xi_w\beta)}{\xi_w\nu_w(1+\beta)}$ is a constant, 
\begin{equation}
\hat{\ell}_{t} = \nu\hat{L}_{t} + \hat{c}_{t} - \hat{w}_t,\label{eq:TE_lambda_w_tilde_hat}  
\end{equation}
and the wage markup shock $\hat{\lambda}_{wt}$ follows the exogenous process
\begin{equation}
\hat{\lambda}_{wt} = \rho_w \hat{\lambda}_{w,t-1} + \varepsilon_{wt},\qquad \varepsilon_{wt}\sim\mathcal{N}(0,\sigma^2_w).\label{eq:TE_lambda_w_hat}
\end{equation}
Hours are given by 
\begin{equation}
\hat{L}_t = \frac{1}{1-\alpha}\left(\frac{y}{y+F}\hat{y}_t - \alpha \hat{k}_t - \hat{a}_t +\left(\frac{\rho k}{y+F}-\alpha\right)\hat{\rho}_t\right)\label{eq:TE_L_hat},
\end{equation}
and the rental rate of capital by
\begin{equation}
\hat{\rho}_t = \hat{w}_t + \hat{L}_t - \hat{k}_t.\label{eq:TE_rho_hat}
\end{equation}
The nominal interest rate follows the interest rate rule
\begin{equation}
\hat{R}_t = \rho_R \hat{R}_{t-1} + (1-\rho_R)\phi_\pi \hat{\pi}_t +\hat{m}_{t},\label{eq:TE_R_hat}
\end{equation}
where the monetary-policy shock follows the exogenous process
\begin{equation}
\hat{m}_{t} = \rho_m \hat{m}_{t-1} + \varepsilon_{m t},\qquad \varepsilon_{mt}\sim\mathcal{N}(0,\sigma^2_m).\label{eq:TE_m_hat}
\end{equation}
Finally, taxes follow the tax rule
\begin{equation}
\hat{\tau}_t = \frac{g}{\tau}\hat{g}_t + \frac{b}{\beta\tau}\left(\hat{R}_{t-1}-\hat{\pi}_t\right).\label{eq:TE_tau_hat}
\end{equation}

\section{Additional Results}\label{app:additional}

\begin{figure}[htbp!]
    \centering
    \begin{minipage}{\linewidth}
        \centering
        \resizebox{\linewidth}{!}{\input{CREE_VD.pgf}}
        \captionof{figure}{Forecast-error variance decomposition.}\label{fig:var_decompose}
        \vspace{0.5em}
        \footnotesize
        \raggedright
        \textit{Notes:} One-dimensional simple models. Variance decomposition is performed at the posterior mode.
    \end{minipage}
\end{figure}

\begin{landscape}
\begin{figure}[htbp!]
    \centering
    \begin{minipage}{\linewidth}
        \centering
        \resizebox{\linewidth}{!}{\input{IRF_w_CI.pgf}}
        \captionof{figure}{Impulse-response functions.}\label{fig:IRFs}
        \vspace{0.5em}
        \footnotesize
        \raggedright
        \textit{Notes:} Responses of endogenous variables (columns) to one-standard-deviation shocks (rows). One-dimensional simple models. The solid line represents the posterior mode. Shaded areas are 68~percent HPDIs computed using Laplace's approximation.  Output, consumption, investment, hours, and real wage measured in percents; inflation and nominal interest rates measured in percentage points. Shocks are normalized to increase output on impact at the posterior mode.
    \end{minipage}
\end{figure}
\end{landscape}

\section{Bayesian Estimation}\label{app:data}
The likelihood is based on the measurement equation
\[
\begin{pmatrix}\Delta Y_t & \Delta C_t & \Delta I_t & L_t & \pi_t & \Delta w_t & R_t\end{pmatrix}'=\begin{pmatrix}\hat{y}_t & \hat{c}_t &\hat{i}_t & \hat{L}_t & \hat{\pi}_t & \hat{w}_t & \hat{R}_t\end{pmatrix}' + \overline{\zeta},
\]
where $\Delta$ denotes the temporal difference operator, $Y$ denotes real GDP per capita, $C$ denotes real consumption per capita, $I$ denotes real investment per capita, $L$ denotes hours worked per capita, $\pi$ denotes the inflation rate, $w$ denotes the real wage index, $R$ denotes the nominal interest rate, and $\overline{\zeta}$ is the vector containing the sample mean of the vector on the left side of above equation. The vector of means $\overline{\zeta}$ is only informative about level variables that are fixed in the estimation step. Therefore, the likelihood can be constructed using the demeaned values of $\Delta Y_t$, $\Delta C_t$, $\Delta I_t$, $L_t$, $\pi_t$, $\Delta w_t$, and $R_t$. 

The data are from the Federal Reserve Economic Database (FRED).  Tables \ref{table:data1} and \ref{table:data2} describe the original data and the transformations used in Bayesian estimation.

\begin{table}[htbp!]
\centering
\caption{Description of data.}\label{table:data1}
\begin{threeparttable}
\resizebox{\linewidth}{!}{
\renewcommand{\arraystretch}{1.43}
\begin{tabular}{l l c c}
\toprule
Data & Mnemonic & Frequency & Transform \\ 
\midrule
Real gross domestic product per capita & A939RX0Q048SBEA & Q & --- \\
Share of GDP: personal consumption expenditures: nondurable goods & DNDGREI1Q156NBEA & Q & --- \\
Share of GDP: personal consumption expenditures: services & DSERREI1Q156NBEA & Q & --- \\
Share of GDP: personal consumption expenditures: durable goods & DDURREI1Q156NBEA & Q & --- \\
Share of GDP: gross private domestic investment & A006REI1Q156NBEA & Q & --- \\
Nonfarm business sector: average weekly hours & PRS85006023 & Q & --- \\
Civilian employment level & CE16OV & M & EoP \\
Civilian non-institutional population & CNP16OV & M & EoP \\
Gross domestic product: implicit price deflator & GDPDEF & Q & --- \\
Non-farm business sector: real hourly compensation for all workers & COMPRNFB & Q & --- \\
Effective federal funds rate & FEDFUNDS & M & Ave \\
\bottomrule
\end{tabular}
} 
\begin{tablenotes}
\footnotesize
\item \textit{Note:} Q: quarterly, M: monthly, EoP: end of period, Ave: quarterly average.
\end{tablenotes}
\end{threeparttable}
\end{table}

\begin{table}[htbp!]
\centering
\caption{Variables used in Bayesian estimation.}\label{table:data2}
\begin{threeparttable}
\resizebox{\linewidth}{!}{
\renewcommand{\arraystretch}{1.43}
\begin{tabular}{l l}
\toprule
Variable & Definition \\ 
\midrule
Real GDP per capita & $Y = 100 \times \log(\text{A939RX0Q048SBEA})$ \\
Real consumption per capita & $C = 100 \times \log((\text{DNDGREI1Q156NBEA} + \text{DSERREI1Q156NBEA}) \times \text{A939RX0Q048SBEA})$ \\
Real investment per capita & $I = 100 \times \log((\text{DDURREI1Q156NBEA} + \text{A006REI1Q156NBEA}) \times \text{A939RX0Q048SBEA})$ \\
Hours worked & $L = 100 \times \log(\text{PRS85006023} \times \text{CE16OV}/\text{CNP16OV})$ \\
Inflation rate & $\pi = 100 \times \log(\text{GDPDEF}/\text{GDPDEF}(-1))$ \\
Real wage & $w=100 \times \log(\text{COMPRNFB})$ \\
Interest rate & $R = \text{FEDFUNDS}/4$ \\
\bottomrule
\end{tabular}
}
\end{threeparttable}
\end{table}

\begin{table}[htbp!]
    \centering
    \caption{Prior densities and posterior estimates.}
    \begin{threeparttable}
    \vspace*{\fill}
    \begin{minipage}[c]{\linewidth}
    \centering
    \renewcommand{\arraystretch}{1.15}
    \resizebox*{!}{\dimexpr\textheight-2\baselineskip\relax}{\begin{tabular}{llcccccc}
\toprule
& & \multicolumn{3}{c}{Prior distribution} & \multicolumn{2}{c}{Posterior mode}\\
\cmidrule(lr){3-5} \cmidrule(lr){6-7}
Coeff. & Description & Distr. & Mean & Std. Dev. & 1d & RE \\
\midrule
$\nu$ & Inverse Frisch elasticity & G & 2.00 & 0.50 & \parbox[t]{20mm}{\centering 1.91 \vspace{-1mm} \\ {\small [1.48, 2.46]}\vspace{2mm}} & \parbox[t]{20mm}{\centering 0.56 \vspace{-1mm} \\ {\small [0.44, 0.71]}\vspace{2mm}} \\
$\alpha$ & Capital share & N & 0.30 & 0.05 & \parbox[t]{20mm}{\centering 0.28 \vspace{-1mm} \\ {\small [0.27, 0.29]}\vspace{2mm}} & \parbox[t]{20mm}{\centering 0.28 \vspace{-1mm} \\ {\small [0.27, 0.29]}\vspace{2mm}} \\
$\lambda_p$ & Steady-state price markup & B & 0.15 & 0.05 & \parbox[t]{20mm}{\centering 0.50 \vspace{-1mm} \\ {\small [0.45, 0.55]}\vspace{2mm}} & \parbox[t]{20mm}{\centering 0.41 \vspace{-1mm} \\ {\small [0.36, 0.47]}\vspace{2mm}} \\
$\lambda_w$ & Steady-state wage markup & B & 0.15 & 0.05 & \parbox[t]{20mm}{\centering 0.11 \vspace{-1mm} \\ {\small [0.07, 0.16]}\vspace{2mm}} & \parbox[t]{20mm}{\centering 0.21 \vspace{-1mm} \\ {\small [0.16, 0.26]}\vspace{2mm}} \\
$\xi_p$ & Calvo, prices & B & 0.50 & 0.10 & \parbox[t]{20mm}{\centering 0.77 \vspace{-1mm} \\ {\small [0.74, 0.79]}\vspace{2mm}} & \parbox[t]{20mm}{\centering 0.65 \vspace{-1mm} \\ {\small [0.61, 0.68]}\vspace{2mm}} \\
$\xi_w$ & Calvo, wages & B & 0.50 & 0.10 & \parbox[t]{20mm}{\centering 0.77 \vspace{-1mm} \\ {\small [0.72, 0.81]}\vspace{2mm}} & \parbox[t]{20mm}{\centering 0.15 \vspace{-1mm} \\ {\small [0.11, 0.19]}\vspace{2mm}} \\
$\rho_R$ & Taylor-rule smoothing & B & 0.60 & 0.20 & \parbox[t]{20mm}{\centering 0.87 \vspace{-1mm} \\ {\small [0.83, 0.90]}\vspace{2mm}} & \parbox[t]{20mm}{\centering 0.54 \vspace{-1mm} \\ {\small [0.48, 0.61]}\vspace{2mm}} \\
$\phi_\pi$ & Taylor rule, inflation & N & 1.50 & 0.20 & \parbox[t]{20mm}{\centering 1.07 \vspace{-1mm} \\ {\small [1.03, 1.10]}\vspace{2mm}} & \parbox[t]{20mm}{\centering 1.67 \vspace{-1mm} \\ {\small [1.57, 1.77]}\vspace{2mm}} \\
$\varsigma_k$ & Capital-adjustment cost & G & 4.00 & 1.00 & \parbox[t]{20mm}{\centering 1.70 \vspace{-1mm} \\ {\small [1.47, 1.98]}\vspace{2mm}} & \parbox[t]{20mm}{\centering 1.06 \vspace{-1mm} \\ {\small [0.87, 1.29]}\vspace{2mm}} \\
$\rho_a$ & Technology shock, AR & B & 0.60 & 0.15 & \parbox[t]{20mm}{\centering 0.93 \vspace{-1mm} \\ {\small [0.90, 0.95]}\vspace{2mm}} & \parbox[t]{20mm}{\centering 0.91 \vspace{-1mm} \\ {\small [0.90, 0.93]}\vspace{2mm}} \\
$\rho_m$ & Monetary-policy shock, AR & B & 0.60 & 0.15 & \parbox[t]{20mm}{\centering 0.30 \vspace{-1mm} \\ {\small [0.24, 0.37]}\vspace{2mm}} & \parbox[t]{20mm}{\centering 0.26 \vspace{-1mm} \\ {\small [0.20, 0.33]}\vspace{2mm}} \\
$\rho_g$ & Government-spending shock, AR & B & 0.60 & 0.15 & \parbox[t]{20mm}{\centering 0.97 \vspace{-1mm} \\ {\small [0.95, 0.98]}\vspace{2mm}} & \parbox[t]{20mm}{\centering 0.97 \vspace{-1mm} \\ {\small [0.96, 0.98]}\vspace{2mm}} \\
$\rho_p$ & Price-markup shock, AR & B & 0.60 & 0.15 & \parbox[t]{20mm}{\centering 0.91 \vspace{-1mm} \\ {\small [0.87, 0.93]}\vspace{2mm}} & \parbox[t]{20mm}{\centering 0.96 \vspace{-1mm} \\ {\small [0.95, 0.97]}\vspace{2mm}} \\
$\rho_w$ & Wage-markup shock, AR & B & 0.60 & 0.15 & \parbox[t]{20mm}{\centering 0.97 \vspace{-1mm} \\ {\small [0.96, 0.98]}\vspace{2mm}} & \parbox[t]{20mm}{\centering 0.97 \vspace{-1mm} \\ {\small [0.96, 0.98]}\vspace{2mm}} \\
$\rho_\psi$ & Preference shock, AR & B & 0.60 & 0.15 & \parbox[t]{20mm}{\centering 0.96 \vspace{-1mm} \\ {\small [0.94, 0.97]}\vspace{2mm}} & \parbox[t]{20mm}{\centering 0.92 \vspace{-1mm} \\ {\small [0.90, 0.94]}\vspace{2mm}} \\
$\rho_\mu$ & Investment shock, AR & B & 0.60 & 0.15 & \parbox[t]{20mm}{\centering 0.87 \vspace{-1mm} \\ {\small [0.85, 0.89]}\vspace{2mm}} & \parbox[t]{20mm}{\centering 0.90 \vspace{-1mm} \\ {\small [0.88, 0.92]}\vspace{2mm}} \\
$\theta_p$ & Price-markup shock, MA & B & 0.50 & 0.20 & \parbox[t]{20mm}{\centering 0.41 \vspace{-1mm} \\ {\small [0.35, 0.48]}\vspace{2mm}} & \parbox[t]{20mm}{\centering 0.22 \vspace{-1mm} \\ {\small [0.14, 0.34]}\vspace{2mm}} \\
$\theta_w$ & Wage-markup shock, MA & B & 0.50 & 0.20 & \parbox[t]{20mm}{\centering 0.64 \vspace{-1mm} \\ {\small [0.57, 0.70]}\vspace{2mm}} & \parbox[t]{20mm}{\centering 0.32 \vspace{-1mm} \\ {\small [0.24, 0.41]}\vspace{2mm}} \\
$\sigma_a$ & Technology shock, SD & IG & 0.50 & 1.00 & \parbox[t]{20mm}{\centering 0.54 \vspace{-1mm} \\ {\small [0.51, 0.57]}\vspace{2mm}} & \parbox[t]{20mm}{\centering 0.56 \vspace{-1mm} \\ {\small [0.53, 0.59]}\vspace{2mm}} \\
$\sigma_m$ & Monetary-policy shock, SD & IG & 0.50 & 1.00 & \parbox[t]{20mm}{\centering 0.22 \vspace{-1mm} \\ {\small [0.21, 0.23]}\vspace{2mm}} & \parbox[t]{20mm}{\centering 0.31 \vspace{-1mm} \\ {\small [0.29, 0.35]}\vspace{2mm}} \\
$\sigma_g$ & Government-spending shock, SD & IG & 0.50 & 1.00 & \parbox[t]{20mm}{\centering 1.53 \vspace{-1mm} \\ {\small [1.46, 1.61]}\vspace{2mm}} & \parbox[t]{20mm}{\centering 1.52 \vspace{-1mm} \\ {\small [1.45, 1.60]}\vspace{2mm}} \\
$\sigma_p$ & Price-markup shock, SD & IG & 0.50 & 1.00 & \parbox[t]{20mm}{\centering 0.26 \vspace{-1mm} \\ {\small [0.24, 0.27]}\vspace{2mm}} & \parbox[t]{20mm}{\centering 0.23 \vspace{-1mm} \\ {\small [0.20, 0.27]}\vspace{2mm}} \\
$\sigma_w$ & Wage-markup shock, SD & IG & 0.50 & 1.00 & \parbox[t]{20mm}{\centering 0.42 \vspace{-1mm} \\ {\small [0.39, 0.45]}\vspace{2mm}} & \parbox[t]{20mm}{\centering 0.91 \vspace{-1mm} \\ {\small [0.72, 1.15]}\vspace{2mm}} \\
$\sigma_\psi$ & Preference shock, SD & IG & 0.50 & 1.00 & \parbox[t]{20mm}{\centering 0.56 \vspace{-1mm} \\ {\small [0.53, 0.59]}\vspace{2mm}} & \parbox[t]{20mm}{\centering 1.45 \vspace{-1mm} \\ {\small [1.21, 1.72]}\vspace{2mm}} \\
$\sigma_\mu$ & Investment shock, SD & IG & 0.50 & 1.00 & \parbox[t]{20mm}{\centering 5.74 \vspace{-1mm} \\ {\small [4.92, 6.70]}\vspace{2mm}} & \parbox[t]{20mm}{\centering 4.23 \vspace{-1mm} \\ {\small [3.54, 5.05]}\vspace{2mm}} \\
\bottomrule
\end{tabular}}
    \end{minipage}
    \vspace*{\fill}
    \begin{tablenotes}
    \footnotesize
    \item \textit{Notes:} B: beta, G: gamma, IG: inverse gamma, N: normal. 68~percent HPDIs computed using Laplace's approximation in brackets.
    \end{tablenotes}
    \end{threeparttable}      
\end{table}

\newpage
\fontsize{10}{10}\selectfont\baselineskip0.36cm
\bibliographystyle{te}
\bibliography{misspec.bib}

\newpage 
\fontsize{11.5}{11.5}\selectfont\baselineskip0.615cm

\section*{Online Appendices}

\section{Weighted Mean-Squared Forecast Error}\label{app:MSE}
The agent's time-$t$ one-step-ahead forecast error given model $\theta$ is defined as 
\[
e_t(\theta) \equiv y_{t+1}-E^\theta_t[y_{t+1}],
\]
where $E^\theta_t$ denotes the agent's subjective expectation conditional on her information at time $t$ and given model $\theta$. The weighted average of mean-squared forecast errors given a symmetric weight matrix $W\in\mathbb{R}^{n\times n}$ is defined as
\[
\text{MSE}_W(\theta)=\mathbb{E}\left[e'_t(\theta) W e_t(\theta)\right].
\]

Instead of assuming that the agent uses a model that minimizes the KLDR, one can assume that she makes her forecasts using a model $\theta$ that minimizes $\text{MSE}_W(\theta)$ for some matrix $W$. Using the mean-squared forecast error as the notion of fit has two disadvantages relative to the KLDR. First, the choice of matrix $W$ introduces additional degrees of freedom when the observable is not a scalar. Second, the minimizer of the weighted mean-squared error is in general not invariant to linear transformations of the vector of observable (unless if the weight matrix $W$ is transformed accordingly). However, the following proposition establishes that mean-squared forecast-error minimization coincides with KLDR minimization under the appropriate choice of the weighting matrix $W$:
\begin{propapp}
Let $\theta$ denote a pseudo-true $d$-state model, and let $\hat{\Sigma}_y^{\theta}$ denote the implied subjective variance of $y_{t+1}$ conditional on the agent's information at time $t$. If $W$ is equal to the inverse of $\hat{\Sigma}_y^{\theta}$, then
$\theta\in\arg\min_{\theta\in \Theta_d}\text{MSE}_W(\theta)$.
\end{propapp}
The proof of the proposition is standard, and so, is omitted. 

\section{Exponential Ergodicity}\label{app:exp_erg}
This appendix provides a set of sufficient conditions for a process to be exponentially ergodic and discusses the relationship between those conditions and the notion of full information. 
\begin{propapp}
\label{propapp:sufficient_for_exp_erg}
Consider a process $\mathbb{P}$ that can be represented as
\begin{equation}\label{eq:true_model_VAR_app}
\begin{aligned}
    & f_{t} = Ff_{t-1} + \epsilon_t,\\
    & y_t = H'f_t,
\end{aligned}
\end{equation}
where $f_t\in\mathbb{R}^m$, $\epsilon_t$ is a zero mean i.i.d. shock with a finite variance-covariance matrix, $F\in\mathbb{R}^{m\times m}$ is a convergent matrix, $H\in\mathbb{R}^{m\times n}$, and the variance-covariance of $f_t$ is normalized to be the identity matrix. If $H$ is a rank-$m$ matrix and $\big\Vert \frac{F+F'}{2}\big\Vert_2=\Vert F\Vert_2$, where $\Vert\cdot\Vert_2$ denotes the spectral norm, then $\mathbb{P}$ is exponentially ergodic.
\end{propapp}
The assumption that the process has a representation of the form \eqref{eq:true_model_VAR_app} is without loss of generality. The Wold representation theorem implies that any mean zero, covariance stationary, and purely non-deterministic process has a representation of this form (possibly with $m=\infty$). The assumption that the variance-covariance of $f_t$ equals identity is also without loss of generality. It can always be arranged to hold by an appropriate normalization of $f_t$.\footnote{See Lemma \ref{lem:Akaike} of the Online Appendix and its proof for how this can be done.} The assumption on matrix $F$ rules out a severe form of defectiveness by guaranteeing that the largest eigenvalue of the symmetric part of $F$ coincides with the largest singular value of $F$. It is satisfied, for example, if $F$ is diagonal or symmetric, but it is much weaker than symmetry. 

The most substantial assumption of the proposition is the requirement that $H$ is a rank-$m$ matrix. 
This assumption can be seen as a full-information (or spanning) assumption: If the agent observes an observable of the form \eqref{eq:true_model_VAR_app} with a full-rank matrix $H$, then she has enough information to forecast the observable as well as in the full-information rational-expectations benchmark---even if she fails to do so due to the constraint on her set of models. The following proposition shows that this assumption, in general, cannot be dispensed with:

\begin{propapp}\label{propapp:incomplete_info}
Suppose the observable is one-dimensional, and the true process $\mathbb{P}$ can be represented as in \eqref{eq:true_model_VAR_app} for some $f_t\in\mathbb{R}^m$, $\epsilon_t\sim\mathcal{N}(0,\Sigma)$, diagonal divergent matrix $F\in\mathbb{R}^{m\times m}$, diagonal matrix $\Sigma\in\mathbb{R}^{m\times m}$, and matrix $H\in \mathbb{R}^{m\times n}$. If the representation in \eqref{eq:true_model_VAR_app} is minimal and $m>1$, then the $s$-period-ahead forecast of an agent who uses a pseudo-true one-state model $\theta$ is given by
\begin{equation*}
E^{\theta}_t[y_{t+s}]= {a}^{s}(1-\eta)\sum_{\tau=0}^\infty {a}^\tau {\eta}^\tau y_{t-\tau}
\end{equation*}
for some $a\in(-1,1)$ and $\eta\in(0,1)$.\footnote{The representation in \eqref{eq:true_model_VAR_app} of a process is \emph{minimal} if there exists no representation for the process of the same form in which the dimension of $f_t$ is strictly smaller.}
\end{propapp}

\section{Partial Equilibrium and General Equilibrium}\label{app:PE_GE}
In this appendix, I argue that the implications of the general framework are largely unchanged in a general equilibrium setting where the observable's law of motion depends on agents' choices. I consider a stylized general equilibrium (GE) economy in which observables are linear functions of exogenous shocks and agents' actions. Specifically, I assume that, in equilibrium, the vector of observables $y_t\in\mathbb{R}^n$ can be written as
\begin{equation}
y^{\text{GE}}_t = \tilde{H}'f_t + g x^{\text{GE}}_t,\label{eq:GE_y}
\end{equation}
where $x_t\in\mathbb{R}$ is agents' time-$t$ action, $f_t\in\mathbb{R}^m$ is the vector of exogenous shocks, $\tilde{H}\in\mathbb{R}^{m\times n}$ is a rank-$m$ matrix, and $g\in\mathbb{R}^n$ is a vector that parameterizes the strength of the GE feedback from agents' actions to the aggregate observable. Agents' best-response functions are given by
\begin{equation}\label{eq:GE_x}
x^{\text{GE}}_t = b'y^{\text{GE}}_t + E_t\left[\sum_{s=1}^\infty\beta^s c'y^{\text{GE}}_{t+s}\right]. 
\end{equation}
For simplicity, I assume that the shocks follow $m$ independent AR(1) processes: 
\begin{equation}\label{eq:GE_f}
    f_t = F f_{t-1}+\epsilon_t,\qquad \epsilon_t\sim\mathcal{N}(0,\Sigma),
\end{equation}
where $F=\text{diag}(\alpha_1,\dots,\alpha_ m)$ and $\Sigma=\text{diag}(\sigma_1^2,\dots,\sigma_m^2)$. Equations \eqref{eq:GE_y}--\eqref{eq:GE_f} together with the specification of agents' subjective expectations fully characterize the (general)  equilibrium of the economy.

I contrast this economy with a partial equilibrium (PE) economy in which
\begin{align}
    & y^{\text{PE}}_t = H'f_t,\label{eq:PE_y}\\
    & x^{\text{PE}}_t = b'y^{\text{PE}}_t + E_t\left[\sum_{s=1}^\infty\beta^s c'y^{\text{PE}}_{t+s}\right]\label{eq:PE_x},
\end{align}
and $f_t$ follows \eqref{eq:GE_f}. The term ``partial equilibrium'' is inspired by the following hypothetical scenario: Suppose we considered the economy described by equations \eqref{eq:GE_y}--\eqref{eq:GE_f} but ignored the fact that agents' actions affect the observable, which in turn affects agents' actions, and so on. Then the response of the GE economy to shocks would be described by equations \eqref{eq:PE_y}--\eqref{eq:PE_x}. The following result establishes an observational equivalence between the GE and PE economies:
\begin{propapp}
Consider the general equilibrium economy \eqref{eq:GE_y}--\eqref{eq:GE_f} and the partial equilibrium economy \eqref{eq:GE_f}--\eqref{eq:PE_x}, and suppose that, in each economy, agents use pseudo-true Markovian $d$-state models to forecast the observable. If 
\[
\tilde{H} = 
H\left(I-\left(b + \sum_{k=1}^d \frac{\alpha_k \beta}{1-\alpha_k\beta} H^\dagger e_ke_k'Hc\right)g'\right),
\]
then the linear equilibria of the two economies are observationally equivalent.
\end{propapp}

Several remarks are in order. First, the result is a corollary of the linear-invariance result (Theorem \ref{thm:linear_invariance}) and the fact that agents' actions are linear in the observable. Second, the proposition covers the rational-expectations case by setting $d=m$. Third, when $\beta=0$, the effect of going from PE to GE is to amplify the response of observables to shocks, as measured by matrix $H'$, by the GE multiplier $(I-gb')^{-1}$. When $\beta>0$, the multiplier has an additional term, which captures the general-equilibrium effect of the updating of expectations by agents.

Last but not least, the distinctions between exogenous and endogenous variables, on one hand, and PE and GE, on the other, are largely inconsequential in this framework. Agents' expectations of endogenous variables are consistent with their expectations of exogenous variables and the structural equations of the economy, the GE economy is just the PE economy with a linearly transformed $H$ matrix, and agents' expectations in the GE economy are just linear transformations of their expectations in the PE economy. 

\newpage
\section{Proofs}

\subsection*{Proof of Theorem \ref{thm:linear_invariance}}

As a preliminary step, I fix an arbitrary $d$-state model $\theta=(A,B,Q,R)$ for the agent and compute her forecasts and the KLDR of her model from the true process. If the support of $P^\theta$ does not coincide with $\mathcal{W}$, the support of the true process, then $\text{KLDR}(\theta)=+\infty$. In what follows, I assume that $P^\theta$ is supported on $\mathcal{W}$. 

Note that minimizing the KLDR over the set $\Theta_d$ of $d$-state models is equivalent to minimizing the KLDR over the set $\Theta^m_0\cup \Theta^m_1\cup\dots\cup\Theta^m_d$, where $\Theta^m_k$ denotes the set of models whose minimal realization requires $k$ state variables. Therefore, in the proofs, I assume without loss of generality that the $d$-state model $\theta$ is minimal, i.e., that there exists no $d'$-state model with $d'<d$ that is observationally equivalent to $\theta$.

\paragraph*{The Kullback--Leibler divergence rate.}
Since the entropy rate of the true process is finite, the KLDR of $\theta$ from the true process is given by 
\[
\text{KLDR}(\theta) = \lim_{t\to\infty}\frac{1}{t}\mathbb{E}\left[-\log f^\theta(y_1,\dots,y_t)\right]+\text{constant}.
\]
Furthermore, by the chain rule,
\[
\lim_{t\to\infty}\frac{1}{t}\mathbb{E}\left[-\log f^\theta(y_1,\dots,y_t)\right] = \lim_{t\to\infty}\frac{1}{t}\sum_{\tau=1}^t \mathbb{E}\left[-\log f^\theta(y_\tau|y_{\tau-1},\dots,y_1)\right].
\]
Since $P^\theta$ and $\mathbb{P}$ are both stationary,
\[
\mathbb{E}\left[-\log f^\theta(y_\tau|y_{\tau-1},\dots,y_1)\right] = \mathbb{E}\left[-\log f^\theta(y_0|y_{-1},\dots,y_{1-\tau})\right].
\]
On the other hand, since $P^\theta$ is a stationary ergodic Gaussian process and $\mathbb{E}[\Vert y_t\Vert^2]<\infty$, the sequence $\{-\log f^\theta(y_0|y_{-1},\dots,y_{1-\tau})\}_\tau$ is uniformly bounded by an integrable function for any $\theta$. Thus, by the dominated convergence theorem,
\[
\lim_{\tau\to\infty}\mathbb{E}\left[-\log f^\theta(y_\tau|y_{\tau-1},\dots,y_1)\right] = \mathbb{E}\left[-\log f^\theta(y_0|y_{-1},\dots)\right] =\mathbb{E}\left[-\log f^\theta(y_{t+1}|y_{t},\dots)\right],
\]
where the second equality uses the stationarity of $P^\theta$, and the fact that $\log f^\theta(y_{t+1}|y_{t},\dots)$ is well defined is a consequence of the assumption that $A$ is convergent and $Q$ is positive definite for any $\theta=(A,B,Q,R)$. The above display implies that the Ces\`{a}ro sum also converges:
\[
\lim_{t\to\infty}\frac{1}{t}\sum_{\tau=1}^t \mathbb{E}\left[-\log f^\theta(y_\tau|y_{\tau-1},\dots,y_1)\right]\to \mathbb{E}\left[-\log f^\theta(y_{t+1}|y_{t},\dots)\right].
 \]
Therefore, to compute the KLDR, I only need to compute the subjective distribution of $y_{t+1}$ under model $\theta$ conditional on the history of observations $\{y_t,y_{t-1},\dots\}$.

Let $E_t^\theta[\cdot]$ denote the agent's subjective expectation given model $\theta$ and conditional on history $\{y_\tau\}_{\tau=-\infty}^{t}$, and let $\text{Var}^\theta_t(\cdot)$ denote the corresponding variance-covariance matrix. Let $\hat{z}_{t}\equiv E_t^\theta[z_{t+1}]$ denote the agent's conditional expectation of the subjective state. I can express $\hat{z}_{t}$ recursively using the Kalman filter:
\begin{equation}\label{eq:Kalman_observer}
\hat{z}_{t}=(A-KB')\hat{z}_{t-1}+K y_{t},
\end{equation}
where $K\in \mathbb{R}^{d\times n}$ is the Kalman gain defined as 
\begin{equation}\label{eq:Kalman_gain}
K\equiv A\hat{\Sigma}_z B\left(B'\hat{\Sigma}_z B + R\right)^{\dagger},
\end{equation}
the dagger denotes the Moore--Penrose pseudo-inverse, and $\hat{\Sigma}_{z}\equiv\text{Var}_t^\theta(z_{t+1})$ is the subjective conditional variance of $z_{t+1}$, which solves the following (generalized) algebraic Riccati equation:\footnote{Note that I allow for the possibility that $P^\theta$ is supported on some proper subspace $\mathcal{W}$ of $\mathbb{R}^n$, in which case $B'\hat{\Sigma}_z B + R$ might not be invertible. The Moore--Penrose pseudo-inverse is then the appropriate generalization of matrix inverse in the expression for the Kalman gain. See Chapter 4 of \cite{Anderson_Moore} for a treatment in the non-singular case and \cite{silverman1976discrete} for the case where $B'\hat{\Sigma}_z B + R$ may be singular.} \footnote{The assumptions that the $d$-state model $\theta$ is minimal and $Q$ is positive definite imply that the Riccati equation has a unique positive semidefinite solution and that $A-KB'$ is a convergent matrix.}
\begin{equation}\label{eq:Riccati}
\hat{\Sigma}_z = A\left(\hat{\Sigma}_z-\hat{\Sigma}_z B\left(B'\hat{\Sigma}_z B + R\right)^{\dagger}B'\hat{\Sigma}_z\right)A'+Q.
\end{equation}
Solving equation \eqref{eq:Kalman_observer} backward, I get
\[
\hat{z}_{t} = \sum_{\tau=0}^\infty (A-KB')^{\tau}K y_{t-\tau}.
\]
The agent's subjective conditional expectation of $y_{t+1}$ can be written in terms of her conditional expectation of $z_{t+1}$:
\[
E_t^\theta[y_{t+1}] = B'E_t^\theta[z_{t+1}]=B'\sum_{\tau=0}^\infty (A-KB')^{\tau}K y_{t-\tau}.
\]
Likewise, the subjective conditional variance of $y_{t+1}$ can be expressed in terms of the subjective conditional variance of $z_{t+1}$:
\begin{equation}\label{eq:Sigma_y_hat}
\hat{\Sigma}_y\equiv\text{Var}^\theta_t(y_{t+1}) = B'\hat{\Sigma}_z B + R.
\end{equation}
More generally, the agent's $s$-period-ahead forecast of the vector of observables is given by
\begin{equation}
E^\theta_t[y_{t+s}] = B'A^{s-1}E_t^\theta[z_{t+1}] = B'A^{s-1}\sum_{\tau=0}^\infty (A-KB')^{\tau}K y_{t-\tau}\label{eq:s_ahead_forecast}.
\end{equation}
The Kullback--Leibler divergence rate is thus equal to
\begin{align}
\text{KLDR}(\theta) = &\; -\frac{1}{2}\log\text{det}^*\left(\hat{\Sigma}_y^{\dagger}\right)+\frac{n}{2}\log\left(2\pi\right) + \frac{1}{2}\tr\left(\hat{\Sigma}_y^{\dagger}\Gamma_0\right)\nonumber\\ &  -\frac{1}{2}\sum_{\tau=1}^\infty\tr\left(\hat{\Sigma}_y^{\dagger}\Phi_\tau \Gamma'_\tau \right)-\frac{1}{2}\sum_{\tau=1}^\infty\tr\left(\hat{\Sigma}_y^{\dagger}\Gamma_\tau\Phi'_\tau  \right)\nonumber\\ & + \frac{1}{2}\sum_{s=1}^\infty\sum_{\tau=1}^\infty\tr\left(\hat{\Sigma}_y^{\dagger}\Phi_s\Gamma_{\tau-s}\Phi'_\tau\right)+\text{constant},\label{eq:KLDR}
\end{align}
where $\Gamma_{l}\equiv \mathbb{E}[y_t y'_{t-l}]$ denotes the lag-$l$ autocovariance matrix for the vector of observables under the true process, $\Phi_\tau\equiv B'(A-KB')^{\tau-1}K$, and the constant contains terms that do not depend on $\theta$. Matrix $\hat{\Sigma}_y^{\dagger}$ denotes the Moore--Penrose pseudo-inverse of $\hat{\Sigma}_y$ and $\text{det}^*(\hat{\Sigma}_y^{\dagger})$ denotes its pseudo-determinant.\footnote{The pseudo-determinant is the product of all non-zero eigenvalues of a square matrix.} These objects are the appropriate counterparts of the matrix inverse and the determinant for the case where $\mathcal{W}$ does not equal $\mathbb{R}^n$, and so, the subjective model $\theta$ is degenerate.


\paragraph*{Proof of Theorem \ref{thm:linear_invariance}.} Let $\tilde{n}$ denote the dimension of vector $\tilde{y}_t=T y_t$, let $\widetilde{\mathcal{W}}$ denote the linear subspace of $\mathbb{R}^{\tilde{n}}$ defined as $\widetilde{\mathcal{W}}\equiv \{\tilde{y}\in\mathbb{R}^{\tilde{n}}:\tilde{y}=Ty\;\text{for some}\;y\in \mathcal{W}\}$, let $\widetilde{\Theta}_d$ denote the set of $d$-state models when the vector of observable is $\tilde{y}_t\in\mathbb{R}^{\tilde{n}}$, and let $\widetilde{\text{KLDR}}(\tilde{\theta})$ denote the KLDR of model $\tilde{\theta}\in\widetilde{\Theta}_d$ from the true process $\widetilde{\mathbb{P}}\equiv T(\mathbb{P})$.

Let $\theta\in\Theta_d$ denote an arbitrary pseudo-true $d$-state model when the true process is $\mathbb{P}$ and $\tilde{\theta}\in\widetilde{\Theta}_d$ denote an arbitrary pseudo-true $d$-state model when the true process is $\widetilde{\mathbb{P}}$. I first show that $T(P^{\theta})$ and $P^{\tilde{\theta}}$ are both supported on $\widetilde{\mathcal{W}}$. Note that there always exists a $d$-state model for which the KLDR is finite---one such model is the one according to which $y_t$ is i.i.d. over time and has a variance-covariance matrix that coincides with the true variance-covariance matrix $\Gamma_0$. Therefore, for any pseudo-true $d$-state model, the KLDR is finite. Thus, $P^\theta$ is supported on $\mathcal{W}$, and so, $T(P^\theta)$ is supported on $\widetilde{\mathcal{W}}$. On the other hand, since the true distribution $\mathbb{P}$ is supported on $\mathcal{W}$, the transformed distribution $\widetilde{\mathbb{P}}$ is supported on $\widetilde{\mathcal{W}}$. Consequently, by the above argument, $P^{\tilde{\theta}}$ is also supported on $\widetilde{\mathcal{W}}$. Therefore, I can restrict my attention to models $\theta\in\Theta_d$ such that $P^\theta$ is supported on $\mathcal{W}$ and models $\tilde{\theta}\in\widetilde{\Theta}_d$ such that $P^{\tilde{\theta}}$ is supported on $\widetilde{\mathcal{W}}$. 

For any model $\theta=(A,B,Q,R)\in\Theta_d$, define model $T(\theta)\in\widetilde{\Theta}_d$ as $T(\theta)\equiv\left(A,BT',Q,TRT'\right)$.  I next show that $\widetilde{\text{KLDR}}(T(\theta))=\text{KLDR}(\theta)$, up to an additive constant that does not depend on $\theta$. Fix some model $\theta\in\Theta_d$. Let $\hat{\Sigma}_z\equiv\text{Var}_t^{\theta}(z_{t+1})$ denote the subjective conditional variance of the subjective state under model $\theta$, and let $\widetilde{\hat{\Sigma}}_z\equiv\text{Var}_t^{T(\theta)}(z_{t+1})$ denote the corresponding conditional variance under model $T(\theta)$. Matrices $\hat{\Sigma}_z$ and $\widetilde{\hat{\Sigma}}_z$ solve the following Riccati equations:
\begin{align}
& \hat{\Sigma}_z = A\left(\hat{\Sigma}_z-\hat{\Sigma}_z B\left(B'\hat{\Sigma}_z B + R\right)^{\dagger}B'\hat{\Sigma}_z\right)A'+Q,\label{eq:Riccati_no_tilde}\\
& \widetilde{\hat{\Sigma}}_z = A\left(\widetilde{\hat{\Sigma}}_z-\widetilde{\hat{\Sigma}}_z BT'\left(TB'\widetilde{\hat{\Sigma}}_z BT' + TRT'\right)^{\dagger}TB'\widetilde{\hat{\Sigma}}_z\right)A'+Q.\label{eq:Riccati_tilde}
\end{align}
Since matrix $T$ has full rank, $T^\dagger = (T'T)^{-1}T$ and $T^\dagger T=I$. Therefore, $\widetilde{\hat{\Sigma}}_z=\hat{\Sigma}_z$. Next, let $K$ denote the Kalman gain given model $\theta$, and let denote $\tilde{K}$ denote the Kalman gain given model $T(\theta)$. Note that
\[
\tilde{K}= A\widetilde{\hat{\Sigma}}_z BT'\left(TB'\widetilde{\hat{\Sigma}}_z BT' + TRT'\right)^{\dagger} = KT^\dagger.
\]
Let $\Phi_\tau\equiv B'(A-KB')^{\tau-1}K$, and let $\widetilde{\Phi}_\tau$ denote the corresponding matrix given model $T(\theta)$. Note that
\[
\widetilde{\Phi}_\tau\equiv T B'(A-KT^\dagger T B')^{\tau-1} KT^\dagger=T\Phi_\tau T^\dagger.
\]
Finally, let $\hat{\Sigma}_y\equiv \text{Var}_t^\theta(y_{t+1})$ denote the subjective conditional variance of $y_{t+1}$ given model $\theta$, and let $\widetilde{\hat{\Sigma}}_y\equiv \text{Var}_t^{T(\theta)}(\tilde{y}_{t+1})$ denote the corresponding conditional variance given model $T(\theta)$. Note that
\[
\widetilde{\hat{\Sigma}}_y = TB'\widetilde{\hat{\Sigma}}_z BT'+TRT'=T\hat{\Sigma}_y T'.
\]
One the other hand, $\widetilde{\Gamma}_l\equiv\widetilde{\mathbb{E}}[\tilde{y}_t\tilde{y}_{t-l}']=T\mathbb{E}[y_ty_{t-l}]T'=T\Gamma_l T'$. Therefore, by equation \eqref{eq:KLDR}, 
\begin{align*}
\widetilde{\text{KLDR}}(T(\theta)) = &\; -\frac{1}{2}\log\text{det}^*\left({T^\dagger}'\hat{\Sigma}_y^{\dagger}T^\dagger\right)+\frac{n}{2}\log\left(2\pi\right) + \frac{1}{2}\tr\left({T^\dagger}'\hat{\Sigma}_y^{\dagger}T^\dagger T\Gamma_0 T'\right)\nonumber\\ &  -\frac{1}{2}\sum_{\tau=1}^\infty\tr\left({T^\dagger}'\hat{\Sigma}_y^{\dagger}T^\dagger T\Phi_\tau T^\dagger T\Gamma'_\tau T'\right)-\frac{1}{2}\sum_{\tau=1}^\infty\tr\left({T^\dagger}'\hat{\Sigma}_y^{\dagger}T^\dagger T\Gamma_\tau T' {T^\dagger}'\Phi'_\tau T'\right)\nonumber\\ & + \frac{1}{2}\sum_{s=1}^\infty\sum_{\tau=1}^\infty\tr\left({T^\dagger}'\hat{\Sigma}_y^{\dagger} T^\dagger T\Phi_s T^\dagger T\Gamma_{\tau-s} T'{T^\dagger}'\Phi'_\tau T'\right)+\text{constant}.
\end{align*}
The fact that $T^\dagger T=I$ implies that the above expression is equal to $\text{KLDR}(\theta)$, up to an additive constant that does not depend on $\theta$.

Likewise, for any model $\tilde{\theta}=(\tilde{A},\tilde{B},\tilde{Q},\tilde{R})\in\widetilde{\Theta}_d$, define $T^{-1}(\tilde{\theta})\equiv (\tilde{A},\tilde{B}{T^\dagger}',\tilde{Q},T^\dagger\tilde{R}{T^\dagger}')\in \Theta_d$. By an argument similar to the one in the previous paragraph, $\text{KLDR}(T^{-1}(\tilde{\theta}))=\widetilde{\text{KLDR}}(\tilde{\theta})$, up to an additive constant that does not depend on $\tilde{\theta}$.

Therefore, the mapping $T$ defines an isomorphism between the set of models $\Theta_d$ and the set of models  $\widetilde{\Theta}_d$: Any model $\theta\in\Theta_d$ can be identified with a model $T(\theta)\in\widetilde{\Theta}_d$ such that the KLDR of $P^\theta$ from the process $\mathbb{P}$ is equal to the KLDR of $P^{T(\theta)}$ from $T(\mathbb{P})$, and any model $\tilde{\theta}\in\widetilde{\Theta}_d$ can be identified with a model $T^{-1}(\tilde{\theta})\in \Theta_d$ such that the KLDR of $P^{T^{-1}(\tilde{\theta})}$ from the process $\mathbb{P}$ is equal to the KLDR of $P^{\tilde{\theta}}$ from the process $T(\mathbb{P})$. This conclusion immediately implies that the set of pseudo-true $d$-state models under true process $\mathbb{P}$ is identified with the set of pseudo-true $d$-state models under true process $T(\mathbb{P})$. 

It only remains to show that $P^{T(\theta)}=T(P^\theta)$ for any model $\theta\in\Theta_d$. Since $P^{T(\theta)}$ and $T(P^\theta)$ are both zero mean, stationary, and normal distributions over $\{\tilde{y}_t\}_{t=-\infty}^\infty$, it is sufficient to show that the autocovariance matrices of $\tilde{y}_t$ are identical at all lags under the two distributions. But this follows the definitions of distributions $P^{T(\theta)}$ and $T(P^\theta)$.\hfill\qed

\subsection*{Proof of Theorem \ref{thm:1_state_general}}
Before establishing the theorem, I state and prove a lemma that underpins all the characterization results of the paper:
\begin{lemmaapp}\label{lem:d-state-characterization}
Model $\theta=(A,B,Q,R)$ is a pseudo-true $d$-state model given true autocovariance matrices $\{\Gamma_l\}_l$ with $\Gamma_0$ invertible if and only if $A=M$, $B={D}'{N}^{-1}$, $Q=I- M \left(I- {D}'D\right){M}'$, and $R = {{N}^{-1}}'\left(I-D{D}'\right){N}^{-1}$, where $(M,D,N)$ is a tuple that minimizes
\begin{align}
\text{KLDR}(\tilde{M},\tilde{D},\tilde{N}) \equiv &\; -\frac{1}{2}\log\text{det}\left(\tilde{N}\tilde{N}'\right)+ \frac{1}{2}\tr\left(\tilde{N}'\Gamma_0 \tilde{N}\right)-\sum_{\tau=1}^\infty\tr\left(\left(\tilde{M}  \left(I-\tilde{D}'\tilde{D}\right)\right)^{\tau-1} \tilde{M} \tilde{D}'\tilde{N}' \Gamma'_\tau \tilde{N}\tilde{D}\right)\nonumber\\ & + \frac{1}{2}\sum_{s=1}^\infty\sum_{\tau=1}^\infty\tr\left(\tilde{D}\left(\tilde{M}  \left(I-\tilde{D}'\tilde{D}\right)\right)^{s-1} \tilde{M} \tilde{D}'\tilde{N}'\Gamma_{\tau-s}\tilde{N} \tilde{D} \tilde{M}'\left(\left(I-\tilde{D}'\tilde{D}\right)\tilde{M}'\right)^{\tau-1}\tilde{D}'\right),\label{eq:def_H_M_D_N}
\end{align}
subject to the constraints that $\tilde{M}$ is a $d\times d$ convergent matrix, $\tilde{D}$ is an $n\times d$ diagonal matrix with elements in the $[0,1]$ interval, $\tilde{N}$ is an $n\times n$ invertible matrix, and $\Vert \tilde{M}(I-\tilde{D}\tilde{D}')\tilde{M}'\Vert_2<1$.
\end{lemmaapp}
\begin{proof}
The assumption that $Q$ is positive definite implies that the solution $\hat{\Sigma}_z$ to the Riccati equation \eqref{eq:Riccati} is invertible. On the other hand, since $\Gamma_0$ is invertible, I can restrict attention to subjective models for which $\hat{\Sigma}_y$ is non-singular.\footnote{Since the variance-covariance matrix $\Gamma_0$ of the true process is invertible, $\text{KLDR}(\theta)=+\infty$ for any subjective model $\theta$ with a singular $\hat{\Sigma}_y$. Note that, in light of Theorem \ref{thm:linear_invariance}, the restriction to true processes with invertible variance-covariance matrices is without loss of generality.} The pseudo-inverses and pseudo-determinants in equations \eqref{eq:Riccati} and \eqref{eq:KLDR} thus reduce to matrix inverses and determinants. 

I start by expressing $\hat{\Sigma}_y^{\frac{-1}{2}}B'\hat{\Sigma}_z^{\frac{1}{2}}$ as its singular value decomposition:
\begin{equation}
\hat{\Sigma}_y^{\frac{-1}{2}}B'\hat{\Sigma}_z^{\frac{1}{2}} = UD V',\label{eq:SVD}
\end{equation}
where $U\in\mathbb{R}^{n\times n}$ and $V\in\mathbb{R}^{d\times d}$ are orthogonal matrices, and $D\in\mathbb{R}^{n\times d}$ is a rectangular diagonal matrix with singular values of $\hat{\Sigma}_z^{\frac{1}{2}}B\hat{\Sigma}_y^{\frac{-1}{2}}$ on the diagonal. Note that
\begin{equation}\label{eq:VDDV}
VD'DV' = \hat{\Sigma}_z^{\frac{1}{2}}B\left(B'\hat{\Sigma}_z B+R\right)^{-1}B'\hat{\Sigma}_z^{\frac{1}{2}}.   
\end{equation}
Since $R$ is a symmetric positive semidefinite matrix and $V$ is orthogonal, diagonal elements of $D$ are weakly smaller than $1$ (strictly so if $R$ is positive definite). Next, define $M \equiv V^{-1}\hat{\Sigma}_z^{\frac{-1}{2}}A\hat{\Sigma}_z^{\frac{1}{2}}V$. Then,
\begin{align}
    & A = \hat{\Sigma}_z^{\frac{1}{2}} V M V^{-1} \hat{\Sigma}_z^{\frac{-1}{2}},\label{eq:A_norm}\\
    & B = \hat{\Sigma}_z^{\frac{-1}{2}}VD'U'\hat{\Sigma}_y^{\frac{1}{2}},\label{eq:B_norm}\\
    & K = \hat{\Sigma}_z^{\frac{1}{2}}V M D'U'\hat{\Sigma}_y^{\frac{-1}{2}}\label{eq:K_norm},
\end{align}
and so
\begin{align*}
& KB' = \hat{\Sigma}_z^{\frac{1}{2}} VM  D'DV'\hat{\Sigma}_z^{\frac{-1}{2}},\\
& \Phi_\tau = \hat{\Sigma}_y^{\frac{1}{2}}UD\left(M  \left(I-D'D\right)\right)^{\tau-1} M  D'U'\hat{\Sigma}_y^{\frac{-1}{2}}.
\end{align*}
Note that since $A$ is a convergent matrix, so is $M$. Substituting in \eqref{eq:Riccati} for $A$ from equation \eqref{eq:A_norm} and for $B$ from \eqref{eq:B_norm}, I get
\begin{align}
    Q & = \hat{\Sigma}_z-A\left(\hat{\Sigma}_z-\hat{\Sigma}_z B\left(B'\hat{\Sigma}_z B + R\right)^{-1}B'\hat{\Sigma}_z\right)A'\nonumber\\ & =  \hat{\Sigma}_z-\hat{\Sigma}_z^{\frac{1}{2}} V M V^{-1} \hat{\Sigma}_z^{\frac{-1}{2}}\left(\hat{\Sigma}_z-\hat{\Sigma}_z^{\frac{1}{2}} VD'DV\hat{\Sigma}_z^{\frac{1}{2}}\right)\hat{\Sigma}_z^{\frac{-1}{2}}VM'V^{-1}\hat{\Sigma}_z^{\frac{1}{2}}\nonumber\\ & =  \hat{\Sigma}_z-\hat{\Sigma}_z^{\frac{1}{2}} V M \left(I- D'D\right)M'V^{-1}\hat{\Sigma}_z^{\frac{1}{2}}.\label{eq:Riccati_transformed}
\end{align}
Therefore, since $Q$ is positive definite, the eigenvalues of $V M \left(I- D'D\right)M'V^{-1}$ must all lie inside the unit circle. This implies that $\rho(M \left(I- D'D\right)M')=\Vert M \left(I- D'D\right)M'\Vert_2<1$, where $\rho(\cdot)$ denotes the spectral radius, and I am using the facts that the spectral radius is invariant to similarity transformations and equal to the spectral norm for symmetric matrices.

I can further reduce the number of parameters in the agent's model by transforming $\hat{\Sigma}_y^{\frac{-1}{2}}$ using the orthogonal matrix $U$. Define $N \equiv \hat{\Sigma}_y^{\frac{-1}{2}}U$. Since $\hat{\Sigma}_y^{\frac{-1}{2}}$ and $U$ are invertible matrices, so is $N$. Because $\hat{\Sigma}_y^{\frac{-1}{2}}$ is symmetric, $UN'= NU' = \hat{\Sigma}_y^{\frac{-1}{2}}$, so $\hat{\Sigma}^{-1}_y = NU'UN' = NN'$, and
\[
\tr\left(\hat{\Sigma}^{-1}_y\Gamma_0\right) = \tr\left(\hat{\Sigma}_y^{\frac{-1}{2}}\Gamma_0 \hat{\Sigma}_y^{\frac{-1}{2}}\right) = \tr\left(UN'\Gamma_0 NU'\right) = \tr\left(N'\Gamma_0 N\right).
\]
On the other hand,
\begin{align*}
    \tr\left(\hat{\Sigma}_y^{-1}\Phi_\tau \Gamma'_\tau \right) & = \tr\left(\hat{\Sigma}_y^{\frac{-1}{2}}UD\left(M  \left(I-D'D\right)\right)^{\tau-1} M  D'U'\hat{\Sigma}_y^{\frac{-1}{2}}\Gamma'_\tau \right) = \tr\left(\left(M  \left(I-D'D\right)\right)^{\tau-1} M D'N' \Gamma'_\tau ND\right),
\end{align*}
and
\begin{align*}
    \tr\left(\hat{\Sigma}_y^{-1}\Phi_s\Gamma_{\tau-s}\Phi'_\tau\right) & = \tr\left(D\left(M  \left(I-D'D\right)\right)^{s-1} M D'N'\Gamma_{\tau-s}N D M'\left(\left(I-D'D\right)M'\right)^{\tau-1}D'\right).
\end{align*}
Therefore, the KLDR can be expressed in terms of matrices $M$, $D$, and $N$ as
\[
\text{KLDR}(\theta)=\text{KLDR}(M,D,N)+\text{constant},
\]
where $\text{KLDR}(M,D,N)$ is as in the statement of the lemma. 

It only remains to show that, for any $(\hat{M},\hat{D},\hat{N})$ such that $\hat{M}$ is a $d\times d$ convergent matrix, $\hat{D}$ is an $n\times d$ diagonal matrix with elements in the $[0,1]$ interval, $\hat{N}$ is an $n\times n$ invertible matrix, and $\Vert \hat{M}(I-\hat{D}\hat{D}')\hat{M}'\Vert_2<1$, one can construct a corresponding $(A,B,Q,R)$ such that $A$ is convergent, $Q$ is positive definite, and $R$ is positive semidefinite. Given such a tuple $(\hat{M},\hat{D},\hat{N})$, let $A = \hat{M}$, $B = {\hat{D}}'{\hat{N}}^{-1}$, $Q = I- \hat{M} \left(I- {\hat{D}}'\hat{D}\right){\hat{M}}'$, and $R = \mbox{${\hat{N}}^{-1}$}'\left(I-\hat{D}{\hat{D}}'\right){\hat{N}}^{-1}$. Since $\hat{M}$ is convergent, so is $A$. Since $\Vert \hat{M}(I-\hat{D}\hat{D}')\hat{M}'\Vert_2<1$, matrix $Q$ is positive definite. And since $\hat{D}$ is a diagonal matrix with elements in the $[0,1]$ interval, $R$ is positive semidefinite. It is easy to verify that then $\hat{\Sigma}_z=I$ is then the solution to the Riccati equation \eqref{eq:Riccati}, and so, $\hat{\Sigma}_y = (\hat{N}\hat{N}')^{-1}$. Therefore, I can choose $U=(\hat{N}\hat{N}')^{\frac{-1}{2}}\hat{N}$, $D=\hat{D}$, and $V=I$ in equation \eqref{eq:SVD}. Substituting in the expressions for $M$ and $N$, I get $M=\hat{M}$ and $N=\hat{N}$. This completes the proof of the lemma.

For future reference, I also compute several other objects in terms of the $M$, $D$, and $N$ matrices. The matrix of Kalman gain is given by
\begin{equation}
K = MD'N'.\label{eq:kalman_M_D_N}
\end{equation}
The subjective forecasts can then be found by substituting for $A$, $B$, and $K$ in \eqref{eq:s_ahead_forecast}:
\begin{equation}
E^\theta_t[y_{t+s}] = {N'}^{-1}DM^{s-1}\sum_{\tau=0}^\infty \left(M  \left(I-D'D\right)\right)^{\tau}M D'N' y_{t-\tau}.\label{eq:s_ahead_forecast_simplified}
\end{equation}
The subjective variance of $y_{t+1}$ conditional on the information available to the agent at time $t$ is given by $\hat{\Sigma}_y = \left(NN'\right)^{-1}$. The unconditional subjective variance of $y$ is given by
\[
\text{Var}^\theta(y) = B'\text{Var}^\theta(z) B + R,
\]
where $\text{Var}^\theta(z)$ solves the discrete Lyapunov equation
\[
\text{Var}^\theta(z) = A\text{Var}^\theta(z)A' + Q.
\]
Solving the above equation forward, I get
\[
\text{Var}^\theta(z) = I +  \sum_{\tau=1}^\infty M^\tau D'D{M'}^\tau.
\]
Therefore,
\begin{equation}\label{eq:d-state-uncond-var}
\text{Var}^\theta(y) = B'\sum_{\tau=0}^\infty A^\tau Q {A'}^\tau B + R = {N^{-1}}'\left(I+ \sum_{\tau=1}^\infty DM^\tau D'D{M'}^\tau D'\right)N^{-1}.
\end{equation}
\end{proof}

I can now establish Theorem \ref{thm:1_state_general}.

\paragraph*{Proof of Theorem \ref{thm:1_state_general}.}
Let $M$, $D$, and $N$ be as in Lemma \ref{lem:d-state-characterization}. When $d=1$, then $M = a$ for some $a\in [-1,1]$ and $D = d_1 e_1$ for some $d_1 \in[0,1]$, where $e_1$ denotes the first coordinate vector. Define $\eta \equiv 1-d_1^2$ and $S \equiv \Gamma_0^{\frac{1}{2}}N$. Then $\text{KLDR}$, defined in \eqref{eq:def_H_M_D_N}, can be written (with slight abuse of notation) as a function of $a$, $\eta$, and $S$:
\begin{align*}
\text{KLDR}(a,\eta,S) = &\; -\frac{1}{2}\log\text{det}\left(SS'\right)+ \frac{1}{2}\tr\left(S'S\right)-\frac{1}{2}e_1'S'\Omega(a,\eta) Se_1+\text{constant},
\end{align*}
where
\begin{align*}
    \Omega(a,\eta) & \equiv a(1-\eta)\sum_{\tau=1}^\infty (a\eta)^{\tau-1} \Gamma_0^{\frac{-1}{2}}(\Gamma_\tau+\Gamma'_\tau)\Gamma_0^{\frac{-1}{2}}-a^2(1-\eta)^2\sum_{s=1}^\infty\sum_{\tau=1}^\infty (a\eta)^{s+\tau-2}\Gamma_0^{\frac{-1}{2}}\Gamma_{\tau-s}\Gamma_0^{\frac{-1}{2}}.
\end{align*}
I can simplify the second term of $\Omega(a,\eta)$ further:
\begin{align*}
    \sum_{s=1}^\infty\sum_{\tau=1}^\infty (a\eta)^{s+\tau-2}\Gamma_0^{\frac{-1}{2}}\Gamma_{\tau-s}\Gamma_0^{\frac{-1}{2}} &  = \sum_{s=1}^\infty\sum_{\tau=s+1}^\infty (a\eta)^{s+\tau-2}\Gamma_0^{\frac{-1}{2}}\left(\Gamma_{\tau-s}+\Gamma'_{\tau-s}\right)\Gamma_0^{\frac{-1}{2}} + \sum_{s=1}^\infty (a\eta)^{2(s-1)}I\\ & = \sum_{s=1}^\infty\sum_{\tau=1}^\infty (a\eta)^{2(s-1)+\tau}\Gamma_0^{\frac{-1}{2}}\left(\Gamma_{\tau}+\Gamma'_{\tau}\right)\Gamma_0^{\frac{-1}{2}} + \sum_{s=1}^\infty (a\eta)^{2(s-1)}I\\ &  = \left(\sum_{s=1}^\infty (a\eta)^{2(s-1)}\right)\left(I + \sum_{\tau=1}^\infty (a\eta)^\tau \Gamma_0^{\frac{-1}{2}}(\Gamma_\tau+\Gamma'_\tau)\Gamma_0^{\frac{-1}{2}}\right)\\ &  = \frac{1}{1-a^2\eta^2}\left(I + a\eta\sum_{\tau=1}^\infty (a\eta)^{\tau-1} \Gamma_0^{\frac{-1}{2}}(\Gamma_\tau+\Gamma'_\tau)\Gamma_0^{\frac{-1}{2}}\right).
\end{align*}
Therefore,
\begin{equation}
    \Omega(a,\eta) = -\frac{a^2(1-\eta)^2}{1-a^2\eta^2}I + \frac{(1-\eta)(1-a^2\eta)}{1-a^2\eta^2}\sum_{\tau=1}^\infty a^\tau\eta^{\tau-1} \Gamma_0^{\frac{-1}{2}}(\Gamma_\tau+\Gamma'_\tau)\Gamma_0^{\frac{-1}{2}}.\label{eq:Omega}
\end{equation}
By Lemma \ref{lem:d-state-characterization}, minimizing the KLDR with respect to $A$, $B$, $Q$, and $R$ is equivalent to minimizing $\text{KLDR}(M,D,N)$ with respect to $M$, $D$, and $N$. But for any $a$, $\eta$, and $S$, one can construct a corresponding $M$, $D$, and $N$, and vice versa. Therefore, I can instead minimize $\text{KLDR}(a,\eta,S)$ with respect to $a$, $\eta$, and $S$.

I first minimize $\text{KLDR}(a,\eta,S)$ with respect to $S$ taking $a$ and $\eta$ as given. The first-order optimality condition with respect to $S$ is given by $S^{-1} = S' - e_1e_1' S'\Omega(a,\eta)$, which implies that 
\begin{equation}\label{eq:FOC}
S'S - e_1e_1'S'\Omega(a,\eta) S = I.
\end{equation}
Therefore, for any solution to the problem of minimizing $\text{KLDR}(a,\eta,S)$,
\[
n=\tr(I)=\tr\left(S'S\right)-\tr\left(e_1e_1'S'\Omega(a,\eta) S\right)=\tr\left(S'S\right)-e_1'S'\Omega(a,\eta) Se_1.
\]
Thus, minimizing $\text{KLDR}(a,\eta,S)$ with respect to $a$, $\eta$, and $S$ is equivalent to solving the following program:
\[
\max_{a,\eta}\;\; \text{det}\left(S(a,\eta){S}'(a,\eta)\right),
\]
where
\begin{equation}\label{eq:S_star}
S(a,\eta) \in \argmin_{S}\; -\frac{1}{2}\log\text{det}\left(SS'\right)+\frac{1}{2}\tr\left(S'S\right)-\frac{1}{2}e_1'S'\Omega(a,\eta) S e_1.
\end{equation}
I proceed by first characterizing $S(a,\eta)$. Note that the necessary first-order optimality conditions for problem \eqref{eq:S_star} are given by matrix equation \eqref{eq:FOC}. 
\begin{claimapp}\label{claim:FOC_S}
For any matrix $S$ that solves equation \eqref{eq:FOC}, the necessary first-order optimality condition for problem \eqref{eq:S_star},
\begin{enumerate}[(i)]
\item $S e_1 = \displaystyle\frac{1}{\sqrt{1-\lambda}}u$,
\item ${S'}^{-1}e_1 = \sqrt{1-\lambda}u$,
\item $SS'=I+\frac{\lambda}{1-\lambda}uu'$,
\end{enumerate}
where $\lambda$ is an eigenvalue of the real symmetric matrix $\Omega(a,\eta)$ and $u$ is a corresponding eigenvector normalized such that $u'u=1$.
\end{claimapp}
I return to proving the claim toward the end of the proof. Equation \eqref{eq:FOC} in general has multiple solutions, with each solution corresponding to a local extremum of problem \eqref{eq:S_star}. The global optimum of problem \eqref{eq:S_star} is given by the solution to equation \eqref{eq:FOC} that results in the largest value for $\det(SS')$. But by part (iii) of Claim \ref{claim:FOC_S}, $\det(SS')=(1-\lambda)^{-1}$. Thus, for any pseudo-true one-state model, $a$ and $\eta$ maximize $\lambda_{\text{max}}(\Omega(a,\eta))$ and $S$ satisfies parts (i)--(iii) of Claim \ref{claim:FOC_S}, with $\lambda=\lambda_{\text{max}}(\Omega)$ and $u=u_{\text{max}}(\Omega)$ the corresponding eigenvector.

I next find parameters $A$, $B$, $Q$, and $R$ representing the $a$, $\eta$, and $S$ that minimize $\text{KLDR}(a,\eta,S)$. First, note that $M = a$, $D = \sqrt{1-\eta}e_1$, and $N = \Gamma_0^{\frac{-1}{2}}S$. The representation in Lemma \ref{lem:d-state-characterization} is thus given by $A = a$, $B = \sqrt{1-\eta}e_1'S^{-1}\Gamma_0^{\frac{1}{2}}$, $Q = 1- a^2\eta$, and $R = \Gamma_0^{\frac{1}{2}}{S^{-1}}'\left(I-(1-\eta)e_1e_1'\right)S^{-1}\Gamma_0^{\frac{1}{2}}$.\footnote{For this $(A,B,Q,R)$ tuple to represent a one-state model, I need $A$ to be convergent, $Q$ to be positive definite, and $R$ to be positive semidefinite. That $R$ is always positive semidefinite is immediate. Showing that $A$ is convergent and $Q$ is positive definite takes more work. I do so in Lemma \ref{lemapp:1-state-stationary-ergodic}.} By Claim \ref{claim:FOC_S} and the argument above,
\begin{align*}
& e_1'S^{-1} = \sqrt{1-\lambda_{\text{max}}(\Omega)}u'_{\text{max}}(\Omega),\\
& {S^{-1}}'S^{-1}=\left(SS'\right)^{-1} = I-\lambda_{\text{max}}(\Omega) u_{\text{max}}(\Omega)u'_{\text{max}}(\Omega).
\end{align*}
Thus,
\begin{align*}
    & B = \sqrt{\left(1-\eta\right)\left(1-\lambda_{\text{max}}(\Omega)\right)}u'_{\text{max}}(\Omega)\Gamma_0^{\frac{1}{2}},
\end{align*}  
and
\begin{align*}
    R & = \Gamma_0^{\frac{1}{2}}\left(I-\lambda_{\text{max}}(\Omega) u_{\text{max}}(\Omega)u'_{\text{max}}(\Omega)\right)\Gamma_0^{\frac{1}{2}}-(1-\eta)\left(1-\lambda_{\text{max}}(\Omega)\right)\Gamma_0^{\frac{1}{2}}u_{\text{max}}(\Omega)u'_{\text{max}}(\Omega)\Gamma_0^{\frac{1}{2}}\\ & = \Gamma_0^{\frac{1}{2}}\left[I - \left(1-\eta +\eta\lambda_{\text{max}}(\Omega)\right)u_{\text{max}}(\Omega)u'_{\text{max}}(\Omega)\right]\Gamma_0^{\frac{1}{2}}.
\end{align*}
Finally, note that $M=a$, $D=\sqrt{1-\eta}e_1$, and $N=\Gamma_0^{\frac{-1}{2}}S$. Therefore, by equation \eqref{eq:s_ahead_forecast_simplified}, the subjective forecasts are given by
\begin{equation}
E^\theta_t[y_{t+s}] = a^{s}(1-\eta){\Gamma}_0^{\frac{1}{2}}{S'}^{-1}e_1e_1'S'{\Gamma}_0^{\frac{-1}{2}}\sum_{\tau=0}^\infty a^\tau\eta^\tau y_{t-\tau}.\label{eq:s_ahead_forecast_simplified_1_state}
\end{equation}
Using Claim \ref{claim:FOC_S} to substitute for the optimal $S$, I get
\[
E^\theta_t[y_{t+s}] = a^{s}(1-\eta)\Gamma_0^{\frac{1}{2}}u_{\text{max}}(\Omega) u_{\text{max}}'(\Omega)\Gamma_0^{\frac{-1}{2}}\sum_{\tau=0}^\infty a^\tau\eta^\tau y_{t-\tau},
\]
where $u_{\text{max}}(\Omega)$ is a unit-norm eigenvector of $\Omega$ with eigenvalue $\lambda_{\text{max}}(\Omega)$. The theorem then follows by the definitions of $p$ and $q$.\hfill\qed

\paragraph{Proof of Claim \ref{claim:FOC_S}.}
The first-order optimality condition with respect to $S$ is given by 
\begin{equation}
S'S-e_1e_1'S'\Omega S = I. \label{eq:FOC_1_d_S_1}
\end{equation}
Multiplying the transpose of the above equation from right by $e_1$ and from left by ${S'}^{-1}$, I get
\begin{equation}
Se_1 - \Omega Se_1 = {S'}^{-1}e_1.\label{eq:FOC_1_d_S_2}
\end{equation}
On the other hand, multiplying equation \eqref{eq:FOC_1_d_S_1} from left by $S$ and from right by $S^{-1}$, I get
\begin{equation}
    SS' = I + Se_1e_1'S'\Omega.\label{eq:FOC_1_d_S_3}
\end{equation}
By the Sherman--Morrison formula,
\[
{S'}^{-1}S^{-1} = I - \frac{Se_1e_1'S'\Omega}{1+e_1'S'\Omega Se_1}.
\]
Multiplying the above equation from right by $S e_1$, I get
\begin{equation}
   {S'}^{-1}e_1 =  \frac{1}{1+e_1'S'\Omega Se_1}Se_1.\label{eq:FOC_1_d_S_4}
\end{equation}
Substituting for $ {S'}^{-1}e_1$ from the above equation in \eqref{eq:FOC_1_d_S_2}  and rearranging the terms, I get
\begin{equation}
\Omega Se_1 = \frac{e_1'S'\Omega Se_1}{1+e_1'S'\Omega Se_1}Se_1.\label{eq:FOC_1_d_S_5}
\end{equation}
That is, $Se_1$ is an eigenvector of $\Omega$. Let $\lambda$ denote the corresponding eigenvalue and let $u=Se_1/\sqrt{e_1'S'Se_1}$. Then equation \eqref{eq:FOC_1_d_S_5} implies
\[
\lambda = \frac{\lambda e_1'S'Se_1}{1+\lambda e_1'S'Se_1}.
\]
I separately consider the cases $\lambda\neq0$ and $\lambda=0$. If $\lambda\neq 0$, then $e_1'S'Se_1 = 1/(1-\lambda)$ and  $Se_1 = u/\sqrt{1-\lambda}$. Equation \eqref{eq:FOC_1_d_S_4} then implies that ${S'}^{-1}e_1 = \sqrt{1-\lambda}u$, and equation \eqref{eq:FOC_1_d_S_3} implies that
\[
SS' = I + \frac{\lambda}{1-\lambda}uu'.
\]
If $\lambda=0$, then equation \eqref{eq:FOC_1_d_S_2} implies that $Se_1=S'^{-1}e_1$, and so, $Se_1$ and $ S'^{-1}e_1$ are both multiples of $u$. Furthermore, $e_1'S^{-1}S e_1=e_1'e_1 = 1$. Therefore, $Se_1=S'^{-1}e_1=u$. On the other hand, equation \eqref{eq:FOC_1_d_S_3} implies that $SS'=I$. This completes the proof of the claim.\hfill\qed

\subsection*{Proof of Theorem \ref{thm:1-state-covariance}}
I first prove two useful lemmas:

\begin{lemmaapp}\label{lem:spectral_radius_of_autocorrelation}
For any purely non-deterministic, stationary ergodic, and non-degenerate process with autocorrelation matrices $\{C_l\}_l$, the spectral radii of autocorrelation matrices satisfy $\rho(C_l)\leq 1$ for any $l$ with the inequality strict for $l=1$.
\end{lemmaapp}

\begin{proof}
Let $\lambda_l$ denote an eigenvalue of $C_l$ largest in magnitude and let $u_l$ denote the corresponding eigenvector normalized such that $u_l'u_l=1$. Define the process $\omega^{(l)}_t \equiv u_l'\Gamma_0^{\frac{-1}{2}}y_t\in\mathbb{R}$. Since $y_t$ is a purely non-deterministic, stationary ergodic, and non-degenerate process, so is $\omega^{(l)}_t$ for any $l$. I first show that $\lambda_l$ is the autocorrelation of process $\omega^{(l)}_t$ at lag $l$. Note that
\begin{align*}
    \mathbb{E}[\omega^{(l)}_t\omega^{(l)}_{t-l}] = u_l'\Gamma_0^{\frac{-1}{2}}\mathbb{E}[y_ty_{t-l}']\Gamma_0^{\frac{-1}{2}}u_l = u_l'\Gamma_0^{\frac{-1}{2}}\Gamma_l \Gamma_0^{\frac{-1}{2}}u_l = u_l'\Gamma_0^{\frac{-1}{2}}\left(\frac{\Gamma_l+\Gamma_l'}{2}\right) \Gamma_0^{\frac{-1}{2}}u_l = u_l'C_l u_l = \lambda_l.
\end{align*}
Furthermore,
\[
\mathbb{E}[\omega^{(l)}_t\omega^{(l)}_t]=u_l'\Gamma_0^{\frac{-1}{2}}\mathbb{E}[y_ty_{t}']\Gamma_0^{\frac{-1}{2}}u_l=u_l'\Gamma_0^{\frac{-1}{2}}\Gamma_0 \Gamma_0^{\frac{-1}{2}}u_l = u_l'u_l=1.
\]
Therefore, since $\omega^{(l)}_t$ is purely non-deterministic, stationary ergodic, and non-degenerate,
\[
\rho(C_l) = |\lambda_l| = \frac{\mathbb{E}[\omega^{(l)}_t\omega^{(l)}_{t-l}]}{\mathbb{E}[\omega^{(l)}_t\omega^{(l)}_t]}\leq 1.
\]
Next, toward a contradiction suppose that $\rho(C_1)=1$. Then $\omega_t^{(1)}$ is perfectly correlated with $\omega_{t-1}^{(1)}$, and so, with $\omega_{t-l}^{(1)}$ for every $l$, contradicting the assumption that $\omega_t^{(1)}$ is purely non-deterministic, stationary ergodic, and non-degenerate.
\end{proof}

\begin{lemmaapp}
\label{lemapp:1-state-stationary-ergodic}
If $\mathbb{P}$ is purely non-deterministic and stationary ergodic, then so is $P^{\theta}$ for any pseudo-true one-state model $\theta$.
\end{lemmaapp} 

\begin{proof}
Define
\begin{align}
    & C(a,\eta) \equiv \sum_{\tau=1}^\infty a^\tau\eta^{\tau-1}C_\tau.\label{eq:def-C}
\end{align}
Then
\begin{equation}\label{eq:lambda_max_Omega}
\lambda_{\text{max}}(\Omega(a,\eta)) = -\frac{a^2(1-\eta)^2}{1-a^2\eta^2} + \frac{2(1-\eta)(1-a^2\eta)}{1-a^2\eta^2}\lambda_{\text{max}}(C(a,\eta)),
\end{equation}
where $\lambda_{\text{max}}(C(a,\eta))$ denotes the largest eigenvalue of $C(a,\eta)$. To simplify the exposition, I prove the result under the assumption that the largest eigenvalue of $C(a,\eta)$ is simple at the point $(a^*,\eta^*)$ that maximizes $\lambda_{\text{max}}(\Omega(a,\eta))$.\footnote{The argument can easily be adapted to the case where the largest eigenvalue of $C(a^*,\eta^*)$ is not necessarily simple by replacing the gradient of $\lambda_{\text{max}}(C(a,\eta))$ with its subdifferential and replacing the usual first-order optimality condition with the condition that the zero vector belongs to the subdifferential.\label{fn:subdiff}}
The partial derivatives of $\lambda_{\text{max}}(\Omega(a,\eta))$ with respect to $a$ and $\eta$ are given by
\begin{alignat}{3}
& \frac{\partial \lambda_{\text{max}}(\Omega(a,\eta))}{\partial a} = && \; \frac{-2a(1-\eta)^2}{\left(1-a^2\eta^2\right)^2}-\frac{4a\eta(1-\eta)^2}{\left(1-a^2\eta^2\right)^2}\lambda_\text{max}(C)\nonumber\\ &&& + \frac{2(1-\eta)(1-a^2\eta)}{1-a^2\eta^2} u'_{\text{max}}(C)\frac{\partial C}{\partial a}u_{\text{max}}(C),\label{eq:FOC_a}\\
& \frac{\partial \lambda_{\text{max}}(\Omega(a,\eta))}{\partial \eta} =  &&\; \frac{2a^2(1-\eta)(1-a^2\eta)}{\left(1-a^2\eta^2\right)^2}-\frac{2\left(1+a^4\eta^2+a^2(1-4\eta+\eta^2)\right)}{\left(1-a^2\eta^2\right)^2}\lambda_\text{max}(C)\nonumber\\ &&& + \frac{2(1-\eta)(1-a^2\eta)}{1-a^2\eta^2} u'_{\text{max}}(C)\frac{\partial C}{\partial \eta}u_{\text{max}}(C)\label{eq:FOC_eta},
\end{alignat}
where $u_{\text{max}}(C)$ denotes the eigenvector of $C$ with eigenvalue $\lambda_{\text{max}}(C)$, normalized such that $u'_{\text{max}}(C)u_{\text{max}}(C)=1$, and
\begin{alignat*}{3}
    & \frac{\partial C}{\partial a} &&  = \sum_{\tau=1}^\infty \tau a^{\tau-1}\eta^{\tau-1} C_\tau,\\
    & \frac{\partial C}{\partial \eta} && = \sum_{\tau=1}^\infty (\tau-1)a^{\tau}\eta^{\tau-2} C_\tau.
\end{alignat*}
Note that
\begin{equation}
\eta u'_{\text{max}}(C)\frac{\partial C}{\partial \eta}u_{\text{max}}(C) + \lambda_{\text{max}}(C) = a u'_{\text{max}}(C)\frac{\partial C}{\partial a}u_{\text{max}}(C)\label{eq:FOC_identity}
\end{equation}
for any $a$ and $\eta$.

Let $a^*$ and $\eta^*$ be scalars in the $[-1,1]$ and $[0,1]$ intervals, respectively, that maximize $\lambda_{\text{max}}(\Omega(a,\eta))$. I separately consider the cases $\eta^*=1$ and $\eta^*<1$. If $\eta^*=1$, then $B=0$ in the representation in the proof of Theorem \ref{thm:1_state_general}, the pseudo-true one-state model is i.i.d., and $A=a^*$ can be chosen arbitrarily to satisfy $|a^*|<1$.\footnote{The pseudo-true one-state model then has a zero-state minimal representation.}

In the rest of the proof, I assume that $\eta^*<1$ and show that this implies $a^*\neq 1$---by a similar argument $a^*\neq -1$. Toward a contradiction, suppose $a^*=1$. Setting $a=1$ in the partial derivatives of $\lambda_{\text{max}}(\Omega(a,\eta))$, I get
\begin{alignat*}{3}
& \frac{\partial \lambda_{\text{max}}(\Omega(a,\eta))}{\partial a}\bigg|_{a=1} = && \; \frac{2(1-\eta)^2}{\left(1-\eta^2\right)^2}\left[-1-2\eta\lambda_\text{max}(C)+(1-\eta^2)u'_{\text{max}}(C)\frac{\partial C}{\partial a}u_{\text{max}}(C)\right],\\
& \frac{\partial\lambda_{\text{max}}(\Omega(a,\eta))}{\partial \eta}\bigg|_{a=1} =  &&\; \frac{2(1-\eta)^2}{\left(1-\eta^2\right)^2}\left[1-2\lambda_{\text{max}}(C)+(1-\eta^2)u'_{\text{max}}(C)\frac{\partial C}{\partial \eta}u_{\text{max}}(C)\right],
\end{alignat*}
where $C=C(1,\eta)$ and its partial derivatives are computed at $a=1$. Multiplying the second equation above by $\eta$ and subtracting from it the first equation, I get
\begin{align*}
   & \eta \frac{\partial \lambda_{\text{max}}(\Omega(a,\eta))}{\partial \eta}\bigg|_{a=1}-\frac{\partial \lambda_{\text{max}}(\Omega(a,\eta))}{\partial a}\bigg|_{a=1}\\ &\quad = \frac{2(1-\eta)^2}{\left(1-\eta^2\right)^2}\left[1+\eta+(1-\eta^2)\left(\eta u'_{\text{max}}(C)\frac{\partial C}{\partial \eta}u_{\text{max}}(C)-u'_{\text{max}}(C)\frac{\partial C}{\partial a}u_{\text{max}}(C)\right)\right]\\ & \quad = \frac{2(1-\eta)^2}{\left(1-\eta^2\right)^2}\left[1+\eta-(1-\eta^2)\lambda_{\text{max}}(C)\right],
\end{align*}
where in the second equality I am using identity \eqref{eq:FOC_identity}. Therefore,
\begin{align*}
\frac{\partial \lambda_{\text{max}}(\Omega(a,\eta))}{\partial a}\bigg|_{a=1}=\eta \frac{\partial \lambda_{\text{max}}(\Omega(a,\eta))}{\partial \eta}\bigg|_{a=1}-\frac{2(1-\eta)^2}{\left(1-\eta^2\right)^2}\left(1+\eta-(1-\eta^2)\lambda_{\text{max}}(C(1,\eta))\right).
\end{align*}
Note that 
\[
\lambda_{\text{max}}(C(1,\eta)) \leq \sum_{\tau=1}^\infty \eta^{\tau-1}\lambda_{\text{max}}(C_\tau)< \sum_{\tau=1}^\infty\eta^{\tau-1} = \frac{1}{1-\eta},
\]
where the second inequality is by Lemma \ref{lem:spectral_radius_of_autocorrelation}. Therefore,
\[
-\frac{2(1-\eta)^2}{\left(1-\eta^2\right)^2}\left(1+\eta-(1-\eta^2)\lambda_{\text{max}}(C(1,\eta))\right)< \frac{2(1-\eta)^2}{\left(1-\eta^2\right)^2}(1+\eta - 1-\eta)=0.
\]
On the other hand, by the optimality of $a^*=1$ and $\eta^*<1$,
\[
\frac{\partial \lambda_{\text{max}}(\Omega(a,\eta))}{\partial \eta}\bigg|_{a^*=1,\eta=\eta^*}\leq 0.
\]
Thus,
\[
\frac{\partial \lambda_{\text{max}}(\Omega(a,\eta))}{\partial a}\bigg|_{a^*=1,\eta=\eta^*}<0,
\]
a contradiction to the assumption of optimality of $a^*=1$ and $\eta^*<1$. This proves that $a^*<1$ and establishes that the one-state model with $a=a^*$ and $\eta=\eta^*$ is purely non-deterministic and stationary ergodic. 
\end{proof}

I can now prove the theorem.

\paragraph*{Proof of Theorem \ref{thm:1-state-covariance}.} Setting $M=a$, $D=\sqrt{1-\eta}e_1$, and $N=\Gamma_0^{\frac{-1}{2}}S$ in equation \eqref{eq:d-state-uncond-var}, I get
\[
\text{Var}^\theta(y) = \Gamma_0^{\frac{1}{2}}\left[I+\frac{1}{1-a^2}\left[a^2 (1 - \eta)^2 - \left(1 -2a^2\eta+a^2\eta^2\right)\lambda\right]uu'\right]\Gamma_0^{\frac{1}{2}},
\]
where $a$, $\eta$, $\lambda=\lambda_{\text{max}}(\Omega(a,\eta))$, and $u$ are as in Theorem \ref{thm:1_state_general}. Substituting for $\lambda_{\text{max}}(\Omega(a,\eta))$ from equation \eqref{eq:lambda_max_Omega} in the above equation, I get
\begin{equation}\label{eq:1-state-uncond_var_expanded}
\text{Var}^{\theta}(y_t) = \Gamma_0^{\frac{1}{2}}\left[I +\frac{2(1-\eta)(1-a^2\eta)}{(1-a^2)(1-a^2\eta^2)} \left(a^2(1-\eta) - (1-2a^2\eta+a^2\eta^2)\lambda_{\text{max}}(C)\right)uu'\right]\Gamma_0^{\frac{1}{2}}.
\end{equation}
Let ${a^*}$ and ${\eta^*}$ be scalars in the $[-1,1]$ and $[0,1]$ intervals, respectively, that maximize $\lambda_{\text{max}}(\Omega(a,\eta))$. I separately consider the cases ${\eta^*}=1$ and ${\eta^*}<1$. If ${\eta^*}=1$, then the right-hand side of equation \eqref{eq:1-state-uncond_var_expanded} is equal to $\Gamma_0$.

Next suppose ${\eta^*}<1$. By the argument in the proof of Lemma \ref{lemapp:1-state-stationary-ergodic}, the first-order optimality condition with respect to $a$ must hold with equality at $a={a^*}$ and $\eta={\eta^*}<1$. Setting $\partial \lambda_{\text{max}}(\Omega(a,\eta))/\partial a=0$ in \eqref{eq:FOC_a} and multiplying both sides of the equation by ${a^*}$, I get, using \eqref{eq:FOC_identity},
\begin{align}
& \frac{2{a^*}^2(1-{\eta^*})^2}{\left(1-{a^*}^2{\eta^*}^2\right)^2}+\frac{4{a^*}^2{\eta^*}(1-{\eta^*})^2}{\left(1-{a^*}^2{\eta^*}^2\right)^2}\lambda_\text{max}(C)\nonumber\\ & \quad =\frac{2(1-{\eta^*})(1-{a^*}^2{\eta^*})}{1-{a^*}^2{\eta^*}^2}\lambda_{\text{max}}(C)+\frac{2(1-{\eta^*})(1-{a^*}^2{\eta^*})}{1-{a^*}^2{\eta^*}^2}{\eta^*} u'_{\text{max}}(C)\frac{\partial C}{\partial \eta}u_{\text{max}}(C).\label{eq:FOC_a_mult_a}
\end{align}
Setting ${\eta^*}=0$ in the above equation, I get ${a^*}^2=\lambda_{\text{max}}(C)$. Setting ${a^*}^2=\lambda_{\text{max}}(C)$ in equation \eqref{eq:1-state-uncond_var_expanded} then establishes that $\text{Var}^{\theta}(y_t)=\Gamma_0$ in the case where ${\eta^*}=0$.

Finally, I consider the case where ${\eta^*}\in (0,1)$. Then additionally the first-order optimality condition with respect to $\eta$ must hold with equality. Setting $\partial \lambda_{\text{max}}(\Omega(a,\eta))/\partial \eta=0$ in equation \eqref{eq:FOC_eta}, multiplying it by ${\eta^*}$, solving for ${\eta^*} u'_{\text{max}}(C)\frac{\partial C}{\partial \eta}u_{\text{max}}(C)$, and substituting in equation \eqref{eq:FOC_a_mult_a}, I get
\begin{align*}
& \frac{2{a^*}^2(1-{\eta^*})^2}{\left(1-{a^*}^2{\eta^*}^2\right)^2}+\frac{4{a^*}^2{\eta^*}(1-{\eta^*})^2}{\left(1-{a^*}^2{\eta^*}^2\right)^2}\lambda_\text{max}(C)\\ & \quad =\frac{2(1-{\eta^*})(1-{a^*}^2{\eta^*})}{1-{a^*}^2{\eta^*}^2}\lambda_{\text{max}}(C)-\frac{2{a^*}^2{\eta^*}(1-{\eta^*})(1-{a^*}^2{\eta^*})}{\left(1-{a^*}^2{\eta^*}^2\right)^2}\\ &\qquad +\frac{2{\eta^*}\left(1+{a^*}^4{\eta^*}^2+{a^*}^2(1-4{\eta^*}+{\eta^*}^2)\right)}{\left(1-{a^*}^2{\eta^*}^2\right)^2}\lambda_\text{max}(C).
\end{align*}
Simplifying the above expression leads to
\[
{a^*}^2(1-{\eta^*}) = \left(1-2{a^*}^2{\eta^*}+{a^*}^2{\eta^*}^2\right)\lambda_{\text{max}}(C).
\]
Combining the above identity with equation \eqref{eq:1-state-uncond_var_expanded} implies that $\text{Var}^{\theta}(y_t)=\Gamma_0$ and finishes the proof of the theorem.\hfill\qed

\subsection*{Proof of Theorem \ref{thm:1-state-exp-erg-closed-form}}
Let $\lambda$ denote the eigenvalue of $C_1$ largest in magnitude.\footnote{The proof does not assume that $\lambda$ is unique. I allow for the possibility that $\lambda$ and $\lambda'=-\lambda$ are both eigenvalues of $C_1$ and $|\lambda|=|\lambda'|=\rho(C_1)$.}  If $\rho(C_1)=0$, then $\rho(C_\tau)=0$ for all $\tau\geq 1$. Since $C_\tau$ are symmetric matrices, this implies that $C_\tau=0$ for all $\tau \geq 1$. Therefore,
\[
\lambda_{\text{max}}(\Omega(a,\eta)) = -\frac{a^2(1-\eta)^2}{1-a^2\eta^2}.
\]
The above expression is maximized by setting $(1-\eta)a=0$. Therefore, by Theorem \ref{thm:1_state_general},  for any pseudo-true one-state model, $E^\theta_t[y_{t+s}] = a^{s}(1-\eta)qp'\sum_{\tau=0}^\infty a^\tau\eta^\tau y_{t-\tau}=0$. On the other hand, if $\rho(C_1)=0$, then $\lambda=0$. Therefore, the theorem holds in the case $\rho(C_1)=0$.

In the rest of the proof, I assume $\rho(C_1)> 0$. Define
\begin{align*}
\overline{f}(a,\eta) & \equiv -\frac{a^2(1-\eta)^2}{1-a^2\eta^2} + \frac{2(1-\eta)(1-a^2\eta)}{1-a^2\eta^2}\sum_{\tau=1}^\infty |a|^\tau \eta^{\tau-1} \rho(C_1)^\tau\\ & =  -\frac{a^2(1-\eta)^2}{1-a^2\eta^2} + \frac{2(1-\eta)(1-a^2\eta)}{1-a^2\eta^2}\frac{|a|\rho(C_1)}{1-\eta|a|\rho(C_1)},
\end{align*}
where in the second equality I am using the fact that $\rho(C_\tau)<1$, established in Lemma \ref{lem:spectral_radius_of_autocorrelation}. Function $\overline{f}(a,\eta)$ has two maximizers given by $(\overline{a}^*,\overline{\eta}^*)=(-\rho(C_1),0)$ and $(\overline{a}^*,\overline{\eta}^*)=(\rho(C_1),0)$ with the maximum given by $\overline{f}^*=\rho(C_1)^2$.
I establish the theorem by showing that $\lambda_{\text{max}}(\Omega(a,\eta))\leq \overline{f}(a,\eta)$ for all $a$ and $\eta$, $\lambda_{\text{max}}(\Omega(\lambda,0))=\overline{f}(\lambda,0)=\overline{f}^*$, and $\lambda_{\text{max}}(\Omega(-\lambda,0))\leq \overline{f}(-\lambda,0)=\overline{f}^*$ with the inequality strict if $-\lambda$ is not an eigenvalue of $C_1$. This establishes that $(a^*,\eta^*)=(\lambda,0)$ is the unique maximizer of $\lambda_{\text{max}}(\Omega(a,\eta))$ if $-\lambda$ is not eigenvalue of $C_1$ and that $(a^*,\eta^*)=(\lambda,0)$ and $(a^*,\eta^*)=(-\lambda,0)$ are the only maximizers of $\lambda_{\text{max}}(\Omega(a,\eta))$ if $\lambda$ and $-\lambda$ are both eigenvalues of $C_1$.

As the first step in doing so, I show that for all $a$ and $\tau$,
\[
\lambda_{\text{max}}\left(a^\tau C_\tau\right)\leq |a|^\tau\rho(C_1)^\tau,
\]
by considering four disjoint cases: If $a\leq 0$ and $\lambda_{\text{min}}(C_\tau)\leq 0$, then
\[
\lambda_{\text{max}}\left(a^\tau C_\tau\right)=a^\tau\lambda_{\text{min}}(C_\tau)=|a|^\tau \left|\lambda_{\text{min}}(C_\tau)\right|\leq |a|^\tau\rho(C_1)^\tau.
\]
If $a\leq 0$ and $\lambda_{\text{min}}(C_\tau)>0$, then
\[
\lambda_{\text{max}}\left(a^\tau C_\tau\right)=a^\tau\lambda_{\text{min}}(C_\tau)\leq 0 \leq |a|^\tau\rho(C_1)^\tau.
\]
If $a>0$ and $\lambda_{\text{max}}(C_\tau)\leq 0$, then
\[
\lambda_{\text{max}}\left(a^\tau C_\tau\right)=a^\tau\lambda_{\text{max}}(C_\tau)\leq 0 \leq |a|^\tau\rho(C_1)^\tau.
\]
Finally, if $a>0$ and $\lambda_{\text{max}}(C_\tau)> 0$, then
\[
\lambda_{\text{max}}\left(a^\tau C_\tau\right)=a^\tau\lambda_{\text{max}}(C_\tau)=|a|^\tau \left|\lambda_{\text{max}}(C_\tau)\right|\leq |a|^\tau\rho(C_1)^\tau.
\]
Thus, $\lambda_{\text{max}}\left(a^\tau C_\tau\right)\leq |a|^\tau\rho(C_1)^\tau$ regardless of the value of $a$ and the eigenvalues of $C_1$. Therefore,
\begin{align*}
    \lambda_{\text{max}}\left(\sum_{\tau=1}^\infty a^\tau \eta^{\tau-1} C_\tau\right) & \leq \sum_{\tau=1}^\infty \eta^{\tau-1}\lambda_{\text{max}}\left(a^\tau  C_\tau\right) \leq \sum_{\tau=1}^\infty \eta^{\tau-1}|a|^\tau\rho(C_1)^\tau  = \frac{|a|\rho(C_1)}{1-\eta|a|\rho(C_1)},
\end{align*}
where the first inequality is using the fact that $\eta^{\tau-1}\geq 0$ for all $\tau\geq 1$ and Weyl's inequality. Consequently, 
\[
\lambda_{\text{max}}(\Omega(a,\eta))\leq \overline{f}(a,\eta)< \rho(C_1)^2
\]
for any $a,\eta$ such that $(|a|,\eta)\neq (\rho(C_1),0)$.

I finish the proof by arguing that $\lambda_{\text{max}}(\Omega(\lambda,0))=\rho(C_1)^2$ and $\lambda_{\text{max}}(\Omega(-\lambda,0))\leq \overline{f}(-\lambda,0)=\rho(C_1)^2$ with the inequality strict if $-\lambda$ is not an eigenvalue of $C_1$. To see this, first note that
\[
\lambda_{\text{max}}(\Omega(a,0)) = -a^2 + 2\lambda_{\text{max}}(a C_1) = \begin{dcases}-a^2 + 2a\lambda_{\text{min}}(C_1) \qquad &\text{if}\qquad a< 0 ,\\ -a^2 + 2a\lambda_{\text{max}}( C_1) \qquad &\text{if}\qquad a\geq  0.\end{dcases}
\]
Thus,
\[
\max_{a\in[-1,1]}\lambda_{\text{max}}(\Omega(a,0))= \begin{dcases}\lambda_{\text{min}}(C_1)^2  \qquad &\text{if}\qquad |\lambda_{\text{min}}(C_1)|> \lambda_{\text{max}}(C_1) ,\\ \lambda_{\text{max}}(C_1)^2\qquad &\text{if}\qquad |\lambda_{\text{min}}(C_1)|\leq  \lambda_{\text{max}}(C_1),\end{dcases}
\]
and 
\[
\arg \max_{a\in[-1,1]}\lambda_{\text{max}}(\Omega(a,0))= \begin{dcases}\left\{\lambda_{\text{min}}(C_1)\right\}  \qquad &\text{if}\qquad |\lambda_{\text{min}}(C_1)|> \lambda_{\text{max}}(C_1) ,\\\left\{\lambda_{\text{min}}(C_1),\lambda_{\text{max}}(C_1)\right\}\qquad &\text{if}\qquad |\lambda_{\text{min}}(C_1)|=\lambda_{\text{max}}(C_1),\\ \left\{\lambda_{\text{max}}(C_1)\right\}\qquad &\text{if}\qquad |\lambda_{\text{min}}(C_1)|< \lambda_{\text{max}}(C_1).\end{dcases}
\]
Since $C_1$ is a symmetric matrix, the eigenvalues of $C_1$ are all real, and so,
\[
\rho(C_1) = \begin{dcases} -\lambda_{\text{min}}(C_1) \qquad & \text{if}\qquad  |\lambda_{\text{min}}(C_1)|> \lambda_{\text{max}}(C_1),\\ \lambda_{\text{max}}(C_1)\qquad & \text{if}\qquad   |\lambda_{\text{min}}(C_1)|\leq  \lambda_{\text{max}}(C_1). \end{dcases}
\]
This establishes that, in any pseudo-true one-state model, $\eta=0$, $a=\lambda$, and
\[
\Omega(a,\eta) = -\lambda^2 I + 2\lambda C_1.
\]
By Theorem \ref{thm:1_state_general}, $u$ is an eigenvector of $\Omega(a,\eta)$ with eigenvalue $\lambda_{\text{max}}(\Omega(a,\eta))=\lambda^2$ and $u'u=1$. Therefore, $u$ is also an eigenvector of $ C_1$, but with eigenvalue $\lambda$. This completes the proof of the theorem.\hfill\qed

\subsection*{Proof of Proposition \ref{prop:Markovian}}
I first state and prove a lemma, which is used in the proof of the proposition.
\begin{lemmaapp}\label{lemapp:Markovian}
Any Markovian model $\theta$ has a representation as in Lemma \ref{lem:d-state-characterization} for which $D'D=I$.
\end{lemmaapp}

\begin{proof}
Fix a Markovian model $\theta$, and let $M$, $D$, and $N$ be as in Lemma \ref{lem:d-state-characterization}. By \eqref{eq:s_ahead_forecast_simplified}, the $s$-step-ahead forecast under model $\theta$ is given by
\begin{equation*}
E^\theta_t[y_{t+s}] = {N'}^{-1}DM^{s-1}\sum_{\tau=0}^\infty \left(M  \left(I-D'D\right)\right)^{\tau}M D'N' y_{t-\tau}.
\end{equation*}
Since $\theta$ is Markovian and $N$ is invertible, $D\left(M  \left(I-D'D\right)\right)^{\tau}M D'=\mathbf{0}$ for all $\tau\geq 1$. As the first step of the proof, I use this identity and an inductive argument to show that $DM^sD'=(DMD')^s$ for all $s\geq 2$. The following equation establishes the induction base:
\[
\mathbf{0}=D\left(M  \left(I-D'D\right)\right)M D'=DM^2D'-DMD'DMD'.
\]
As the induction hypothesis, suppose $DM^s D'=(DMD')^s$ for some $s\geq 2$. Note that
\begin{align*}
    D\left(M  \left(I-D'D\right)\right)^sM D' & = D\left(M  \left(I-D'D\right)\right)^{s-1}M(I-D'D)M D'\\ & = D\left(M  \left(I-D'D\right)\right)^{s-1}M^2D' - D\left(M  \left(I-D'D\right)\right)^{s-1}MD'DMD'\\ & = D\left(M  \left(I-D'D\right)\right)^{s-1}M^2D',
\end{align*}
where the last equality follows the fact that $D\left(M  \left(I-D'D\right)\right)^{s-1}MD'=\mathbf{0}$ for any $s\geq2$. By a similar argument,
\[
D\left(M  \left(I-D'D\right)\right)^{s}M D' = D\left(M  \left(I-D'D\right)\right)^{s-2}M^3D' = \dots = D\left(M  \left(I-D'D\right)\right)M^{s}D'.
\]
Therefore,
\begin{align*}
    D\left(M  \left(I-D'D\right)\right)^{s}M D'=DM^{s+1}D'-DMD'DM^s D'=DM^{s+1}D'-(DMD')^{s+1},
\end{align*}
where the last equality follows the induction hypothesis. The assumption that $D\left(M  \left(I-D'D\right)\right)^{s}M D'=\mathbf{0}$ then proves the induction step.

I next find a model $\tilde{\theta}$, represented by matrices $\tilde{M}$, $\tilde{D}$, and $\tilde{N}$, that is observationally equivalent to $\theta$ and for which $\tilde{D}'\tilde{D}=I$. Since $D\in\mathbb{R}^{n\times d}$ is a rectangular diagonal matrix and $d\leq n$,
\[
DMD' = \begin{pmatrix} M_1 & \mathbf{0} \\ \mathbf{0} & \mathbf{0}\end{pmatrix}
\]
for some $d\times d$ matrix $M_1$. Let $\tilde{M}=M_1$, $\tilde{D}\in\mathbb{R}^{n\times d}$ be the rectangular diagonal matrix with its diagonal elements equal to one, and $\tilde{N}=N$. Then $\tilde{D}'\tilde{D}=I$. Furthermore,
\[
DM^sD' = (DMD')^s = \begin{pmatrix} M^s_1 & \mathbf{0} \\ \mathbf{0} & \mathbf{0}\end{pmatrix}=\tilde{D}\tilde{M}^s\tilde{D}'.
\]
By equation \eqref{eq:s_ahead_forecast_simplified}, the forecasts are identical under the two models:
\[
E^\theta_t[y_{t+s}] = {N'}^{-1}DM^sD'N' y_{t}=\tilde{N}^{\prime^{\scriptstyle-1}}\tilde{D}\tilde{M}^s\tilde{D}'\tilde{N}' y_{t}=E^{\tilde{\theta}}_t[y_{t+s}].
\]
By equation \eqref{eq:d-state-uncond-var}, the unconditional variance of the observable is also identical under the two models:
\[
\text{Var}^\theta(y) = {N}^{\prime^{\scriptstyle-1}}\left(I+ \sum_{\tau=1}^\infty DM^\tau D'D{M'}^\tau D'\right)N^{-1}=\tilde{N}^{\prime^{\scriptstyle-1}}\left(I+ \sum_{\tau=1}^\infty \tilde{D}\tilde{M}^\tau \tilde{D}'\tilde{D}\tilde{M}^{\prime^\tau} \tilde{D}'\right)\tilde{N}^{-1}=\text{Var}^{\tilde{\theta}}(y).
\]
On the other hand,
\[
E^\theta[y_{t+s}y_t']=E^\theta[E_t^\theta[y_{t+s}]y_t']={N'}^{-1}DM^sD'N'E^\theta[y_ty_t']={N'}^{-1}DM^sD'N'\text{Var}^\theta(y),
\]
and similarly for $E^{\tilde{\theta}}[y_{t+s}y_t']$. Therefore, $E^\theta[y_{t+s}y_t']=E^{\tilde{\theta}}[y_{t+s}y_t']$ for all $s$; that is, $P^\theta$ and $P^{\tilde{\theta}}$ also have identical autocovariance matrices at all lags. This conclusion, together with the fact that $P^\theta$ and $P^{\tilde{\theta}}$ are both zero-mean Gaussian distributions, implies that they are observationally equivalent.
\end{proof}

I can now prove the proposition.
\paragraph*{Proof of Proposition \ref{prop:Markovian}.}

By equation \eqref{eq:Sigma_y_hat}, 
\[
\text{Var}^\theta_t(y_{t+1}) = B'\hat{\Sigma}_z B + R,
\]
where
$\hat{\Sigma}_{z}$ solves the algebraic Riccati equation \eqref{eq:Riccati}. The equation can be written as 
\begin{equation}\label{eq:Riccati_appendix}
\hat{\Sigma}_z = A\hat{\Sigma}_z^{\frac{1}{2}}\left(I-\hat{\Sigma}_z^{\frac{1}{2}} B\left(B'\hat{\Sigma}_z B + R\right)^{-1}B'\hat{\Sigma}_z^{\frac{1}{2}}\right)\hat{\Sigma}_z^{\frac{1}{2}}A'+Q.   
\end{equation}
Since $R$ is a positive semidefinite matrix, so is $I-\hat{\Sigma}_z^{\frac{1}{2}} B\left(B'\hat{\Sigma}_z B + R\right)^{-1}B'\hat{\Sigma}_z^{\frac{1}{2}}$. Therefore, $\hat{\Sigma}_z \succeq Q$, and so, $\text{Var}^\theta_t(y_{t+1}) \succeq B'QB + R$. On the other hand,
\[
\text{Var}^\theta(y_{t+1}|z_t) = B'\text{Var}^\theta(z_{t+1}|z_t)B + R = B'QB + R,
\]
where I am using the assumption that $w_t$ is i.i.d. $\mathcal{N}(0,Q)$, $v_t$ is i.i.d. $\mathcal{N}(0,R)$, and $w_t$ and $v_t$ are independent. This proves the first part of the proposition.

To prove the second part, first assume that $\text{Var}^\theta_t(y_{t+1})=B'QB + R$. Together with equation \eqref{eq:Riccati_appendix}, this implies that 
\[
B'A\hat{\Sigma}_z^{\frac{1}{2}}\left(I-\hat{\Sigma}_z^{\frac{1}{2}} B\left(B'\hat{\Sigma}_z B + R\right)^{-1}B'\hat{\Sigma}_z^{\frac{1}{2}}\right)\hat{\Sigma}_z^{\frac{1}{2}}A'B=\mathbf{0}.
\]
Since $\left(I-\hat{\Sigma}_z^{\frac{1}{2}} B\left(B'\hat{\Sigma}_z B + R\right)^{-1}B'\hat{\Sigma}_z^{\frac{1}{2}}\right)$ is a symmetric positive semidefinite matrix, the above equation implies that 
\[
B'A\hat{\Sigma}_z^{\frac{1}{2}}\left(I-\hat{\Sigma}_z^{\frac{1}{2}} B\left(B'\hat{\Sigma}_z B + R\right)^{-1}B'\hat{\Sigma}_z^{\frac{1}{2}}\right)=\mathbf{0}.
\]
On the other hand, by equation \eqref{eq:s_ahead_forecast}, the one-step-ahead forecast under model $\theta$ is given by
\[
E^\theta_t[y_{t+1}]=B'\sum_{\tau=0}^\infty (A-KB')^{\tau}K y_{t-\tau}.
\]
Substituting for $K$ from equation \eqref{eq:Kalman_gain}, I get
\[
B'(A-KB') = B'\left(A-A\hat{\Sigma}_z B\left(B'\hat{\Sigma}_z B + R\right)^{-1}B'\right)=B'A\hat{\Sigma}_z^{\frac{1}{2}}\left(I-\hat{\Sigma}_z^{\frac{1}{2}} B\left(B'\hat{\Sigma}_z B + R\right)^{-1}B'\hat{\Sigma}_z^{\frac{1}{2}}\right)\hat{\Sigma}_z^{\frac{-1}{2}}=\mathbf{0}.
\]
Therefore,
\[
E^\theta_t[y_{t+1}]=B'\sum_{\tau=0}^\infty (A-KB')^{\tau}K y_{t-\tau}=B'Ky_t.
\]
On the other hand, $\text{Var}^\theta_t(y_{t+1})=B'\hat{\Sigma}_z B + R$. Under model $\theta$, the mean and variance of $y_{t+1}$ conditional on $\{y_{\tau}\}_{\tau\leq t}$ are both independent of $\{y_{\tau}\}_{\tau< t}$. Furthermore, $P^\theta$ is Gaussian. Therefore, it is Markovian.

Next, suppose $P^\theta$ is Markovian. Then by Lemma \ref{lemapp:Markovian}, model $\theta$ has a representation as in Lemma \ref{lem:d-state-characterization} for which $D'D=I$. By equation \eqref{eq:VDDV}, then
\[
\hat{\Sigma}_z^{\frac{1}{2}}B\left(B'\hat{\Sigma}_z B+R\right)^{-1}B'\hat{\Sigma}_z^{\frac{1}{2}}=VD'DV' = I,
\]
where the second equality follows the facts that $D'D=I$ and $V$ is orthogonal.
Substituting in equation \eqref{eq:Riccati_appendix}, I get $\hat{\Sigma}_z=Q$. Therefore,
\[
\text{Var}^\theta_t(y_{t+1}) = B'\hat{\Sigma}_z B + R=B'QB + R.
\]
This completes the proof of the proposition.\hfill\qed

\subsection*{Proof of Theorem \ref{thm:d-state-m-i-o}}
By Lemma \ref{lem:d-state-characterization}, the agent's model can be represented in terms of matrices $M$, $D$, and $N$. Since the agent is restricted to the set of Markovian models, by Lemma \ref{lemapp:Markovian}, I can set $D=(I \; \mathbf{0})'$. Let $S\equiv\Gamma_0^{\frac{1}{2}}N$ and $\Gamma\equiv \Gamma_0^{\frac{-1}{2}}\Gamma_1\Gamma_0^{\frac{-1}{2}}$. The expression for the KLDR in \eqref{eq:def_H_M_D_N} then simplifies to
\begin{align*}
\text{KLDR}(M,S,D) = &\; -\frac{1}{2}\log\text{det}\left(SS'\right)+ \frac{1}{2}\tr\left(S'S\right)-\tr\left( M D'S' \Gamma SD\right) + \frac{1}{2}\tr\left(MD'S'SDM'\right)+\text{constant}.
\end{align*}
Write $S=(S_1\; S_2)$, where $S_1\in\mathbb{R}^{n\times d}$ and $S_2\in\mathbb{R}^{n\times (n-d)}$. The above expression can then be written as
\[
-\frac{1}{2}\log\text{det}\left(S_1S_1'+S_2S_2'\right)+\frac{1}{2}\tr\left(S_1'S_1\right)+\frac{1}{2}\tr\left(S_2'S_2\right)-\tr\left( M S_1' \Gamma S_1\right) + \frac{1}{2}\tr\left(M S_1'S_1M'\right)+\text{constant}.
\]

I next optimize the above expression with respect to $M$, $S_1$, and $S_2$. The first-order optimality condition with respect to $S_2$ is given by $\left(S_1S_1' + S_2S_2'\right)^{-1}S_2=S_2$, which can be rewritten as $S_1S_1'S_2 + S_2(S_2'S_2-I)=\mathbf{0}$. Let $b_0$ be an arbitrary vector in $\mathbb{R}^{n-d}$, $b_1\equiv S_1'S_2b_0\in \mathbb{R}^d$, and $b_2 \equiv (S_2'S_2-I)b_0\in \mathbb{R}^{n-d}$. The above equation then implies that 
\[
0=\left(S_1S_1'S_2 + S_2(S_2'S_2-I)\right)b_0 = S_1b_1 + S_2 b_2 = Sb,
\]
where $b\equiv \begin{pmatrix}b_1' & b_2'\end{pmatrix}'\in\mathbb{R}^n$. Since $S$ is an invertible matrix, it must be that $b=0$. Therefore, $b_1=0$ and $b_2=0$. Since $b_0$ was arbitrary, $S_2'S_2=I$ and $S_1'S_2=\mathbf{0}$. On the other hand,
\[
\log\text{det}\left(S_1S_1'+S_2S_2'\right) = \log\text{det}(SS')=\log\text{det}(S'S)=\log\text{det}\begin{pmatrix}S_1'S_1 & S_1'S_2 \\ S_2'S_1 & S_2'S_2\end{pmatrix}.
\]
Therefore,
\[
\log\text{det}\left(S_1S_1'+S_2S_2'\right)=\log\text{det}\begin{pmatrix}S_1'S_1 & \mathbf{0} \\ \mathbf{0} & I\end{pmatrix}=\log\text{det}(S_1'S_1).
\]
The KLDR can thus be written only as a function of $M$ and $S_1$ as
\begin{equation}
\text{KLDR}(M,S_1)=-\frac{1}{2}\log\text{det}\left(S_1'S_1\right)+\frac{1}{2}\tr\left(S_1'S_1\right)-\tr\left( M S_1' \Gamma S_1\right) + \frac{1}{2}\tr\left(M S_1'S_1M'\right)+\text{constant}.\label{eq:KLDR_M_S_1}
\end{equation}
The first-order optimality conditions with respect to $M$ and $S_1$ are then given by
\begin{align}
    & -S_1'\Gamma'S_1 + MS_1'S_1 = 0,\label{eq:FOC_d_M}\\
    & -{S_1^\dagger}' + S_1 - \Gamma S_1 M - \Gamma' S_1 M' + S_1M'M = 0\label{eq:FOC_d_S_1}.
\end{align}
Since $S_1'S_1$ is invertible, \eqref{eq:FOC_d_M} can be solve for $M$ to get $M=S_1'\Gamma'S_1(S_1'S_1)^{-1}$. Substituting in \eqref{eq:FOC_d_S_1}, I get
\begin{equation}
S_1(S_1'S_1)^{-1} = S_1 - \Gamma S_1S_1'\Gamma'S_1(S_1'S_1)^{-1} - \Gamma'S_1(S_1'S_1)^{-1}S_1'\Gamma S_1 + S_1(S_1'S_1)^{-1}S_1'\Gamma S_1S_1'\Gamma'S_1(S_1'S_1)^{-1},\label{eq:FOC_big_S_1}
\end{equation}
where I am using the fact that $S_1^\dagger=(S_1'S_1)^{-1}S_1'$. Next consider the singular-value decomposition of $S_1$:
\begin{equation}\label{eq:S_1_SVD}
S_1 = U\Sigma V',
\end{equation}
where $U\in\mathbb{R}^{n\times n}$ and $V\in\mathbb{R}^{d\times d}$ are orthogonal matrices, and $\Sigma\in\mathbb{R}^{n\times d}$ is a rectangular diagonal matrix. Substituting for $S_1$ in \eqref{eq:FOC_big_S_1} from \eqref{eq:S_1_SVD} and multiplying the result from left and right by $U'$ and $V\Sigma'$, respectively, I get
\begin{equation}\label{eq:FOC_SVD}
\Sigma(\Sigma'\Sigma)^{-1}\Sigma' = \Sigma\Sigma' - X \Sigma \Sigma'X'\Sigma(\Sigma'\Sigma)^{-1}\Sigma'-X'\Sigma(\Sigma'\Sigma)^{-1}\Sigma'X\Sigma\Sigma' + \Sigma(\Sigma'\Sigma)^{-1}\Sigma'X\Sigma\Sigma'X'\Sigma(\Sigma'\Sigma)^{-1}\Sigma',
\end{equation}
where $X \equiv U'\Gamma U$. Note that $\Sigma = \begin{pmatrix}\Sigma_1\\ \mathbf{0}\end{pmatrix}$ for some diagonal matrix $\Sigma_1\in\mathbb{R}^{n\times d}$. Moreover, since $S_1'S_1$ is invertible, so is $\Sigma_1$. Therefore,
\begin{align*}
    & \Sigma(\Sigma'\Sigma)^{-1}\Sigma' = \begin{pmatrix}I & \mathbf{0}\\ \mathbf{0} & \mathbf{0}\end{pmatrix},\\
    & \Sigma\Sigma' = \begin{pmatrix} \Sigma_1^2 & \mathbf{0}\\ \mathbf{0} & \mathbf{0}\end{pmatrix}.
\end{align*}
Write $X=\begin{pmatrix} X_{11} & X_{12}\\ X_{21} & X_{22}\end{pmatrix}$, where $X_{11}\in\mathbb{R}^{d\times d}$, $X_{12}\in\mathbb{R}^{d\times (n-d)}$, $X_{21}\in\mathbb{R}^{(n-d)\times d}$, and $X_{22}\in\mathbb{R}^{(n-d)\times (n-d)}$. Equation \eqref{eq:FOC_SVD} then implies
\begin{align}
   &  X_{11}'X_{11} = I - \Sigma_1^{-2},\label{eq:FOC_SVD_1}\\
   &  X_{21}\Sigma_1^2X_{11}' + X_{12}'X_{11}\Sigma_1^2=0.\label{eq:FOC_SVD_2}
\end{align}
These equations fully characterize the set of all (local) extrema of the KLDR.

I next use these equations to show that, as long as either $d$ is equal to one or $\Gamma_1$ is symmetric, and for any $i=1,\dots,d$, the $i$th coordinate vector $e_i\in\mathbb{R}^n$ is an eigenvector of $(X+X')/2$ with eigenvalue $e_i'Xe_i$.\footnote{With slight abuse of notation, I use $e_i$ to denote the $i$th coordinate vector both in $\mathbb{R}^n$ and in $\mathbb{R}^d$. Whether $e_i\in\mathbb{R}^n$ or $e_i\in\mathbb{R}^d$ will be clear from the context.} If $e_i'Xe_i=0$, then trivially $e_i$ is an eigenvector of $(X+X')/2$ with eigenvalue $e_i'Xe_i=0$. So in the rest of the proof, I consider the case where $e_i'Xe_i\neq 0$. First, suppose $d=1$. Then $i=1$ and $X_{11}'=X_{11}=e_1'Xe_1\neq 0$. On the other hand, $\Sigma_1$ is a non-zero scalar. Equation \eqref{eq:FOC_SVD_2} then implies that $X_{21}+X_{12}'=0$. Therefore,
\[
\left(\frac{X+X'}{2}\right)e_1 = \frac{1}{2}\begin{pmatrix}2X_{11} & X_{12}+X_{21}' \\ X_{21}+X_{12}' & X_{22}+X_{22}'\end{pmatrix}\begin{pmatrix}1 \\ \mathbf{0}\end{pmatrix}=\begin{pmatrix}X_{11}\\\mathbf{0}\end{pmatrix}=e_1'Xe_1e_1,
\]
proving that $e_1$ is an eigenvector of $(X+X')/2$ with eigenvalue $e_1Xe_1$. Next, suppose $\Gamma_1$ is symmetric. This implies that $\Gamma$, and by extension, $X$ are symmetric matrices. Equation \eqref{eq:FOC_SVD_1} then implies that $X_{11}$ is a diagonal matrix. Since $\Sigma_1$ is also diagonal, it commutes with $X_{11}$. Equation \eqref{eq:FOC_SVD_2} then implies that
\begin{equation}
(X_{21}+X_{12}')X_{11}=2X_{21}X_{11}=\mathbf{0},\label{eq:SVD_intermediate}  
\end{equation}
where I am using the fact that $\Sigma_1$ is non-singular and $X$ is symmetric. But since $X_{11}$ is a diagonal matrix, it can be written as 
\[
X_{11}=\sum_{k=1}^d e_k'X_{11}e_ke_ke_k'.
\]
Substituting in \eqref{eq:SVD_intermediate}, I get
\[
\sum_{k=1}^dX_{21} e_k'X_{11}e_ke_ke_k'=\mathbf{0}.
\]
In particular, it must be the case that $X_{21}e_i'X_{11}e_ie_i=0$. But since $e_i'X_{11}e_i=e_i'Xe_i\neq 0$, it must be that $X_{21}e_i=0$. Therefore,
\[
\left(\frac{X+X'}{2}\right)e_i = \begin{pmatrix}X_{11} & X_{12}\\ X_{21} & X_{22}\end{pmatrix}\begin{pmatrix}e_i \\ \mathbf{0}\end{pmatrix}=\begin{pmatrix}X_{11}e_i\\ X_{21}e_i\end{pmatrix}=\begin{pmatrix}e_i'X_{11}e_ie_i\\ 0\end{pmatrix}=e_i'Xe_ie_i,
\]
where the third equality relies on the fact that $X_{11}$ is diagonal. This proves that $e_i$ is an eigenvector of $(X+X')/2$ with eigenvalue $e_i'Xe_ie_i$.

I next show that any matrices $M$ and $S_1$ that satisfy the first-order optimality conditions \eqref{eq:FOC_d_M} and \eqref{eq:FOC_d_S_1} must be of the form 
\begin{align}
    & M = \sum_{i=1}^d a_iv_iv_i',\label{eq:M_solution}\\
    & S_1=\sum_{i=1}^d \frac{1}{\sqrt{1-a_i^2}}u_iv_i',\label{eq:S_1_solution}
\end{align}
where $\{a_i\}_{i=1}^d$ are eigenvalues of $C_1$, $u_i\in\mathbb{R}^n$ denotes an eigenvector with eigenvalue $a_i$ normalized such that $u_i'u_k=\mathbbm{1}_{\{i=k\}}$ for all $i,k\in\{1,\dots,d\}$, and $\{v_i\}_{i=1}^d$ is an orthonormal basis for $\mathbb{R}^d$. To see this, first note that equation \eqref{eq:S_1_SVD} can be written as
\[
S_1 = U\Sigma V'=U\sum_{i=1^d}\sigma_ie_ie_i'V',
\]
where $\sigma_i$ denotes the $i$th diagonal element of $\Sigma\in\mathbb{R}^{n\times d}$. I let $u_i\equiv Ue_i$ and $v_i\equiv Ve_i$. Since $U$ and $V$ are orthogonal matrices, $\{u_i\}_{i=1}^d$ is a set of orthonormal vectors and $\{v_i\}_{i=1}^d$ is an orthonormal basis for $\mathbb{R}^d$. Therefore, to show that $S_1$ takes the form given in \eqref{eq:S_1_solution}, I only need to show that $u_i$ is an eigenvector of $C_1$ with eigenvalue $a_i$ and $\sigma_i=1/\sqrt{1-a_i^2}$. Note that
\[
C_1u_i=\frac{1}{2}\left(\Gamma+\Gamma'\right)Ue_i=\frac{1}{2}UU'\left(\Gamma+\Gamma'\right)Ue_i=U\left(\frac{X+X'}{2}\right)e_i=Ue_i'Xe_ie_i=e_i'Xe_iu_i,
\]
where the fourth equality uses the fact that $e_i$ is an eigenvector of $(X+X')/2$. Therefore, $u_i$ is an eigenvector of $C_1$. On the other hand, multiplying equation \eqref{eq:FOC_SVD_1} from left and right by $e_i'$ and $e_i$, respectively, for $i\in\{1,\dots,d\}$ and using the fact that $X_{11}$ is diagonal, I get
\[
\left(e_i'X_{11}e_i\right)^2 = 1- \sigma_i^{-2}.
\]
But 
\[
e_i'X_{11}e_i = e_i'Xe_i = e_i'\left(\frac{X+X'}{2}\right)e_i= e_i'U\left(\frac{\Gamma+\Gamma'}{2}\right)Ue_i=u_i'C_1u_i=u_i'a_iu_i=a_i,
\]
where $a_i$ denotes the eigenvalue of $C_1$ with eigenvector $u_i$. Therefore, $\sigma_i=1/\sqrt{1-a_i^2}$. Finally, recall that $M = S_1'\Gamma'S_1(S_1'S_1)^{-1}$. By assumption, either $d=1$, and so, $S_1$ is a vector in $\mathbb{R}^n$ or $\Gamma$ is symmetric. Either way $S_1'\Gamma'S_1=S_1'(\Gamma+\Gamma')S_1/2=S_1'C_1S_1$. Therefore,
\begin{align*}
M & = S_1'C_1S_1(S_1'S_1)^{-1}=\left(\sum_{i,k=1}^d \frac{1}{\sqrt{1-a_i^2}}v_iu_i'C_1\frac{1}{\sqrt{1-a_k^2}}u_kv_k'\right)\left(\sum_{i,k=1}^d \frac{1}{\sqrt{1-a_i^2}}v_iu_i'\frac{1}{\sqrt{1-a_k^2}}u_kv_k'\right)^{-1}\\ & = \left(\sum_{i=1}^d \frac{1}{1-a_i^2}v_ia_iv_i'\right)\left(\sum_{i=1}^d \frac{1}{1-a_i^2}v_iv_i'\right)^{-1}=\sum_{i=1}^d a_iv_iv_i',
\end{align*}
where I am using the facts that $u_i$ is an eigenvector of $C_1$ with eigenvalue $a_i$ and that $\{u_i\}_{i=1}^d$ and $\{v_i\}_{i=1}^d$ are orthonormal sets of vectors. 

Although any $M$ and $S_1$ of the forms \eqref{eq:M_solution} and \eqref{eq:S_1_solution} satisfy the necessary optimality condition, not all such candidates are global minimizers of the KLDR. To find the global optima, I substitute the solutions to the first-order optimality conditions in the KLDR and select the solutions that minimize the KLDR. Multiplying equation \eqref{eq:FOC_d_S_1} from left by $S_1'$, I get
\[
I = S_1'S_1 - S_1'\Gamma S_1 M - S_1'\Gamma'S_1M' + S_1'S_1M'M.
\]
Computing the trace of the above equation and substituting the result in \eqref{eq:KLDR_M_S_1}, I get
\begin{align*}
\text{KLDR}(M,S_1)& =-\frac{1}{2}\log\text{det}\left(S_1'S_1\right)+\frac{1}{2}\tr\left(S_1'S_1\right)-\tr\left( M S_1' \Gamma S_1\right) + \frac{1}{2}\tr\left(M S_1'S_1M'\right)+\text{constant}\\ & = -\frac{1}{2}\log\text{det}\left(S_1'S_1\right)+\frac{1}{2}\tr(I)+\text{constant}.
\end{align*}
Therefore, the $M$ and $S_1$ pairs that minimize the KLDR are the ones that maximize the determinant of $S_1'S_1$. But since $S_1'S_1$ is a symmetric matrix with eigenvalues $\{1/(1-a_i^2)\}_{i=1}^d$, its determinant is equal to $\prod_{i=1}^d\frac{1}{1-a_i^2}$. Therefore, any $M$ and $S_1$ pair that minimize the KLDR are of the forms \eqref{eq:M_solution} and \eqref{eq:S_1_solution} with $\{a_i\}_{i=1}^d$ the top $d$ eigenvalues of $C_1$ in magnitude (with the possibility that some of the $a_i$ are equal). 

With the expressions for the pseudo-true $M$ and $S_1$ in hand, I can prove the theorem. 

\paragraph*{Part (a).} The forecasts given a model parameterized by matrices $M$, $D$, and $N$ are given by equation \eqref{eq:s_ahead_forecast_simplified}. Using the definition of  $S\equiv\Gamma_0^{\frac{1}{2}}N$ and the fact that $D'D$ can be taken to be identity matrix, I can write equation \eqref{eq:s_ahead_forecast_simplified} as follows:
\begin{equation*}
E^\theta_t[y_{t+s}] = \Gamma_0^{\frac{1}{2}}{S'}^{-1}DM^{s} D'S'\Gamma_0^{\frac{-1}{2}} y_{t-\tau}.
\end{equation*}
Note that for any matrix $S=(S_1\; S_2)$ that satisfies the first-order optimality condition with respect to $S_2$,
\[
S^{-1}=\begin{pmatrix}(S_1'S_1)^{-1}S_1'\\S_2'\end{pmatrix}.
\]
Therefore,
\[
S'^{-1}=\begin{pmatrix}S_1(S_1'S_1)^{-1} & S_2\end{pmatrix},
\]
and so
\begin{equation}
    S'^{-1}D=S_1(S_1'S_1)^{-1}.\label{eq:S_prime_inverse_D}
\end{equation}
The forecasts can thus be written only in terms of matrices $M$ and $S_1$ as follows:
\begin{align*}
     E^\theta_t[y_{t+s}] & = \Gamma_0^{\frac{1}{2}}S_1(S_1'S_1)^{-1}M^s S_1' \Gamma_0^{\frac{-1}{2}}.
\end{align*}
Substituting for $M$ and $S_1$ using \eqref{eq:M_solution} and \eqref{eq:S_1_solution}  and simplifying the resulting expression, I get
\[
E^\theta_t[y_{t+s}] = \Gamma_0^{\frac{1}{2}}\sum_{i=1}^da_i^su_iu_i'\Gamma_0^{\frac{-1}{2}}.
\]
Letting $p_i\equiv \Gamma_0^{\frac{-1}{2}}u_i$ and $q_i\equiv \Gamma_0^{\frac{1}{2}}u_i$ completes the proof of part (a).

\paragraph*{Part (b).} Equation  \eqref{eq:d-state-uncond-var} gives the variance-covariance matrix under a model parameterized by matrices $M$, $D$, and $N$. Using the definition of $S$ and setting $D'D=I$, equation \eqref{eq:d-state-uncond-var} can be written as follows:
\[
\text{Var}^\theta(y) = \Gamma_0^{\frac{1}{2}}\left({S'}^{-1}S^{-1}+{S^{-1}}'D\sum_{\tau=1}^\infty M^\tau {M'}^\tau D'S^{-1}\right)\Gamma_0^{\frac{1}{2}}.
\]
To prove part (b), I need to show that the terms in parentheses add up to the identity matrix. I start with the first term:
\begin{equation}
    {S'}^{-1}S^{-1} = (SS')^{-1}=(S_1S_1' + S_2S_2')^{-1}.\label{eq:SS_prime_inverse}
\end{equation}
The fact that $S_2'S_2=I$ implies that $S_2$ can be written as
\[
S_2 = \sum_{i=d+1}^n u_iw_i',
\]
where $u_i\in\mathbb{R}^n$ and $w_i\in\mathbb{R}^{n-d}$ for $i=d+1,\dots,n$, $\{u_i\}_{i=d+1}^n$ are orthonormal vectors, and $\{w_i\}_{i=d+1}^n$ constitutes an orthonormal basis for $\mathbb{R}^{n-d}$. On the other hand, the fact that $S_1'S_2=\mathbf{0}$ implies that $u_i'u_k=0$ for any $i\in\{1,\dots,d\}$ and $k\in\{d+1,\dots,n\}$. Therefore, $\{u_i\}_{i=1}^n$ constitutes an orthonormal basis for $\mathbb{R}^n$. Substituting for $S_1$ and $S_2$ in \eqref{eq:SS_prime_inverse}, I get
\[
{S'}^{-1}S^{-1} = \left(\sum_{i=1}^d\frac{1}{1-a_i^2}u_iu_i'+\sum_{i=d+1}^nu_iu_i'\right)^{-1}=\sum_{i=1}^d(1-a_i^2)u_iu_i'+\sum_{i=d+1}^nu_iu_i',
\]
where the second equality uses the fact that $\{u_i\}_{i=1}^n$ are orthonormal. Next consider the second term:
\begin{align*}
    {S^{-1}}'D\sum_{\tau=1}^\infty M^\tau {M'}^\tau D'S^{-1} & = S_1(S_1'S_1)^{-1}\sum_{\tau=1}^\infty M^\tau {M'}^\tau(S_1'S_1)^{-1}S_1'\\ & = \sum_{i=1}^d\sqrt{1-a_i^2}u_iv_i'\sum_{\tau=1}^\infty\sum_{k=1}^da_k^{2\tau}v_kv_k'\sum_{l=1}^d\sqrt{1-a_l^2}v_lu_l'\\ & = \sum_{i=1}^d(1-a_i^2)u_iu_i'\sum_{\tau=1}^\infty a_i^{2\tau}\\ &  = \sum_{i=1}^da_i^2u_iu_i',
\end{align*}
where the first equality uses \eqref{eq:S_prime_inverse_D} and the second equality is by \eqref{eq:M_solution} and \eqref{eq:S_1_solution}. Putting everything together,
\[
{S'}^{-1}S^{-1}+{S^{-1}}'D\sum_{\tau=1}^\infty M^\tau {M'}^\tau D'S^{-1}=\sum_{i=1}^d(1-a_i^2)u_iu_i'+\sum_{i=d+1}^nu_iu_i'+\sum_{i=1}^da_i^2u_iu_i'=\sum_{i=1}^nu_iu_i'=I,
\]
where the last equality follows the fact that $\{u_i\}_{i=1}^n$ is an orthonormal basis for $\mathbb{R}^n$.\hfill\qed

\subsection*{Proof of Proposition \ref{prop:decomposition}}
Recall that I have assumed (without loss of generality) that $\Gamma_{0}$ is non-singular. Since $C_1$ is symmetric, $\{u_i\}_{i=1}^d$ constitutes an orthonormal basis for $\mathbb{R}^n$, and so, $\Gamma_{0}^{\frac{-1}{2}}y_t$ can be expressed as 
\[
\Gamma_{0}^{\frac{-1}{2}}y_t = \sum_{i=1}^n \omega_{it}u_i,
\]
where $\omega_{it}\equiv u_i'\Gamma_{0}^{\frac{-1}{2}}y_t$. Therefore,
\[
y_t = \Gamma_{0}^{\frac{1}{2}}\sum_{i=1}^n \omega_{it}u_i = \sum_{i=1}^n \Gamma_{0}^{\frac{1}{2}}u_i u_i'\Gamma_{0}^{\frac{-1}{2}}y_t = \sum_{i=1}^n y^{(i)}_{t}q_i,
\]
where the last equality uses the definitions of $y^{(i)}_{t}$ and $q_i$.

The lag-one autocovariance of $y^{(i)}_t$ is given by
\begin{align*}
\mathbb{E}\left[ y^{(i)}_ty^{(i)}_{t-1}\right]={p_i}'\mathbb{E}[y_ty_{t-1}]p_i = {p_i}'\Gamma_1 p_i={p_i}'\left(\frac{\Gamma_1+\Gamma_1'}{2}\right) p_i={p_i}'\Gamma_0^{\frac{1}{2}}C_1 \Gamma_0^{\frac{1}{2}}p_i=u_i'C_1u_i,
\end{align*}
where the first equality uses the definition $y^{(i)}_{t}$, and the last equality uses the definition of $p_i$. The fact that $u_i$ is an eigenvector of $C_1$ implies $u_i'C_1u_i = a_iu_i'u_i = a_i$, where $a_i$ is the $i$th largest (in magnitude) eigenvalue of $C_1$. Moreover, $\mathbb{E}\left[ {y^{(i)}_t}^2\right]={p_i}'\Gamma_0 p_i=u_i'u_i=1$. Therefore,
\[
\rho_i\equiv \mathbb{E}\left[ y^{(i)}_ty^{(i)}_{t-1}\right]/\sqrt{\mathbb{E}\left[ {y^{(i)}_t}^2\right]}=a_i.
\]
The proposition follows the fact that $a_i$ is the $i$th largest eigenvalue of $C_1$ in magnitude.\hfill\qed

\subsection*{Proof of Proposition \ref{prop:d-state-comovement}}
I prove the result under the assumption that the top $D$ eigenvalues of the first autocorrelation matrix, $C_1$, are all distinct. This assumption is true for generic true processes. By Theorem \ref{thm:d-state-m-i-o}(a) (or Theorem \ref{thm:1-state-exp-erg-closed-form}), the forecasts of an agent who uses a pseudo-true $d$-state model $\theta$ are given by
\begin{equation}\label{eq:forecasts_comovement_proof}
E^{\theta}_t[y_{t+s}] = \sum_{i=1}^d{a_i}^{s}q_i{p_i}'y_t,
\end{equation}
where $a_i$ is the $i$th largest eigenvalue of $C_1$, $u_i$ denotes the corresponding eigenvector, normalized to have unit norm, $p_i\equiv\Gamma_0^{\frac{-1}{2}}u_i$, and $q_i\equiv\Gamma_0^{\frac{1}{2}}u_i$. Since the eigenvalues of $C_1$ are all distinct, the corresponding eigenvectors are unique (up to multiplicative constants). Therefore, all agents use the same values of $\{(a_i,p_i,q_i)\}_i$ to forecast.

Consider agent $j$ who is constrained to models of dimension $d_j$. The agent's optimal action given her pseudo-true $d$-state model is given by
\begin{align*}
x_{jt} & = E_t^{\theta_j}\left[\sum_{s=1}^\infty c_{js}'y_{t+s}\right] = \sum_{s=1}^\infty c_{js}'E_t^{\theta_j}\left[y_{t+s}\right]\\ & =\sum_{s=1}^\infty c_{js}'\sum_{i=1}^{d_j}{a_i}^{s}q_i{p_i}'y_t = \sum_{i=1}^{d_j}g_{ji}y_t^{(i)},
\end{align*}
where $\theta_j$ denotes agent $j$'s pseudo-true model, $y_{t}^{(i)}\equiv p_i'y_t$ as before, $g_{ji}\equiv \sum_{s=1}^\infty {a_i}^{s}c_{js}'q_i$ is a constant, which is a finite since $\{c_{js}\}_s$ is absolutely summable. Using vector notation, $x_t\equiv (x_{1t},\dots,x_{Jt})'\in\mathbb{R}^{J}$, I can write the above expression as $x_t = G y_{t}^{(1:D)}$, where $G\equiv \begin{pmatrix} g_1' & g_2' & \cdots &g_{J}'\end{pmatrix}'\in\mathbb{R}^{J\times D}$, $g_j \equiv \begin{pmatrix}g_{j1} & g_{j2} & \dots g_{jd_j} & 0 & \dots & 0\end{pmatrix}\in\mathbb{R}^{1\times D}$, and $y_{t}^{(1:D)} \equiv \begin{pmatrix} y_t^{(1)} & y_t^{(2)} & \cdots & y_t^{(D)}\end{pmatrix}'\in \mathbb{R}^D$.\hfill\qed

\subsection*{Proof of Proposition \ref{prop:d-state-over-under-reaction}}
I start by taking $v$ to be an arbitrary $n$-dimensional vector and computing the autocovariances of $v'y_t$ under the pseudo-true and true models. Define $w\equiv \Gamma_0^{\frac{1}{2}}v$. Under a pseudo-true $d$-state model~$\theta$,
\begin{align*}
    E^{\theta}\left[v'y_tv'y_{t-l}\right] & =v'E^{\theta}\left[y_ty_{t-l}'\right]v = v'E^{\theta}\left[E^{\theta}_{t-l}\left[y_t\right]y_{t-l}'\right]v\\ & = v'\sum_{i=1}^d{a_i}^lq_ip_i' E^{\theta}\left[y_{t-l}y_{t-l}'\right]v =\sum_{i=1}^d{a_i}^lv'q_ip_i'\Gamma_0v\\ & =\sum_{i=1}^d{a_i}^lv'\Gamma_0^{\frac{1}{2}}u_iu_i'\Gamma_0^{\frac{1}{2}}v=\sum_{i=1}^d{a_i}^lw'u_iu_i'w,
\end{align*}
where the first equality is by Theorem \ref{thm:linear_invariance}, the second equality follows the fact that the agent's subjective model satisfies the law of iterated expectations, the third and fifth equalities are by Theorem \ref{thm:d-state-m-i-o}(a) (or Theorem \ref{thm:1-state-exp-erg-closed-form} depending on the assumptions), and the fourth equality is by Theorem \ref{thm:d-state-m-i-o}(b) (or Theorem \ref{thm:1-state-covariance}). On the other hand, under the true model,
\begin{align*}
    \mathbb{E}[v'y_tv'y_{t-l}] & =v'\mathbb{E}[y_ty_{t-l}']v= v'\Gamma_lv = v'\left(\frac{\Gamma_l+\Gamma'_l}{2}\right)v=v'\Gamma_0^{\frac{1}{2}}C_l\Gamma_0^{\frac{1}{2}}v=w'C_lw.
\end{align*}

To prove part (a), set $v=p_1$, which implies $v'y_t=y_{t}^{(1)}$ and $w=\Gamma_0^{\frac{1}{2}}p_1=u_1$. Therefore,
\[
\left|E^{\theta}\left[y_{t}^{(1)}y_{t-l}^{(1)}\right]\right|=\left|\sum_{i=1}^d{a_i}^lu_1'u_iu_i'u_1\right|=\left|{a_1}^l\right| = |a_1|^l,
\]
for any pseudo-true model $\theta$. Furthermore,
\[
\left|\mathbb{E}\left[y_{t}^{(1)}y_{t-l}^{(1)}\right]\right|=\left|u_1'C_lu_1\right|\leq \rho(C_l) u_1'u_1 = \rho(C_l)\leq \rho(C_1)^l=|a_1|^l,
\]
where the second inequality is using the assumption that the true process is exponentially ergodic, and the last equality is due to the fact that $a_1$ is the eigenvalue of $C_1$ largest in magnitude. On the other hand, by Theorem \ref{thm:d-state-m-i-o}(b) (or Theorem \ref{thm:1-state-covariance}), the variance of $y_{t}^{(1)}$ is the same under the true and pseudo-true $d$-state models. Therefore, the agent overestimates the magnitude of $y_{t}^{(1)}$'s autocorrelation at all lags.

In part (b), set $v=p_n$, which implies $v'y_t=y_{t}^{(n)}$ and $w=\Gamma_0^{\frac{1}{2}}p_n=u_n$. Thus,
\[
\left|E^{\theta}\left[y_{t}^{(n)}y_{t-l}^{(n)}\right]\right|=\left|\sum_{i=1}^d{a_i}^lu_n'u_iu_i'u_n\right|=0,
\]
for any pseudo-true model $\theta$, where I am using the fact that $\{u_i\}_{i=1}^n$ is an orthonormal basis and the assumption that $d<n$. Hence, the agent underestimates the magnitude of $y_{t}^{(n)}$'s autocorrelation at all lags, regardless of the true autocorrelation of $y_{t}^{(n)}$.\hfill\qed

\subsection*{Proof of Proposition \ref{propapp:sufficient_for_exp_erg}}

I first prove a useful lemma, which offers a canonical representation of the autocorrelation matrices for stochastic processes that can be represented as in \eqref{eq:true_model_VAR_app}:\footnote{Versions of this result have previously appeared in the control and time-series literatures. For early examples, see \cite{ho1966effective} and \cite{Akaike}.}

\begin{lemmaapp}\label{lem:Akaike}
Suppose $\{C_l\}_{l}$ are the autocorrelation matrices of a non-degenerate $n$-dimensional stationary ergodic process that can be represented as in \eqref{eq:true_model_VAR_app} with $f_t\in\mathbb{R}^m$. There exists a convergent $m\times m$ matrix $\mathbb{F}$ with $\Vert \mathbb{F}\Vert_2 \leq 1$ and a semi-orthogonal $m\times n$ matrix $\mathbb{H}$ such that
\begin{equation}
C_l = \mathbb{H}'\left(\frac{\mathbb{F}^{l}+{\mathbb{F}'}^l}{2}\right)\mathbb{H}.\label{eq:C-state-space}
\end{equation}
Conversely, for any positive integers $m\geq n$, $m\times m$ convergent matrix $\mathbb{F}$ with $\Vert \mathbb{F}\Vert_2 \leq 1$, and semi-orthogonal $m\times n$ matrix $\mathbb{H}$, there exists an $n$-dimensional stationary ergodic process with autocorrelation matrices $\{C_l\}_{l}$ of the form \eqref{eq:C-state-space}, which can be represented as in \eqref{eq:true_model_VAR_app}.\footnote{Matrix $\mathbb{H}\in\mathbb{R}^{m\times n}$ is semi-orthogonal if $\mathbb{H}'\mathbb{H}=I$, where $I$ denotes the $n\times n$ identity matrix.}
\end{lemmaapp}

\begin{proof}
The assumption that the process is non-degenerate requires $m\geq n$, an assumption I maintain throughout the first part of the proof. Given representation \eqref{eq:true_model_VAR_app}, the autocovariance matrices are given by
\[
\Gamma_l = \mathbb{E}\left[y_t y'_{t-l}\right] = H'F^l\mathbb{E}\left[f_{t-l} f'_{t-l}\right]H=H'F^lVH,
\]
where $V \equiv \mathbb{E}\left[f_tf'_t\right]$ is the unique solution to the following discrete-time Lyapunov equation:
\begin{equation}\label{eq:Lyapunov_true}
V = FVF' + \Sigma,
\end{equation}
and $\Sigma$ is the variance-covariance matrix of $\epsilon_t$. Therefore, 
\[
C_l = \left(H'VH\right)^{\frac{-1}{2}}\left(\frac{H'F^lVH+H'V{F'}^lH}{2}\right)\left(H'VH\right)^{\frac{-1}{2}}.
\]
Matrix $V$ is positive semidefinite; it is positive definite if the representation in \eqref{eq:true_model_VAR_app} is minimal.\footnote{See, for instance, \cite{Akaike}.} Without loss of generality, I assume that that is the case. Define $\mathbb{H}' \equiv \left(H'VH\right)^{\frac{-1}{2}}H'V^{\frac{1}{2}}$ and $\mathbb{F} \equiv V^{\frac{-1}{2}}F V^{\frac{1}{2}}$. Then
\begin{equation}
C_l = \mathbb{H}'\left(\frac{\mathbb{F}^{l}+{\mathbb{F}'}^l}{2}\right)\mathbb{H}.\label{eq:C_tau_true_VAR}
\end{equation}
Note that since $F$ is a convergent matrix, so is $\mathbb{F}$. Substituting $\mathbb{F}=V^{\frac{-1}{2}}F V^{\frac{1}{2}}$ in equation \eqref{eq:Lyapunov_true}, I get
\[
1- \mathbb{F}\mathbb{F}' = V^{\frac{-1}{2}}\Sigma V^{\frac{-1}{2}}.
\]
Therefore, since $\Sigma$ is positive semidefinite, the spectral radius of $\mathbb{F}\mathbb{F}'$ is weakly smaller than one. This implies that $\Vert \mathbb{F}\Vert_2\leq 1$. On the other hand,
\[
\mathbb{H}'\mathbb{H} = \left(H'VH\right)^{\frac{-1}{2}}H'VH\left(H'VH\right)^{\frac{-1}{2}}=I.
\]
That is, $\mathbb{H}$ is a (full-rank) semi-orthogonal matrix. This proves the first part of the lemma.

I next argue that given a convergent matrix $\hat{\mathbb{F}}\in\mathbb{R}^{m\times m}$ with $\Vert \hat{\mathbb{F}}\Vert_2\leq 1$ and a semi-orthogonal matrix $\hat{\mathbb{H}}\in\mathbb{R}^{m\times n}$ with $m\geq n$, there exists a stationary ergodic process such that the corresponding autocorrelation matrices are given by \eqref{eq:C_tau_true_VAR} with $\mathbb{F}=\hat{\mathbb{F}}$ and $\mathbb{H}=\hat{\mathbb{H}}$. Given any such $\hat{\mathbb{F}}$ and $\hat{\mathbb{H}}$, let $F=\hat{\mathbb{F}}$, $H=\hat{\mathbb{H}}$, and $\Sigma=I-\hat{\mathbb{F}}\hat{\mathbb{F}}'$. The solution to the Lyapunov equation \eqref{eq:Lyapunov_true} is then given by $V=I$. Therefore, $\mathbb{F}=F=\hat{\mathbb{F}}$ and $\mathbb{H}=\hat{\mathbb{H}}(\hat{\mathbb{H}}'\hat{\mathbb{H}})^{\frac{-1}{2}}=\hat{\mathbb{H}}$, where in the last equality I am using the assumption of semi-orthogonality of $\hat{\mathbb{H}}$. By construction, then the autocorrelation matrices of a process of the form \eqref{eq:true_model_VAR_app} with matrices $F$, $H$, and $\Sigma$ as above are given by \eqref{eq:C_tau_true_VAR} with $\mathbb{F}=\hat{\mathbb{F}}$ and $\mathbb{H}=\hat{\mathbb{H}}$.
\end{proof}

\paragraph*{Proof of Proposition \ref{propapp:sufficient_for_exp_erg}.} By Lemma \ref{lem:Akaike}, $C_l = \frac{1}{2}\mathbb{H}'\left(\mathbb{F}^{l}+{\mathbb{F}'}^l\right)\mathbb{H}$, where $\mathbb{H}' \equiv \left(H'VH\right)^{\frac{-1}{2}}H'V^{\frac{1}{2}}$, $\mathbb{F} \equiv V^{\frac{-1}{2}}F V^{\frac{1}{2}}$, and $V \equiv \mathbb{E}\left[f_tf'_t\right]$ is the variance-covariance of $f_t$. Note that since the variance-covariance of $f_t$ is normalized to be the identity matrix, $V=I$, $\mathbb{F}=F$, and $\mathbb{H}=H$. Recall that vector $y_t$ does not contain any redundant observables (which are linear combinations of other observables). This assumption, together with the assumption that $H$ is a rank-$m$ matrix, ensures that $H$ is an invertible $m\times m$ matrix. Therefore, by Lemma \ref{lem:Akaike}, $\mathbb{H}=H$ is an orthogonal matrix. Thus,
\begin{equation}
\rho(C_l) = \rho\left(\mathbb{H}'\left(\frac{\mathbb{F}^l+{\mathbb{F}'}^l}{2}\right)\mathbb{H}\right)=\rho\left(\frac{\mathbb{F}^l+{\mathbb{F}'}^l}{2}\right)= \rho\left(\frac{F^l+{F'}^l}{2}\right)\label{eq:app_rho_C_l}
\end{equation}
for all $l$. But since the spectral radius of a symmetric matrix equals its spectral norm,
\begin{equation}
\rho\left(\frac{F^{l}+{F'}^l}{2}\right)= \left\Vert\frac{F^{l}+{F'}^l}{2}\right\Vert_2\leq \frac{1}{2}\left\Vert F^l\right\Vert_2 + \frac{1}{2}\left\Vert {F'}^l\right\Vert_2 = \left\Vert F^l\right\Vert_2\leq \left\Vert F\right\Vert_2^l.\label{eq:bound_rho_F_F_proof}
\end{equation}
Therefore, $\rho(C_l)\leq \left\Vert F\right\Vert_2^l$. On the other hand, by equations \eqref{eq:app_rho_C_l} and \eqref{eq:bound_rho_F_F_proof}, $\rho(C_1) = \frac{1}{2}\left\Vert F+{F'}\right\Vert_2 = \left\Vert F\right\Vert_2$, where the second equality is by assumption. Thus, $\rho(C_l)\leq \left\Vert F\right\Vert_2^l=\rho(C_1)^l$, and the process is exponentially ergodic.\hfill\qed

\subsection*{Proof of Proposition \ref{propapp:incomplete_info}}

I first state and prove a useful lemma:
\begin{lemmaapp}\label{lem:1-state-not-exp-erg}
Suppose $C_1$ has a unique and simple eigenvalue $\lambda$ with $|\lambda|=\rho(C_1)>0$, and let $u$ denote the corresponding eigenvector normalized to have $u'u=1$.\footnote{The assumption that $\lambda$ is unique and simple is not necessary for the result. The result generalizes to arbitrary matrices $C_1$ with $\rho(C_1)\neq 0$ by replacing $u'C_2u$ with the maximum of $u'C_2 u$ over all unit-norm eigenvectors $u$ of $C_1$ with eigenvalues $\lambda$ such that $|\lambda|=\rho(C_1)$.} If $u'C_2 u>\rho(C_1)^2$, then the agent's forecasts in any pseudo-true one-state model are given by \eqref{eq:1-state-s_ahead_forecast_body} with a tuple $(a,\eta,p,q)$ such that $\eta>0$.
\end{lemmaapp}

\begin{proof}
Define $C(a,\eta)$ as in the proof of Lemma \ref{lemapp:1-state-stationary-ergodic}. As in the proof of Lemma \ref{lemapp:1-state-stationary-ergodic}, I present the argument under the assumption that the largest eigenvalue of $C(a,\eta)$ is simple at the point $(a^*,\eta^*)$ that maximizes $\lambda_{\text{max}}(C(a,\eta))$.\footnote{See footnote \ref{fn:subdiff} for how the argument can be generalized.} I start by proposing a candidate solution to the problem of maximizing $\lambda_{\text{max}}(\Omega(a,\eta))$ at which $\eta=0$ and argue that the candidate does not satisfy the necessary first-order optimality conditions. Setting $\eta=0$ in equations \eqref{eq:FOC_a} and \eqref{eq:FOC_eta}, I get
\begin{align*}
    & \frac{\partial \lambda_{\text{max}}(\Omega(a,\eta))}{\partial a}\bigg|_{\eta=0}=-2a+2u'_{\text{max}}(a C_1) C_1u_{\text{max}}(a C_1),\\ 
    & \frac{\partial \lambda_{\text{max}}(\Omega(a,\eta)))}{\partial \eta}\bigg|_{\eta=0} = 2a^2-2(1+a^2)\lambda_{\text{max}}(a C_1)+2a^2u'_{\text{max}}(a C_1) C_2u_{\text{max}}(a C_1),
\end{align*}
where I am using the fact that $ C = a C_1$ when $\eta=0$. Any solution to $\partial \lambda_{\text{max}}(\Omega(a,\eta))/\partial a|_{\eta=0}=0$ satisfies $a=\lambda$, where $\lambda=\lambda_{\text{min}}( C_1)$ if $\lambda_{\text{max}}( C_1)\leq 0$, $\lambda=\lambda_{\text{max}}( C_1)$ if $\lambda_{\text{min}}( C_1)\geq 0$, and $\lambda \in\{\lambda_{\text{max}}( C_1),\lambda_{\text{min}}( C_1)\}$ otherwise. Evaluating $\lambda_{\text{max}}(\Omega(a,\eta))$ at $a=\lambda$ and $\eta=0$, I get $\lambda_{\text{max}}(\Omega(\lambda,0))=\lambda^2$. Therefore, for the solution $(a,\eta)=(\lambda,0)$ to the first-order condition $\partial \lambda_{\text{max}}(\Omega(a,\eta))/\partial a=0$ to be a maximizer of $\lambda_{\text{max}}(\Omega(a,\eta))$, it must be the case that $\lambda$ is the eigenvalue of $ C_1$ largest in magnitude and $u=u_{\text{max}}(a C_1)$ is a corresponding eigenvector normalized such that $u'u=1$. Substituting in the expression for $\partial \lambda_{\text{max}}(\Omega(a,\eta))/\partial \eta|_{\eta=0}$, I get
\begin{align*}
    & \frac{\partial \lambda_{\text{max}}(\Omega(a,\eta))}{\partial \eta}\bigg|_{a=\lambda,\eta=0} = 2\rho( C_1)^2\left(u' C_2u-\rho( C_1)^2\right)>0,
\end{align*}
where the inequality follows the assumption that $u'C_2u>\rho(C_1)^2$. This implies that the pair $\eta=0$ and $a=\lambda$ does not constitute a local maximizer of $\lambda_{\text{max}}(\Omega(a,\eta))$. Since this pair is the only candidate with $\eta=0$ that may satisfy the first-order conditions, in any pseudo-true one-state model, $\eta>0$. This establishes the lemma.
\end{proof}

\paragraph*{Proof of Proposition \ref{propapp:incomplete_info}.} Let $\sigma^2$ denote the variance of $y_t$. By the argument in the proof of Lemma \ref{lem:Akaike}, the lag-$l$ autocorrelation of $y_t$ is given by
\[
C_l = \mathbb{H}' \left(\frac{\mathbb{F}^l+{\mathbb{F}'}^l}{2}\right)\mathbb{H},
\]
where $\mathbb{F}\equiv V^{\frac{-1}{2}}FV^{\frac{1}{2}}$, $\mathbb{H}'\equiv \left(H'VH\right)^{\frac{-1}{2}}H'V^{\frac{1}{2}}$, and $V$ is the solution to the discrete-time Lyapunov equation \eqref{eq:Lyapunov_true}. Since $F$ and $\Sigma$ are diagonal matrices, so is $V$. Therefore, $\mathbb{F}=F$. On the other hand, by Lemma \ref{lem:Akaike}, $\mathbb{H}$ is a semi-orthogonal matrix. Therefore, $\mathbb{H}'\mathbb{H}=1$, and so, $C_l = \sum_{i=1}^m w_i \alpha_i^l$, where $w_i\equiv\mathbb{H}_i^2\geq 0$, $\sum_{i=1}^m w_i=1$, and $\alpha_i$ is the $i$th diagonal element of $F$.  That is, $C_l^{\frac{1}{l}}$ is equal to the weighted $l$-norm of vector $(\alpha_1,\dots,\alpha_m)$ with weights $w=(w_1,\dots,w_m)$.

Since the representation in \eqref{eq:true_model_VAR_app} is minimal, $w_i>0$ for all $i$, and all $\alpha_i$ are distinct. If that were not the case, there would exist some $\tilde{m}<m$ such that $C_l = \sum_{i=1}^{\tilde{m}}\tilde{w}_i {\tilde{\alpha}}_i^l$ for some non-negative weights $\tilde{w}_i$ that sum up to one and some $\tilde{\alpha}_i\in(-1,1)$. Consider the process $\widetilde{\mathbb{P}}$ represented as in \eqref{eq:true_model_VAR_app} with $F=\text{diag}(\tilde{\alpha}_1,\dots,\tilde{\alpha}_{\tilde{m}})$, $\epsilon_t\sim\mathcal{N}(0,\Sigma)$, $\Sigma=I-FF'$, and $H=\sigma\diag(\sqrt{\tilde{w}_1},\dots,\sqrt{\tilde{w}_{\tilde{m}}})$. By the argument in the proof of Lemma \ref{lem:Akaike}, $\widetilde{\mathbb{P}}$ has the same autocorrelation matrices as $\mathbb{P}$. Moreover, both $\mathbb{P}$ and $\widetilde{\mathbb{P}}$ are mean-zero and normal and both have variance $\sigma^2$. Therefore, $\mathbb{P}$ and $\widetilde{\mathbb{P}}$ are observationally equivalent, a contradiction to the assumption that the representation I started with was minimal.

Next, note that, by the generalized mean inequality, $C_l^{\frac{1}{l}}>C_1$ for all $l\geq 2$, where the strictness of the inequality follows the facts that $w_i>0$ for all $i$ and all $\alpha_i$ are distinct. In particular, $u'C_2 u=C_2>C_1^2=\rho(C_1)^2$, where I am using the fact that $y_t$ is a scalar. Thus, by Lemma \ref{lem:1-state-not-exp-erg}, $\eta>0$. To see why $\eta<1$, recall that by Theorem \ref{thm:1_state_general}, the $(a,\eta)$ pair maximizes
\[
\Omega(\tilde{a},\tilde{\eta})=-\frac{\tilde{a}^2(1-\tilde{\eta})^2}{1-\tilde{a}^2\tilde{\eta}^2} + \frac{2(1-\tilde{\eta})(1-\tilde{a}^2\tilde{\eta})}{1-\tilde{a}^2\tilde{\eta}^2}\sum_{\tau=1}^\infty \tilde{a}^\tau\tilde{\eta}^{\tau-1} C_\tau.
\]
But $\Omega(\tilde{a},1)=0<C_1^2=\Omega(C_1,0)$ for any $\tilde{a}$. Therefore, $\eta=1$ cannot be part of the description of a pseudo-true one-state model. Finally, $a\in(1,1)$ by Lemma \ref{lemapp:1-state-stationary-ergodic}. The proposition then follows Theorem \ref{thm:1_state_general} by noting that $q{p}'=1$ whenever $y_t$ is a scalar. \hfill\qed
\end{document}